\documentclass[10pt,pdftex,prd,aps,amsfonts,onecolumn,showpacs,showkeys,byrevtex,nofootinbib]{revtex4}

\usepackage[english]{babel}
\usepackage{amsmath}
\usepackage{amssymb}
\usepackage{bm}
\usepackage[final]{pdfpages}
\usepackage{graphicx}

\usepackage[pdftitle={Relativistic Cosmological Perturbation Theory and the
  Evolution of Small-Scale Inhomogeneities}, pdfauthor={P. G. Miedema},
   pdfsubject={Gauge Invariant Cosmological Density Perturbations,
     Growth of Small-Scale Density Perturbations, Population III Stars},
   pdfstartview={FitH}, pdfkeywords={Cosmological Perturbations,
     Population III stars},
bookmarks, bookmarksnumbered, colorlinks, linkcolor={black},
urlcolor={black}, citecolor={black}]{hyperref}

\newcommand{\nul}{{\scriptscriptstyle(0)}}
\newcommand{\een}{{\scriptscriptstyle(1)}}

\begin{document}

\title{Relativistic Cosmological Perturbation Theory \\ and the Evolution of Small-Scale Inhomogeneities}

\author{P.\ G.\ Miedema}
\email{pieter.miedema@gmail.com}
\affiliation{Netherlands Defence Academy\\
   Hogeschoollaan 2, \\ NL-4818 CR  Breda, The Netherlands}

\date{March 24, 2014}

\begin{abstract}
  It is shown that a first-order relativistic perturbation theory for
  the open, flat or closed Friedmann-Lema\^\i tre-Robertson-Walker
  universe admits one, and only one, gauge-invariant quantity which
  describes the perturbation to the energy density and which becomes
  equal to the usual Newtonian energy density in the non-relativistic
  limit. The same holds true for the perturbation to the particle
  number density.  These facts exclude all definitions of
  gauge-invariant quantities used to describe density perturbations in
  former theories.  Using these two new quantities, a manifestly
  covariant and gauge-invariant cosmological perturbation theory,
  adapted to non-barotropic equations of state for the pressure, has
  been developed.  The new theory is valid for all scales since metric
  gradients do not occur in the final evolution equations.  The new
  theory has an exact non-relativistic limit with a time-independent
  Newtonian potential.  The usual Newtonian perturbation theory is
  inadequate to study the evolution of density perturbations.

  In the radiation-dominated era, perturbations in the particle number
  density are gravitationally coupled to perturbations in the total
  energy density, irrespective of the nature of the particles.  This
  implies that structure formation can commence only after decoupling
  of matter and radiation.

  After decoupling of matter and radiation density perturbations
  evolve diabatically, i.e., they exchange heat with their
  environment.  This heat loss of a perturbation may enhance the
  growth rate of its mass sufficiently to explain stellar formation in
  the early universe, a phenomenon not understood, as yet, without the
  additional assumption of the existence of Cold Dark Matter.  This
  theoretical observation is the main result of this article.

\end{abstract}

\maketitle

\section{\label{sec:introduction}Introduction}

Since measurements of the fundamental parameters of our universe are
very precise today, cosmology is nowadays a mature branch of
astrophysics.  Despite advances in observational as well as
theoretical cosmology, there is, as yet, no correct cosmological
perturbation theory which describes unequivocally the evolution of
density perturbations.  This is mainly due to the fact that the
perturbed Einstein equations and conservation laws have, next to the
physical solutions, also spurious solutions, the so-called gauge modes
which obscure the physics.  There is up till now no unique way to deal
with these gauge modes.  In this article it will be shown that there
is one and only one correct way to get rid of the gauge modes.

Lifshitz~\cite{lifshitz1946} and Lifshitz and Khalatnikov~\cite{c15}
initiated the research of cosmological perturbations.  They simply
discarded the gauge modes.  A better strategy, devised by
Bardeen~\cite{c13}, is to construct so-called \emph{gauge-invariant}
quantities, i.e., quantities that have the same value in \emph{all}
coordinate systems.  This is done by taking linear combinations of
gauge-dependent variables, which occur naturally in the perturbed
Einstein equations and conservation laws.  In his seminal article
Bardeen demonstrated that the use of gauge-invariant quantities in the
construction of a perturbation theory ensures that it is free of
spurious solutions.  Kodama and Sasaki~\cite{kodama1984} elaborated
and clarified the work of Bardeen.  The article of Bardeen has
inspired the pioneering works of Ellis
\textit{et~al.}~\cite{Ellis1,Ellis2,ellis-1998} and Mukhanov
\textit{et~al.}~\cite{mfb1992,Mukhanov-2005}.  These researchers
proposed alternative perturbation theories using gauge-invariant
quantities which differ from the ones used by Bardeen.

The general notion is that the `gauge issue' of cosmology has been
resolved by now and that first-order cosmological perturbations are
nowadays well-understood~\cite{Christopherson:2011ra,Malik:2008im,
  Knobel:2012wa,Peter:2013woa}.  This, however, is not the case.  In
fact, the perturbation theories developed thus far, have four
important problems.  Firstly, none of the gauge-invariant perturbation
theories in the current literature is based on definitions for
gauge-invariant quantities that allow for the non-relativistic limit.
This fact makes it impossible to put a precise physical interpretation
on the gauge-invariant quantities used in these theories.  The usual
limit for the Hubble function
$H\rightarrow0$~\cite{mfb1992,Mukhanov-2005} in an attempt to arrive
at the Poisson equation of the Newtonian Theory of Gravity does not
apply, since this limit violates the background Einstein equations and
conservation laws (\ref{subeq:einstein-flrw}).  Furthermore, the
`limit' $H\rightarrow0$ yields a `Poisson equation' with a
\emph{time-dependent} potential, whereas the gravitational potential
is in the Newtonian Theory of Gravity independent of time,
(\ref{FRW6gi-flat-newt}).  Moreover, the fact that a gauge-invariant
perturbation theory yields the Newtonian equation
(\ref{eq:delta-standard})~\cite{Ellis1,Ellis2,ellis-1998} can
\emph{not} be considered as a non-relativistic limit, since the
standard equation (\ref{eq:delta-standard}) follows from the General
Theory of Relativity and, as a consequence, its general solution
contains gauge modes. This will be demonstrated in detail in
Section~\ref{sec:stand-theory}\@. In the approach of
Bardeen~\cite{c13} two different definitions of gauge-invariant
density perturbations are given, which become equal to each other and
equal to the gauge-dependent density perturbation in the small-scale
limit.  The assumption in Bardeen's theory is that gauge-dependent
quantities become gauge-invariant in the small-scale limit.  As will
be demonstrated in Section~\ref{sec:newt-limit} gauge-dependent
quantities are also gauge-dependent in the non-relativistic limit,
implying that the small-scale limit is \emph{not} equivalent to the
non-relativistic limit.  Secondly, the assumption that on sufficiently
small scales the Newtonian perturbation theory is valid, is not
correct. In contrast to what is asserted in the
literature~\cite{2013JCAP...01..002C}, there is \emph{no} relation
whatsoever between the relativistic perturbation theory,
(\ref{subeq:final-rad}) and (\ref{final-dust}), and the Newtonian
perturbation theory (\ref{eq:delta-standard}).  The only relation
between the General Theory of Relativity and the Newtonian Theory of
Gravity is the non-relativistic limit, Section~\ref{sec:newt-limit}.
Thirdly, both the metric
approach~\cite{c13,kodama1984,mfb1992,Mukhanov-2005} as well as the
covariant formalism~\cite{Ellis1,Ellis2,ellis-1998} have gradient
terms~\cite{2012arXiv1206.1478M} in the governing equations, which
make the evolution equations unnecessarily complicated.  Finally, it
follows from thermodynamics and the conservation laws that density
perturbations in the era \emph{after} decoupling of matter and
radiation are \emph{diabatic} (`non-adiabatic') so that they may
exchange heat with their environment.  The phenomenon of heat
exchange, which will prove to be crucial for structure formation after
decoupling, has not been taken into account in former cosmological
perturbation theories.  In other words, in all current perturbation
theories it is inaccurately assumed that pressure perturbations do not
play a role \emph{after} decoupling, see the remarks below
(\ref{state-mat}) and (\ref{eq:dT-constant}).

In this article a completely new cosmological perturbation theory
(\ref{subeq:final}) for the open ($K=-1$), flat ($K=0$) or closed
($K=+1$) Friedmann-Lema\^\i tre-Robertson-Walker (\textsc{flrw})
universe is presented which does not have the problems of former
theories.  In the derivation of the new perturbation theory, neither
the metric approach nor the covariant formalism will be applied.
Instead, simple linear combinations (\ref{e-n-gi}) are used and it
will be shown that they are the only possible linear combinations.  A
perturbation theory based on these gauge-invariant quantities has an
exact non-relativistic limit, i.e., the energy density perturbation
$\varepsilon^{\text{gi}}_\een$ and the particle number density
perturbation $n^{\text{gi}}_\een$ become equal to their Newtonian
counterparts (\ref{eq:poisson}) and (\ref{eq:newt-ngi}), respectively,
and the perturbed energy constraint equation becomes equal to the
Poisson equation with a potential which is \emph{independent of time}
(\ref{FRW6gi-flat-newt}), as it should be in the Newtonian theory.

Equations (\ref{subeq:final}) are exact, since in their derivation no
assumptions or approximations (other than linearisation of the
Einstein equations and conservation laws) have been made.  The absence
of metric gradients has the effect that equations (\ref{subeq:final})
are valid for \emph{all} scales, so that there is no need to
distinguish between sub-horizon and super-horizon perturbations.

The new theory (\ref{subeq:final}) has the property that both the
evolution equations and their solutions are invariant under general
infinitesimal coordinate transformations (\ref{func}).  In other
words, equations (\ref{subeq:final}) constitute a \emph{manifestly
  covariant} (i.e., the equations take the same \emph{form} in
\emph{all} coordinate systems) and \emph{gauge-invariant} (i.e., the
outcome takes the same \emph{value} in \emph{all} coordinate systems)
perturbation theory.  Consequently, there is no need to choose a
particular coordinate system to solve these equations and interpret
their outcome.

Since a realistic equation of state (\ref{eq:equat-of-state-pressure})
is incorporated in equations (\ref{subeq:final}) and since these
equations are applicable to open, flat or closed \textsc{flrw}
universes they will be referred to as the \emph{generalised
  cosmological perturbation theory}.  The upshot of this theory is a
possible explanation of primeval stars, the so-called (hypothetical)
Population~\textsc{iii} stars, independent of the existence of Cold
Dark Matter (\textsc{cdm}).

\section{\label{sec:our-ins}Outline of the Generalised Cosmological Perturbation Theory}

\subsection{\label{subsec:EoS}Equation of State}

Barotropic equations of state for the pressure $p=p(\varepsilon)$,
where $\varepsilon$ is the energy density, are commonly used in
cosmology~\cite{Ellis-Maartens-MacCallum-2012}.  In particular, the
linear barotropic equation of state $p=w\varepsilon$, with
$w=\tfrac{1}{3}$ in the radiation-dominated era and $w=0$ in the era
after decoupling of matter and radiation, describes very well the
global evolution of an \emph{un}perturbed \textsc{flrw} universe.  As
will become clear in Section~\ref{sec:diabatic} barotropic equations
of state do not take into account pressure perturbations \emph{after}
decoupling of matter and radiation.  Since these pressure
perturbations are important for star formation in the early universe,
Section~\ref{sec:pop-iii-stars}, realistic equations of state are
needed.  From thermodynamics it is known that both the energy density
$\varepsilon$ and the pressure $p$ depend on the independent
quantities $n$ and $T$, i.e.,
\begin{equation}
  \label{eq:es-p-T}
  \varepsilon=\varepsilon(n,T), \qquad
      p=p(n,T),
\end{equation}
where $n$ is the particle number density and $T$ the temperature.
Since $T$ can, in principle, be eliminated from these equations of
state, a computationally more convenient equation of state for the
pressure will be used, namely
\begin{equation}
  \label{eq:equat-of-state-pressure}
    p=p(n,\varepsilon).
\end{equation}
This, non-barotropic, equation of state and its partial derivatives
has been included in the generalised cosmological perturbation theory
(\ref{subeq:final}).

\subsection{\label{subsec:Sys-of-Ref}Choosing a System of Reference}

In order to construct a cosmological perturbation theory a suitable
system of reference must be chosen.  Due to the general covariance of
the Einstein equations and conservation laws, Einstein's gravitational
theory is invariant under a general coordinate transformation
$x^\mu\rightarrow \hat{x}^\mu(x^\nu)$, implying that preferred
coordinate systems do not exist (Weinberg~\cite{weinberg-2008},
Appendix~B).  In particular, the linearised Einstein equations and
conservation laws are invariant under a general linear coordinate
transformation
\begin{equation}
     x^\mu \rightarrow  x^\mu - \xi^\mu(t,\bm{x}), \label{func}
\end{equation}
where $\xi^\mu(t,\bm{x})$ are four arbitrary, first-order
(infinitesimal) functions of the time ($x^0=ct$) and space
[$\bm{x}=(x^1,x^2,x^3)$] coordinates. Since preferred systems of
reference do not exist and since the result of the calculations,
namely the generalised cosmological perturbation theory
(\ref{subeq:final}), is manifestly covariant and gauge-invariant, one
may use any suitable and convenient coordinate system to derive
equations (\ref{subeq:final}).

In order to put an accurate interpretation on the gauge-invariant
quantities (\ref{subeq:gi-quant}) one needs the non-relativistic
limit.  In the Newtonian Theory of Gravity space and time are strictly
separated, implying that in this theory \emph{all} coordinate systems
are essentially \emph{synchronous}.  In view of the non-relativistic
limit, it would, therefore, be convenient to use synchronous
coordinates~\cite{lifshitz1946,c15,I.12} in the background as well as
in the perturbed universe.  In these coordinates the metric of
\textsc{flrw} universes has the form
\begin{equation}
  \label{eq:metric-flrw}
 g_{00}=1, \qquad g_{0i}=0, \qquad g_{ij}=-a^2(t)\tilde{g}_{ij}(\bm{x}), 
\end{equation}
where $a(t)$ is the scale factor of the universe, $g_{00}=1$ indicates
that coordinate time is equal to proper time, $g_{0i}=0$ is the
synchronicity condition (see Landau and Lifshitz~\cite{I.12},
\S~84) and $\tilde{g}_{ij}$ is the metric of the
three-dimensional maximally symmetric subspaces of constant time.  The
functions $\xi^\mu$ of the general infinitesimal
transformation (\ref{func}) become
\begin{equation}
  \label{eq:synchronous}
\xi^0=\psi(\bm{x}), \qquad
     \xi^i=\tilde{g}^{ik}\partial_k \psi(\bm{x})
       \int^{ct} \frac{{\text{d}}\tau}{a^2(\tau)} + \chi^i(\bm{x}),
\end{equation}
if only transformations between synchronous coordinates are allowed.
The four functions $\psi(\bm{x})$ and $\chi^i(\bm{x})$ cannot be fixed
since the four coordinate conditions $g_{00}=1$ and $g_{0i}=0$ have
already exhausted all four degrees of freedom, see Weinberg~\cite{c8},
Section~7.4 on coordinate conditions.

An advantage of a synchronous system of reference is that the
space-space components of the four-dimensional Ricci curvature tensor
$R^\mu{}_\nu$ is broken down into two parts such that one part
contains exclusively all time-derivatives of the space-space
components of the metric tensor and the second part $R^i{}_j$ is
precisely the Ricci curvature tensor of the three-dimensional
subspaces of constant time, see (97.10) in the textbook of Landau and
Lifshitz~\cite{I.12}.  This fact will be fully exploited to derive and
simplify the perturbed Einstein equations for scalar perturbations.

\subsection{\label{subsec:Eli-gauge-modes}Identifying the Key
  Variables to solve the Gauge Problem of Cosmology}

In order to find the gauge-invariant quantities which are the energy
density and particle number density perturbations all gauge-dependent
variables which play a key role in the evolution of density
perturbations should first be identified.  Next, the perturbed
Einstein equations and conservation laws should be rewritten in a form
such that these key variables stand out clearly.  This will now be
done.

The obvious variables are the (gauge-dependent) perturbations to the
energy density $\varepsilon_\een$ and the particle number density
$n_\een$.  Local density perturbations induce locally a space-time
curvature, so that a third quantity that plays a role in the evolution
of density perturbations is the local perturbation $R^\mu_{\een\nu}$
to the global four-dimensional Ricci curvature tensor $R^\mu{}_\nu$.
Since synchronous coordinates are used, the space-space components of
the perturbed four-dimensional Ricci tensor $R^\mu_{\een\nu}$ is
broken down into the perturbed Ricci tensor $R^i_{\een j}$ of the
spaces of constant time and a part which contains all time-derivatives
of the spatial parts of the perturbed metric, $h^i{}_j$, see
(\ref{basis-3}).  Perturbations do not evolve if there is no fluid
flow, i.e., if the \emph{spatial} component $\bm{u}_\een$ of the fluid
four-velocity vanishes.  Therefore, the three perturbed spatial
components $u^i_\een$ of the fluid four-velocity are of importance.
Finally, since the total energy content of the universe influences its
global expansion, a local density perturbation may locally affect the
global expansion.  Therefore, the perturbation $\theta_\een$ to the
expansion scalar $\theta\equiv u^\mu{}_{;\mu}$, where $u^\mu$ is the
fluid four-velocity ($u^\mu u_\mu=1$), is also important.
Consequently, a first step in the derivation of a cosmological
perturbation theory is to rewrite the perturbed Einstein equations and
conservation laws in a form such that $\varepsilon_\een$, $n_\een$,
the perturbations to the spatial curvature $R^i_{\een j}$,
(\ref{eq:ricci-1}), its trace $R_\een$, (\ref{driekrom}), the
perturbation $\theta_\een$, (\ref{fes5}), to the expansion scalar, and
the three spatial components $u^i_\een$ of the fluid four-velocity
stand out clearly.  The result is the set of perturbed Einstein
equations and conservation laws (\ref{subeq:basis}).  In this set the
constraint equations (\ref{basis-1}) and (\ref{basis-2}) contain time
derivatives of the metric no higher than first-order, whereas the
dynamical equations (\ref{basis-3}) contain time derivatives of the
metric no higher than second-order.  This is a mathematical property
of the Einstein equations, see Weinberg~\cite{c8}, Section~7.5 on the
Cauchy problem and Landau and Lifshitz~\cite{I.12}, \S~95 on the
peculiarities of the structure of the Einstein
equations. Consequently, this form must first be obtained before any
further calculations can be done.  For easy manipulation, mixed upper
and lower tensor indexes are introduced.

\subsection{\label{subsec:evo-scalar}Evolution Equations for Scalar Perturbations}

The Einstein equations and conservation laws are now written in a
suitable form (\ref{subeq:basis}) to apply the two
facts~\cite{lifshitz1946,c15,I.12,York1974,SteWa,Stewart} that tensor,
vector and scalar perturbations are, in first-order, independent of
each other and that only scalar perturbations are coupled to density
perturbations, see Section~\ref{sec:decomp-h-u}\@.  As a consequence,
in the study of the evolution of density perturbations only scalar
perturbations need to be considered.  Making full use of the
properties of scalar perturbations, the perturbation equations can
considerably be simplified.  In fact, the perturbation equations can
be rewritten in the same form as the background equations, as follows.

Since only the \emph{irrotational} part of the spatial part of the
fluid four-velocity is coupled to scalar perturbations, one may
replace the spatial part of the fluid four-velocity by its divergence
$\vartheta_\een$, (\ref{fes5}), so that the three momentum
conservation laws (\ref{basis-5}) can be recast in one evolution
equation (\ref{FRW5}) for the divergence $\vartheta_\een$. This, in
turn, implies that the energy constraint equation (\ref{basis-1}) can
be rewritten as an algebraic equation in terms of $\varepsilon_\een$,
$R_\een$, $\theta_\een$ and $\vartheta_\een$ by eliminating
$\dot{h}^k{}_k$ with the help of (\ref{fes5}), namely equation
(\ref{con-sp-1}).  Now, the three perturbed momentum constraint
equations (\ref{basis-2}) can be rewritten as one evolution equation
(\ref{FRW6}) for the local perturbation $R_\een$, (\ref{driekrom}), to
the global spatial curvature $R_\nul$, (\ref{eq:glob-curve}).
Finally, it follows that, just as in the unperturbed case, the trace
of the dynamical equations (\ref{basis-3}), expressed in terms of
$\varepsilon_\een$, $p_\een$, $R_\een$, $\theta_\een$ and
$\vartheta_\een$, is equivalent to the time-derivative of the energy
constraint equation (\ref{con-sp-1}) together with the momentum
constraint equation (\ref{FRW6}) and the conservation laws
(\ref{FRW4})--(\ref{FRW4a}).  Since, in addition, the off-diagonal
dynamical equations are not coupled to scalar perturbations, the full
set of equations (\ref{basis-3}) is not needed anymore.  Consequently,
using the variables $\varepsilon_\een$, $n_\een$, $R_\een$,
$\theta_\een$ and $\vartheta_\een$, the complete set of perturbation
equations (\ref{subeq:basis}) can, for scalar perturbations, be
written as an initial value problem (\ref{subeq:pertub-flrw}) with
four \emph{ordinary} differential equations and one \emph{algebraic}
equation to be obeyed by the initial values. Note the remarkable
similarity of the sets (\ref{subeq:einstein-flrw}) and
(\ref{subeq:pertub-flrw}): the set (\ref{subeq:pertub-flrw}) is
precisely the perturbed counterpart of the system
(\ref{subeq:einstein-flrw}).  The systems of equations
(\ref{subeq:einstein-flrw}) and (\ref{subeq:pertub-flrw}) describe
exclusively the evolution of scalar perturbations in \textsc{flrw}
universes and are, therefore, crucial for the understanding of the
evolution of density perturbations in \textsc{flrw} universes.  They
form the basis of the generalised cosmological perturbation theory
(\ref{subeq:final}) which will be derived in
Section~\ref{sec:flat-pert}\@.

\subsection{\label{subsec:sol-gauge-problem}Solution to the Gauge
  Problem of Cosmology}

Equations (\ref{subeq:pertub-flrw}) have, next to physical solutions,
also the non-physical gauge modes (\ref{subeq:gauge-dep}) as
solutions.  Since the perturbed Einstein equations and conservation
laws (\ref{subeq:basis}) have been rewritten in the form
(\ref{subeq:pertub-flrw}) which exclusively describe the evolution of
scalar perturbations, the gauge problem of cosmology can be solved in
a unique and straightforward way by observing the following important
fact:
\begin{itemize}
\item From the systems (\ref{subeq:einstein-flrw}) and
  (\ref{subeq:pertub-flrw}) it follows that exactly three
  \emph{independent} scalars (\ref{eq:scalars-flrw}), namely the
  energy density $\varepsilon$, the particle number density $n$ and
  the expansion scalar $\theta$, play a role in the evolution of
  density perturbations.
\end{itemize}
As a consequence, the first-order perturbations $\varepsilon_\een$,
$n_\een$ and $\theta_\een$ to the three scalars
(\ref{eq:scalars-flrw}) are the only candidates which can be used to
construct gauge-invariant quantities.  Thus, there exist two, and only
two, unique non-zero gauge-invariant quantities
$\varepsilon^{\text{gi}}_\een$ and $n^{\text{gi}}_\een$, defined by
(\ref{e-n-gi}), which could be the perturbations to the energy density
and particle number density, respectively.

It remains to be shown that $\varepsilon^{\text{gi}}_\een$ and
$n^{\text{gi}}_\een$ are indeed the perturbations to the energy
density and particle number density, respectively.  To that end one
has to investigate the basic equations (\ref{subeq:pertub-flrw}) for
scalar perturbations together with the definitions (\ref{e-n-gi}) in
the so-called \emph{non-relativistic limit}.  In this limit, which
will be defined in exact terms in Section~\ref{sec:newt-limit}, the
set (\ref{subeq:pertub-flrw}) combined with the definitions
(\ref{e-n-gi}) reduce to the Newtonian Theory of Gravity, i.e., the
Newtonian results (\ref{eq:poisson}) and (\ref{eq:newt-ngi}) show up.
Therefore, $\varepsilon^{\text{gi}}_\een$ and $n^{\text{gi}}_\een$ are
the real, physical, energy density perturbation and the particle
number density perturbation, respectively.  This solves unambiguously
the long standing gauge problem of cosmology.

\subsection{\label{subsec:deriv-evo-equa}Evolution Equations for
  Cosmological Density Perturbations}

Having established that $\varepsilon^{\text{gi}}_\een$ and
$n^{\text{gi}}_\een$ are the one and only quantities that are the
energy density and particle number density perturbations,
respectively, the last step will be to derive evolution equations for
these quantities.  To that end, the set (\ref{subeq:pertub-flrw}) is
first rewritten in the set (\ref{subeq:pertub-gi}) for the four
independent variables $\varepsilon_\een$, $n_\een$, $\vartheta_\een$
and $R_\een$ by eliminating $\theta_\een$ from equations
(\ref{FRW6})--(\ref{FRW4a}) using the algebraic equation
(\ref{con-sp-1}). Next, the quantities (\ref{e-n-gi}) are expressed in
these four variables, namely (\ref{subeq:pertub-gi-e-n}).  With the
help of the procedure given in the Appendix, the system
(\ref{subeq:pertub-gi}) for the four independent variables
$\varepsilon_\een$, $n_\een$, $\vartheta_\een$ and $R_\een$ will be
rewritten in the system (\ref{subeq:final}) for the two independent
and gauge-invariant quantities defined by (\ref{eq:contrast}).  Note
that the original system (\ref{subeq:pertub-gi}) is of
\emph{fourth}-order, whereas the new system (\ref{subeq:final}) is of
\emph{third}-order.  This is due to the fact that the gauge modes
(\ref{subeq:gauge-dep}) are, next to physical modes, solutions of the
original system (\ref{subeq:pertub-gi}).  Since gauge-invariant
quantities
$\delta_\varepsilon\equiv\varepsilon^{\text{gi}}_\een/\varepsilon_\nul$
and $\delta_n\equiv n^{\text{gi}}_\een/n_\nul$ are used in the
derivation of the new system (\ref{subeq:final}) from the original
system (\ref{subeq:pertub-gi}), one degree of freedom, namely the
gauge function $\xi^0(t,\bm{x})$ has disappeared altogether.  As a
result, the generalised cosmological perturbation theory
(\ref{subeq:final}) is a \emph{manifestly covariant and
  gauge-invariant} perturbation theory. Consequently, equations
(\ref{subeq:final}), which are valid for \emph{all} scales, describe
unequivocally the evolution of density perturbations in \textsc{flrw}
universes.

\subsection{\label{sec:results}New Results and Implications of the
  Generalised Cosmological Perturbation Theory}

\subsubsection{\label{sec:phys-res}Physical Results}

The generalised cosmological perturbation theory (\ref{subeq:final})
yields that perturbations in the particle number density are
\emph{gravitationally} coupled to perturbations in the total energy
density (\ref{fir-ord}), \emph{irrespective} of the nature of the
particles (i.e., ordinary matter or \textsc{cdm}) and independent of
the scale of the perturbations, Section~\ref{sec:rad-dom-era}\@.  This
prevents \textsc{cdm} perturbations from contracting faster than
ordinary matter in the radiation-dominated era.  As a consequence,
structure formation can commence only \emph{after} decoupling of
matter and radiation.

In the radiation-dominated era of \textsc{flrw} universes density
perturbations evolve adiabatically, Section~\ref{sec:diabatic}\@, and
small-scale density perturbations in a \emph{flat} \textsc{flrw}
universe oscillate with an \emph{increasing} amplitude according to
(\ref{dc-small}).

The definitions (\ref{subeq:gi-quant}) imply that first-order
\emph{local} density perturbations do not affect the \emph{global}
Hubble expansion of the universe.  In other words, in first-order
there is no back-reaction on the global expansion.  However, a local
density perturbation induces a local spatial curvature $R_\een$, as
follows from equations (\ref{subeq:pertub-flrw}).

For an equation of state $p=p(n,\varepsilon)$, the combined First and
Second Law of Thermodynamics and the conservation laws (\ref{FRW2})
and (\ref{FRW2a}) imply that in non-static \textsc{flrw} universes
density perturbations in the era after decoupling of matter and
radiation evolve \emph{diabatically} (`non-adiabatically'), i.e., they
exchange heat with their environment.  In this era heat loss of a
perturbation has, in addition to gravity, a more or less favourable
effect on its growth rate, depending on the scale.  For large-scale
perturbations heat loss is unimportant and gravity is the main cause
of contraction.  Therefore, the generalised cosmological perturbation
theory corroborates the results for large-scale perturbations
(\ref{eq:new-dust-53-adiabatic}) of former perturbation theories in
which heat loss has not been taken into account.  Small-scale
perturbations benefit more from heat loss during their evolution than
large-scale perturbations do.  It is shown in
Section~\ref{sec:hier-struc} that for perturbations with scales around
$6.5\,\text{pc}$ gravity and heat loss combine optimally,
Figure~\ref{fig:collapse}, resulting in fast growing density
perturbations.  This may shed new light on the evolution of
small-scale inhomogeneities in the universe.

\subsubsection{\label{sec:math-imp}Mathematical Implications}

The uniqueness of $\varepsilon^{\text{gi}}_\een$ and
$n^{\text{gi}}_\een$ defined by (\ref{e-n-gi}) combined with the fact
that these quantities become equal to their Newtonian counterparts in
the non-relativistic limit, rule out all definitions of
gauge-invariant quantities used in former perturbation
theories~\cite{Christopherson:2011ra,Malik:2008im,Knobel:2012wa,Peter:2013woa,c13,
  kodama1984,Ellis1,Ellis2,ellis-1998,mfb1992,Mukhanov-2005}.

As will be shown in Section~\ref{sec:stand-theory}, for \emph{linear}
barotropic equations of state, $p=w\varepsilon$, thermodynamics and
the conservation laws for \textsc{flrw} universes require that the
factor of proportionality, $w$, is constant.  In an attempt to take
the pressure perturbations into account Christopherson
\textit{et~al.}~\cite{2013JCAP...01..002C} use the linear barotropic
equation of state with a time-dependent factor of proportionality,
$p=w(\eta)\varepsilon$, where $\eta$ is the conformal time
$c\text{d}t=a(\eta)\text{d}\eta$.  This is not correct.

From the General Theory of Relativity, i.e., (\ref{subeq:pertub-gi}),
it follows that the general solution of the standard perturbation
equation (\ref{eq:delta-standard}) contains gauge modes.
Consequently, (\ref{eq:delta-standard}) does not describe the
evolution of density perturbations.

Metric gradient terms~\cite{2012arXiv1206.1478M}, which are explicitly
present in all former perturbation
theories~\cite{Christopherson:2011ra,Malik:2008im,Knobel:2012wa,
  Peter:2013woa,c13,kodama1984,Ellis1,Ellis2,ellis-1998,mfb1992,Mukhanov-2005},
do not occur in the generalised cosmological perturbation theory
(\ref{subeq:final}), since in the perturbation equations
(\ref{subeq:pertub-flrw}) for scalar perturbations \emph{all metric
  perturbations and their derivatives} are contained in three
quantities, namely $R_\een$, $\theta_\een$ and $\vartheta_\een$, which
are determined by their own propagation equations
(\ref{subeq:pertub-flrw}).  As a consequence, the approach presented
in this article is substantially less complicated than all former
perturbation theories.

The gauge can \emph{not} be fixed completely, since the Newtonian
Theory of Gravity has also a gauge freedom
(\ref{eq:gauge-trans-newt}), namely the freedom to choose spatial
coordinates and to shift time coordinates. This is in accordance with
the fact that with (\ref{eq:metric-flrw}) all degrees of freedom are
used.  The gauge freedom (\ref{eq:gauge-trans-newt}) in the Newtonian
Theory of Gravity follows in a natural way (\ref{eq:non-rel-lim}) from
the gauge freedom (\ref{func}) combined with (\ref{eq:synchronous}) in
Einstein's General Theory of Gravity.  As a result of this residual
gauge freedom, gauge-dependent variables are also gauge-dependent in
the non-relativistic limit. The residual gauge freedom in the
Newtonian Theory of Gravity rules out the use of the so-called
longitudinal, or Newtonian, gauge in which the metric is diagonal and
has only two independent components, since this particular gauge
(which is essentially \emph{synchronous} since $g_{0i}=0$) is supposed
to fix the coordinates completely.

As has already been pointed out by Lifshitz and
Khalatnikov~\cite{lifshitz1946,c15} and Landau and
Lifshitz~\cite{I.12}, \S115, co-moving coordinates can \emph{not} be
used in the perturbed universe, since the spatial part
$\bm{u}_\een(t,\bm{x})$ of the fluid four-velocity is not equal to
zero.  Density perturbations for which $\bm{u}_\een(t,\bm{x})=0$ do
not evolve, (\ref{FRW6gi-flat-newt}).  Only in the non-relativistic
limit the system of reference becomes co-moving.

\section{\label{sec:rev-LK}Derivation of the Generalised Cosmological Perturbation Theory}

In this section the generalised cosmological perturbation theory of
cosmological density perturbations is derived for the open, flat or
closed \textsc{flrw} universe.  The background equations
(\ref{subeq:einstein-flrw}) and first-order perturbation equations
(\ref{subeq:basis}), from which the generalised cosmological
perturbation theory will be derived, can be deduced from the set of
Einstein equations (97.11)--(97.13) from the textbook~\cite{I.12} of
Landau and Lifshitz and the conservation laws $T^{\mu\nu}{}_{;\nu}=0$.
For a detailed derivation of the basic equations
(\ref{subeq:einstein-flrw}) and (\ref{subeq:basis}),
see~\cite{Miedema-Leeuwen-2003}, Sections~II and~III.

\subsection{\label{sec:basic-equations}Basic Equations}

\subsubsection{\label{sec:back-eq}Background Equations}

The complete set of zeroth-order Einstein equations and conservation
laws for an open, flat or closed \textsc{flrw} universe filled with a
perfect fluid with energy-momentum tensor
\begin{equation}
  \label{eq:en-mom-tensor}
  T^{\mu\nu}=(\varepsilon+p)u^\mu u^\nu - p g^{\mu\nu}, \qquad 
p=p(n,\varepsilon),
\end{equation}
is, in synchronous coordinates, given by 
\begin{subequations}
\label{subeq:einstein-flrw}
\begin{alignat}{3}
   3H^2 & =\tfrac{1}{2}R_\nul+\kappa\varepsilon_\nul+\Lambda, \qquad &
             \kappa& =8\pi G_{\text{N}}/c^4, & \label{FRW3}\\
  \dot{R}_\nul & =-2HR_\nul, & & \label{FRW3a}\\
  \dot{\varepsilon}_\nul & = -3H\varepsilon_\nul(1+w), & 
           w & \equiv p_\nul/\varepsilon_\nul, & \label{FRW2} \\
  \vartheta_\nul & =0, & & \\
  \dot{n}_\nul & = -3Hn_\nul. & &  \label{FRW2a}
\end{alignat}
\end{subequations}
The $G_{0i}$ constraint equations and the $G_{ij}$, $i\neq j$,
dynamical equations are identically satisfied. The $G_{ii}$ dynamical
equations are equivalent to the time-derivative of the $G_{00}$
constraint equation (\ref{FRW3}).  Therefore, the $G_{ij}$ dynamical
equations need not be taken into account.  In
equations~(\ref{subeq:einstein-flrw}) $\Lambda$ is the cosmological
constant, $G_{\text{N}}$ the gravitational constant of the Newtonian
Theory of Gravity, $c$ the speed of light.  Notice that $w$ is only a
shorthand notation for the quotient $p_\nul/\varepsilon_\nul$, it does
not mean that the equation of state is barotropic.  An over-dot
denotes differentiation with respect to $ct$ and the sub-index $(0)$
refers to the background, i.e., unperturbed, quantities. Furthermore,
$H\equiv\dot{a}/a$ is the Hubble function which is equal to
$H=\tfrac{1}{3}\theta_\nul$, where $\theta_\nul$ is the background
value of the expansion scalar $\theta\equiv u^\mu{}_{;\mu}$ with
$u^\mu\equiv c^{-1}U^\mu$ the four-velocity, normalised to unity
($u^\mu u_\mu=1$).  A semicolon denotes covariant differentiation with
respect to the background metric $g_{\nul\mu\nu}$.  The \emph{spatial}
part of the background Ricci curvature tensor $R^i_{\nul j}$ and its
trace $R_\nul$ are given by
\begin{equation}
  \label{eq:glob-curve}
    R^i_{\nul j}=-\dfrac{2K}{a^2}\delta^i{}_j,
      \qquad R_\nul=-\dfrac{6K}{a^2}, \qquad  K=-1,0,+1,
\end{equation}
where $R_\nul$ is the global spatial curvature.  The quantity
$\vartheta_\nul$ is the three-divergence of the spatial part of the
four-velocity $u_\nul^\mu$. For an isotropically expanding universe
the four-velocity is $u^\mu_\nul=\delta^\mu{}_0$, so that
$\vartheta_\nul=0$, implying that in the background the coordinate
system is co-moving.

From the system~(\ref{subeq:einstein-flrw}) one may infer that the
evolution of an unperturbed \textsc{flrw} universe is determined by
exactly three independent scalars, namely
\begin{equation}\label{eq:scalars-flrw}
    \varepsilon = T^{\mu\nu} u_\mu u_\nu, \qquad
    n = N^\mu u_\mu, \qquad
    \theta  = u^\mu{}_{;\mu},
\end{equation}
where $N^\mu\equiv nu^\mu$ is the cosmological particle current
four-vector, which satisfies the particle number conservation law
$N^\mu{}_{;\mu}=0$, (\ref{FRW2a}), see Weinberg~\cite{weinberg-2008},
Appendix~B\@.  As will become clear in Section~\ref{sec:unique}, the
quantities (\ref{eq:scalars-flrw}) and their first-order counterparts
play a key role in the evolution of cosmological density
perturbations.

\subsubsection{\label{sec:pert-eq}Perturbation Equations}

The complete set of first-order Einstein equations and conservation
laws for the open, flat or closed \textsc{flrw} universe is, in
synchronous coordinates, given by
\begin{subequations}
\label{subeq:basis}
\begin{align}
  & H\dot{h}^k{}_k +
\tfrac{1}{2}R_\een =
         -\kappa\varepsilon_\een,     \label{basis-1} \\
  & \dot{h}^k{}_{k|i}-\dot{h}^k{}_{i|k} =
         2\kappa(\varepsilon_\nul + p_\nul) u_{\een i}, \label{basis-2} \\
  & \ddot{h}^i{}_j+ 3H\dot{h}^i{}_j+
       \delta^i{}_j H\dot{h}^k{}_k+2R^i_{\een j}=
     -\kappa\delta^i{}_j(\varepsilon_\een-p_\een),
         \label{basis-3} \\
  & \dot{\varepsilon}_\een + 3H(\varepsilon_\een+p_\een)+
       (\varepsilon_\nul +p_\nul)\theta_\een=0,
               \label{basis-4}  \\
  & \frac{1}{c}\frac{{\text{d}}}{{\text{d}} t}
   \Bigl[(\varepsilon_\nul+p_\nul) u^i_\een\Bigr]-
       g^{ik}_\nul p_{\een|k}+5H(\varepsilon_\nul+p_\nul) u^i_\een=0,
            \label{basis-5} \\
  & \dot{n}_\een+3Hn_\een+n_\nul\theta_\een = 0,  \label{basis-6}
\end{align}
\end{subequations}
where $h_{\mu\nu}\equiv -g_{\een\mu\nu}$ with $h_{00}=0$, $h_{0i}=0$
is the perturbed metric, $h^i{}_j=g_\nul^{ik}h_{kj}$, and
$g_\nul^{ij}$ is the unperturbed background metric
(\ref{eq:metric-flrw}) for an open, flat or closed \textsc{flrw}
universe. Quantities with a sub-index $(1)$ are the first-order
counterparts of the background quantities with a sub-index $(0)$.  A
vertical bar denotes covariant differentiation with respect to
$g_{\nul ij}$.

The first-order perturbation to the pressure is given by the perturbed
equation of state
\begin{equation}
\label{eq:p1}
p_\een=p_nn_\een+p_\varepsilon\varepsilon_\een, \qquad
  p_n\equiv\left(\dfrac{\partial p}{\partial n}\right)_\varepsilon, \quad
p_\varepsilon\equiv\left(\dfrac{\partial p}{\partial \varepsilon}\right)_n,
\end{equation}
where $p_n(n,\varepsilon)$ and $p_\varepsilon(n,\varepsilon)$ are the
partial derivatives of the equation of state $p(n,\varepsilon)$.

The first-order perturbation to the spatial part of the Ricci tensor
(\ref{eq:glob-curve}) reads
\begin{equation}
  \label{eq:ricci-1}
     R^i_{\een j}\equiv
     (g^{ik}R_{kj})_\een=
    g^{ik}_\nul R_{\een kj}+\tfrac{1}{3}R_\nul h^i{}_j.
\end{equation}
Using Lifshitz' formula [see Lifshitz and Khalatnikov~\cite{c15},
equation (I.3) and Weinberg~\cite{c8}, equation (10.9.1)]
\begin{equation}
    \Gamma^k_{\een ij}=
   -\tfrac{1}{2} g^{kl}_\nul
     (h_{li|j}+h_{lj|i}-h_{ij|l}), \label{con3pert}
\end{equation}
and the contracted Palatini identities [see~\cite{c15}, equation~(I.5)
and~\cite{c8}, equation (10.9.2)]
\begin{equation}
    R_{\een ij} =
\Gamma^k_{\een ij|k}-\Gamma^k_{\een ik|j},
    \label{palatini}
\end{equation}
one finds, using $g_\nul^{ij}h^k{}_{i|j|k}=g_\nul^{ij}h^k{}_{i|k|j}$,
for the trace of (\ref{eq:ricci-1})
\begin{equation}
   R_\een =
   g_\nul^{ij} (h^k{}_{k|i|j}-h^k{}_{i|k|j}) +
     \tfrac{1}{3}R_\nul h^k{}_k.   \label{driekrom}
\end{equation}
Expression (\ref{driekrom}) is the local perturbation to the global
spatial curvature $R_\nul$ due to a local density perturbation.

Finally, $\theta_\een$ is the first-order perturbation to the
expansion scalar $\theta\equiv u^\mu{}_{;\mu}$.  Using that
$u^\mu_\nul=\delta^\mu{}_0$, one gets
\begin{equation}
  \theta_\een=
  \vartheta_\een-\tfrac{1}{2}\dot{h}^k{}_k, \qquad
  \vartheta_\een\equiv u^k_{\een|k}, 
  \label{fes5}
\end{equation}
where $\vartheta_\een$ is the divergence of the spatial part of the
perturbed four-velocity $u^\mu_\een$.  The quantities (\ref{driekrom})
and (\ref{fes5}) play an important role in the derivation of the
manifestly covariant and gauge-invariant perturbation
theory~(\ref{subeq:final}).  Since 
$u^i_\een(t,\bm{x})\neq0$, the coordinate system can \emph{not} be
co-moving in the perturbed universe.

\subsection{\label{sec:decomp-h-u}Decomposition of the Metric and
  the Spatial Part of the Fluid Four-Velocity}

York~\cite{York1974}, Stewart and Walker~\cite{SteWa} and
Stewart~\cite{Stewart} showed that any symmetric second rank tensor,
and hence the perturbation tensor $h_{ij}$, can uniquely be decomposed
into three parts, i.e.,
\begin{equation}
  h^i{}_j  =h^i_{\parallel j} + h^i_{\perp j} + h^i_{\ast j}, \label{eq:h123}
\end{equation}
where the scalar, vector and tensor perturbations are denoted by
$\parallel$, $\perp$ and $\ast$, respectively.  The constituents have
the properties
\begin{equation}
 h^k_{\perp k}=0, \qquad  h^k_{\ast k}=0, \qquad  h^k_{\ast i|k}=0. \label{eq:prop-hij}
\end{equation}
Moreover, York and Stewart demonstrated that the components
$h^i_{\parallel j}$ can be written in terms of two independent
potentials $\phi(t,\bm{x})$ and $\zeta(t,\bm{x})$, namely
\begin{equation}
  h^i_{\parallel j} =
      \frac{2}{c^2}(\phi\delta^i{}_j+\zeta^{|i}{}_{|j}).
        \label{decomp-hij-par}
\end{equation}
In Section~\ref{sec:newt-limit} it will be shown that, for a
\emph{flat} \textsc{flrw} universe in the non-relativistic limit, the
potential $\phi$ becomes independent of time, the Newtonian potential
is $\varphi(\bm{x})\equiv\phi(\bm{x})/a^2(t_0)$ and the potential
$\zeta(\bm{x},t)$ does not play a role anymore.

Finally, Stewart also proved that the spatial part of the perturbed
four-velocity $\bm{u}_\een$ can uniquely be decomposed into two parts
\begin{equation}
   \bm{u}_{\een} = \bm{u}_{\een\parallel} +
  \bm{u}_{\een\perp}, \label{eq:decomp-u}
\end{equation}
where the constituents have the properties
\begin{equation} 
  \bm{\tilde{\nabla}}\cdot\bm{u}_\een=\bm{\tilde{\nabla}}\cdot\bm{u}_{\een\parallel}, \qquad
\bm{\tilde{\nabla}}\times\bm{u}_\een=\bm{\tilde{\nabla}}\times\bm{u}_{\een\perp}, \label{eq:nabla-u}
\end{equation}
with $\bm{\tilde{\nabla}}^i\equiv\tilde{g}^{ij}\partial_j$, the
generalised vector differential operator.

The three different kinds of perturbations will now be considered
according to the decompositions (\ref{eq:h123}) and
(\ref{eq:decomp-u}), and their properties (\ref{eq:prop-hij}) and
(\ref{eq:nabla-u}).

For tensor perturbations the properties (\ref{eq:prop-hij}) imply that
$R_{\een\ast}=0$, as follows from (\ref{driekrom}).  Using this and
(\ref{eq:prop-hij}), equations (\ref{basis-1})--(\ref{basis-3}) imply
that tensor perturbations are not coupled to $\varepsilon_\een$,
$p_\een$ and $\bm{u}_\een$.

The perturbed Ricci tensor, $R_{\een ij}$, being a symmetric second
rank tensor, should obey the decomposition (\ref{eq:h123}) with the
properties (\ref{eq:prop-hij}), namely $R^k_{\een\perp k}=0$. This
implies with (\ref{driekrom}) that $h^{ij}_\perp$ must obey
$h^{kl}_{\perp |k|l}=0$, in addition to $h^k_{\perp k}=0$.  From
(\ref{basis-1}) and the trace of (\ref{basis-3}) it follows that
vector perturbations are not coupled to $\varepsilon_\een$ and
$p_\een$.  Raising the index $i$ of equations (\ref{basis-2}) with
$g_\nul^{ij}$, one finds that these equations read for vector
perturbations
\begin{equation}\label{basis-2-raise}
  \dot{h}^{kj}_{\perp{|k}}+2Hh^{kj}_{\perp{|k}}=
  2\kappa(\varepsilon_\nul+p_\nul)u^j_{\een},
\end{equation}
where it is used that $\dot{g}^{ij}_\nul=-2Hg^{ij}_\nul$.  Taking the
covariant derivative of (\ref{basis-2-raise}) with respect to the
index $j$ one finds with the additional property $h^{kl}_{\perp
  |k|l}=0$ that equations (\ref{basis-2-raise}) reduce to
$u^j_{\een|j}=0$, implying with (\ref{eq:nabla-u}) that the rotational
part $\bm{u}_{\een\perp}$ is coupled to vector perturbations.

Since both $h^k_{\parallel k}\neq0$ and $R_{\een\parallel}\neq0$,
scalar perturbations are coupled to $\varepsilon_\een$ and $p_\een$.
It will now be demonstrated that $\bm{u}_{\een\parallel}$ is coupled
to scalar perturbations, by showing that equations (\ref{basis-2})
require that the rotation of $\bm{u}_\een$ vanishes, if the metric is
of the form (\ref{decomp-hij-par}).  Differentiating (\ref{basis-2})
covariantly with respect to the index $j$ and substituting
(\ref{decomp-hij-par}) yields
\begin{equation}
    2\dot{\phi}_{|i|j}+\dot{\zeta}^{|k}{}_{|k|i|j}-\dot{\zeta}^{|k}{}_{|i|k|j} =
         \kappa c^2(\varepsilon_\nul + p_\nul) u_{\een i|j}.
    \label{eq:feiko1}
\end{equation}
Interchanging~$i$ and~$j$ and subtracting the result
from~(\ref{eq:feiko1}) one gets
\begin{equation}
    \dot{\zeta}^{|k}{}_{|i|k|j}-\dot{\zeta}^{|k}{}_{|j|k|i} =
         -\kappa c^2(\varepsilon_\nul + p_\nul)
         (u_{\een i|j}-u_{\een j|i}).
    \label{dR0i2-rot}
\end{equation}
By rearranging the covariant derivatives, (\ref{dR0i2-rot}) can be
cast in the form
\begin{align}
   (\dot{\zeta}^{|k}{}_{|i|k|j} & -\dot{\zeta}^{|k}{}_{|i|j|k}) -
    (\dot{\zeta}^{|k}{}_{|j|k|i}-\dot{\zeta}^{|k}{}_{|j|i|k})
   + (\dot{\zeta}^{|k}{}_{|i|j}-\dot{\zeta}^{|k}{}_{|j|i})_{|k} 
        =-\kappa c^2(\varepsilon_\nul + p_\nul)
         (u_{\een i|j}-u_{\een j|i}).
     \label{eq:verwissel}
\end{align}
Using the expressions for the commutator of second order covariant
derivatives (Weinberg~\cite{c8}, Chapter~6, Section~5)
\begin{equation}
\label{eq:commu-Riemann}
   A^i{}_{j|p|q}-A^i{}_{j|q|p} =
      A^i{}_k R^k_{\nul jpq} - A^k{}_j R^i_{\nul kpq}, \qquad 
   B^i{}_{|p|q}-B^i{}_{|q|p} = B^k R^i_{\nul kpq},
\end{equation}
and substituting the background Riemann tensor for the
three-spaces of constant time
\begin{equation}\label{eq:Riemann}
    R^i_{\nul jkl}=K\left(\delta^i{}_k \tilde{g}_{jl}-
         \delta^i{}_l\tilde{g}_{jk}\right), \qquad K=-1,0,+1,
\end{equation}
one finds that the left-hand sides of equations (\ref{eq:verwissel})
vanish identically, implying that the rotation of $\bm{u}_\een$ is
zero.  Therefore, only $\bm{u}_{\een\parallel}$ is coupled to scalar
perturbations.

\subsection{\label{sec:scalar-pert}First-order Equations for Scalar Perturbations}

Since scalar perturbations, i.e., perturbations in $\varepsilon_\een$
and $p_\een$, are only coupled to $h^i_{\parallel j}$ and
$u^i_{\een\parallel}$, one may replace in
(\ref{subeq:basis})--(\ref{fes5}) $h^i{}_j$ by $h^i_{\parallel j}$ and
$u^i_\een$ by $u^i_{\een\parallel}$, to obtain perturbation equations
which exclusively describe the evolution of scalar perturbations. From
now on, only scalar perturbations are considered, and the subscript
$\parallel$ will be omitted.  Using the decompositions (\ref{eq:h123})
and (\ref{eq:decomp-u}) and the properties (\ref{eq:prop-hij}) and
(\ref{eq:nabla-u}), one can rewrite the evolution equations for scalar
perturbations in the form
\begin{subequations}
\label{subeq:pertub-flrw}
\begin{align}
  &2H(\theta_\een-\vartheta_\een)-
       \tfrac{1}{2}R_\een = \kappa\varepsilon_\een,
\label{con-sp-1} \\
   & \dot{R}_\een+
     2HR_\een-
      2\kappa \varepsilon_\nul(1 + w)\vartheta_\een
       +\tfrac{2}{3}R_\nul (\theta_\een-\vartheta_\een)=0,
               \label{FRW6} \\
   &\dot{\varepsilon}_\een + 3H(\varepsilon_\een + p_\een)+
         \varepsilon_\nul(1 + w)\theta_\een=0,  \label{FRW4} \\
   &\dot{\vartheta}_\een+H(2-3\beta^2)\vartheta_\een+
   \frac{1}{\varepsilon_\nul(1+w)}\dfrac{\tilde{\nabla}^2p_\een}{a^2}=0,
   \qquad
      \beta^2\equiv \dfrac{\dot{p}_\nul}{\dot{\varepsilon}_\nul},  \label{FRW5}\\
   &\dot{n}_\een + 3H n_\een + n_\nul\theta_\een=0. \label{FRW4a}
\end{align}
\end{subequations}
The set (\ref{subeq:pertub-flrw}), which is the perturbed counterpart
of the set (\ref{subeq:einstein-flrw}), consists of one algebraic
equation (\ref{con-sp-1}) and four ordinary differential equations
(\ref{FRW6})--(\ref{FRW4a}) for the five unknown quantities
$\varepsilon_\een$, $n_\een$, $\vartheta_\een$, $R_\een$ and
$\theta_\een$.  The quantity $\beta(t)$ is defined by
$\beta^2\equiv\dot{p}_\nul/\dot{\varepsilon}_\nul$.  Using that
$\dot{p}_\nul=p_n\dot{n}_\nul+p_\varepsilon\dot{\varepsilon}_\nul$ and
the conservation laws (\ref{FRW2}) and (\ref{FRW2a}) one gets
\begin{equation}
  \label{eq:beta-matter}
  \beta^2=p_\varepsilon+\dfrac{n_\nul p_n}{\varepsilon_\nul(1+w)}.
\end{equation}
Finally, the symbol $\tilde{\nabla}^2$ denotes the generalised Laplace
operator with respect to the three-space metric $\tilde{g}_{ij}$,
defined by $\tilde{\nabla}^2 f \equiv \tilde{g}^{ij} f_{|i|j}$.

The derivation of the basic equations (\ref{subeq:pertub-flrw}) for
scalar perturbations will now be given.  Eliminating $\dot{h}^k{}_k$
from (\ref{basis-1}) with the help of (\ref{fes5}) yields the
algebraic equation (\ref{con-sp-1}).

Multiplying both sides of equations (\ref{basis-2})
by $g^{ij}_\nul$ and taking the covariant derivative with respect to
the index $j$, one finds
\begin{equation}
  g_\nul^{ij} (\dot{h}^k{}_{k|i|j}-\dot{h}^k{}_{i|k|j}) =
    2\kappa(\varepsilon_\nul+p_\nul) \vartheta_\een,  \label{dR0i4}
\end{equation}
where also (\ref{fes5}) has been used.  The left-hand side of
(\ref{dR0i4}) will turn up as a part of the time-derivative of the
curvature $R_\een$.  In fact, differentiating (\ref{driekrom}) with
respect to time and recalling that the background connection
coefficients $\Gamma^k_{\nul ij}$ are for \textsc{flrw} metrics
(\ref{eq:metric-flrw}) independent of time, one gets, using also
$\dot{g}^{ij}_\nul=-2Hg^{ij}_\nul$ and (\ref{FRW3a}),
\begin{equation}
  \dot{R}_\een = -2HR_\een +
   g_\nul^{ij} (\dot{h}^k{}_{k|i|j}-\dot{h}^k{}_{i|k|j})+
     \tfrac{1}{3}R_\nul \dot{h}^k{}_k.
   \label{dR0i6}
\end{equation}
Combining (\ref{dR0i4}) and (\ref{dR0i6}) and using (\ref{fes5}) to
eliminate $\dot{h}^k{}_k$ yields (\ref{FRW6}).  Thus, the $G^0_{\een
  i}$ momentum constraint equations (\ref{basis-2}) have been recast
into one equation (\ref{FRW6}) for the local spatial curvature due to
a density perturbation.

For $i\not=j$ equations (\ref{basis-3}) need not be considered, since
they are not coupled to scalar perturbations.  Taking the trace of
(\ref{basis-3}) and eliminating the quantity $\dot{h}^k{}_k$ with the
help of (\ref{fes5}), one arrives at
\begin{equation}
   \dot{\theta}_\een-\dot{\vartheta}_\een +
      6H(\theta_\een-\vartheta_\een)-R_\een=
       \tfrac{3}{2}\kappa(\varepsilon_\een-p_\een).
        \label{eq:168a}
\end{equation}
Using (\ref{con-sp-1}) to eliminate the second term of (\ref{eq:168a})
yields for the trace of (\ref{basis-3})
\begin{equation}
  \label{eq:168a-kort}
  \dot{\theta}_\een-\dot{\vartheta}_\een + \tfrac{1}{2}R_\een=
       -\tfrac{3}{2}\kappa(\varepsilon_\een+p_\een).
\end{equation}
This equation is identical to the time-derivative of the constraint
equation (\ref{con-sp-1}), which reads
\begin{equation}
\label{eq:con-sp-1-sum-time}
2\dot{H}(\theta_\een-\vartheta_\een)+2H(\dot{\theta}_\een-\dot{\vartheta}_\een)-
 \tfrac{1}{2}\dot{R}_\een=\kappa\dot{\varepsilon}_\een.
\end{equation}
Eliminating the time-derivatives $\dot{H}$, $\dot{R}_\een$ and
$\dot{\varepsilon}_\een$ with the help of (\ref{FRW3})--(\ref{FRW2}),
(\ref{FRW6}) and (\ref{FRW4}), respectively, yields the dynamical
equation (\ref{eq:168a-kort}).  Consequently, for scalar perturbations the
dynamical equations (\ref{basis-3}) need not be considered.

Finally, taking the covariant derivative of (\ref{basis-5}) with
respect to the metric $g_{\nul ij}$ and using (\ref{fes5}), one gets
\begin{equation}
  \frac{1}{c}\frac{\text{d}}{\text{d}t}
     \Bigl[(\varepsilon_\nul+p_\nul)\vartheta_\een\Bigr]-
    g^{ik}_\nul p_{\een|k|i}+5H(\varepsilon_\nul+p_\nul)\vartheta_\een = 0,  
\label{mom2}
\end{equation}
where it is used that the operations of taking the time-derivative and
the covariant derivative commute, since the background connection
coefficients $\Gamma^k_{\nul ij}$ are independent of time for
\textsc{flrw} metrics.  With (\ref{eq:metric-flrw}),
$\tilde{\nabla}^2f\equiv\tilde{g}^{ij}f_{|i|j}$ and (\ref{FRW2}) one
can rewrite (\ref{mom2}) in the form
\begin{equation}
  \dot{\vartheta}_\een +
 H\left(2-3\frac{\dot{p}_\nul}{\dot{\varepsilon}_\nul}\right)\vartheta_\een+
     \frac{1}{\varepsilon_\nul+p_\nul}\dfrac{\tilde{\nabla}^2 p_\een}{a^2}=0.
                \label{mom3}
\end{equation}
Using the definitions $w\equiv p_\nul/\varepsilon_\nul$ and
$\beta^2\equiv\dot{p}_\nul/\dot{\varepsilon}_\nul$ one arrives at
equation (\ref{FRW5}).

This concludes the derivation of the system
(\ref{subeq:pertub-flrw}). As follows from its derivation, this system
is, for scalar perturbations, equivalent to the full set of
first-order Einstein equations and conservation laws
(\ref{subeq:basis}).

\subsection{\label{sec:unique}Unique Gauge-invariant Cosmological Density Perturbations}

The background equations (\ref{subeq:einstein-flrw}) and the
perturbation equations (\ref{subeq:pertub-flrw}) are both written with
respect to the \emph{same} system of reference. Therefore, these two
sets can be combined to describe the evolution of the five background
quantities $\theta_\nul=3H$, $R_\nul$, $\varepsilon_\nul$,
$\vartheta_\nul=0$, $n_\nul$, and their first-order counterparts
$\theta_\een$, $R_\een$, $\varepsilon_\een$, $\vartheta_\een$,
$n_\een$.  Just as in the background case, one again comes across the
three independent scalars (\ref{eq:scalars-flrw}).  Consequently, the
evolution of cosmological density perturbations is described by the
three \emph{independent} scalars (\ref{eq:scalars-flrw}) and their
first-order perturbations.  A complicating factor is that the
first-order quantities $\varepsilon_\een$ and $n_\een$, which are
supposed to describe the energy density and the particle number
density perturbations, have no physical significance, as will now be
established.

A first-order perturbation to one of the scalars
(\ref{eq:scalars-flrw}) transforms under a general (not necessarily
between synchronous coordinates) infinitesimal coordinate
transformation (\ref{func})~as
\begin{equation}
  \label{eq:trans-scalar}
  S_\een(t,\bm{x}) \rightarrow S_\een(t,\bm{x})+\xi^0(t,\bm{x})\dot{S}_\nul(t),
\end{equation}
where $S_\nul$ and $S_\een$ are the background and first-order
perturbation of one of the three scalars $S=\varepsilon, n,
\theta$. In (\ref{eq:trans-scalar}) the term
$\hat{S}\equiv\xi^0\dot{S}_\nul$ is the so-called \emph{gauge mode}.
The complete set of gauge modes for the system of equations
(\ref{subeq:pertub-flrw}) is given by
\begin{subequations}
\label{subeq:gauge-dep}
\begin{align}
 & \hat{\varepsilon}_\een = 
        \psi\dot{\varepsilon}_\nul, \qquad
     \hat{n}_\een  = 
        \psi\dot{n}_\nul, \qquad
     \hat{\theta}_\een  =
        \psi\dot{\theta}_\nul, \label{tr-theta}\\
   &     \hat{\vartheta}_\een  =-\frac{\tilde{\nabla}^2\psi}{a^2}, \qquad
       \hat{R}_\een  =  
     4H\left[\frac{\tilde{\nabla}^2\psi}{a^2} -
       \tfrac{1}{2}R_\nul\psi\right],
            \label{drie-ijk}
\end{align}
\end{subequations}
where $\xi^0=\psi(\bm{x})$ in synchronous coordinates, see
(\ref{eq:synchronous}).  The quantities (\ref{subeq:gauge-dep}) are
mere coordinate artifacts, which have no physical meaning, since the
gauge function $\psi(\bm{x})$ is an arbitrary (infinitesimal)
function. Equations (\ref{subeq:pertub-flrw}) are invariant under
coordinate transformations (\ref{func}) combined with
(\ref{eq:synchronous}), i.e., the gauge modes (\ref{subeq:gauge-dep})
are solutions of the set (\ref{subeq:pertub-flrw}). This property
combined with the \emph{linearity} of the perturbation equations,
implies that a solution set $(\varepsilon_\een, n_\een, \theta_\een,
\vartheta_\een, R_\een)$ can be augmented with the corresponding gauge
modes (\ref{subeq:gauge-dep}) to obtain a new solution set. Therefore,
the solution set $(\varepsilon_\een, n_\een, \theta_\een,
\vartheta_\een, R_\een)$ has no physical significance, since the
general solution of the set (\ref{subeq:pertub-flrw}) can be modified
by an infinitesimal coordinate transformation.  This is the notorious
gauge problem of cosmology.

In this article, the cosmological gauge problem has been solved. To
that end, the perturbation equations (\ref{subeq:basis}) have first
been rewritten into the form (\ref{subeq:pertub-flrw}) in order to
isolate the scalar perturbations from the vortexes and gravitational
waves.  The fact that the system of equations
(\ref{subeq:pertub-flrw}) describes exclusively the evolution of
\emph{scalar} perturbations has the following important consequence:
\begin{itemize}
\item From the background equations (\ref{subeq:einstein-flrw}) and
  the perturbation equations (\ref{subeq:pertub-flrw}) for scalar
  perturbations it follows that only the three \emph{independent}
  scalars (\ref{eq:scalars-flrw}) and their first-order perturbations
  play a role in the evolution of density perturbations.
\end{itemize}
This fact reduces the number of possible gauge-invariant quantities
substantially, since one needs to consider only the three independent
scalars (\ref{eq:scalars-flrw}).  Since scalar perturbations transform
under the general infinitesimal transformation (\ref{func}) according
to (\ref{eq:trans-scalar}), one can combine two independent scalars to
eliminate the gauge function $\xi^0(t,\bm{x})$. With the three
independent scalars (\ref{eq:scalars-flrw}), one can make
${3\choose2}=3$ different sets of three gauge-invariant quantities.
In each of these sets exactly one gauge-invariant quantity vanishes.
As will be shown in Section~\ref{sec:newt-limit}, the only set for
which the corresponding perturbation theory yields the Newtonian
results (\ref{eq:poisson}) and (\ref{eq:newt-ngi}) in the
non-relativistic limit is given~by
\begin{subequations}
  \label{subeq:gi-quant}
  \begin{align}
   & \varepsilon_\een^{\text{gi}}
        =\varepsilon_\een-\dfrac{\dot{\varepsilon}_\nul}
        {\dot{\theta}_\nul}\theta_\een, \qquad
   n_\een^{\text{gi}}=n_\een-\dfrac{\dot{n}_\nul}
        {\dot{\theta}_\nul}\theta_\een, \label{e-n-gi} \\
      &   \theta_\een^{\text{gi}}  =\theta_\een-\dfrac{\dot{\theta}_\nul}
        {\dot{\theta}_\nul}\theta_\een\equiv0. \label{theta-gi}
  \end{align}
\end{subequations}
It follows from the general transformation rule
(\ref{eq:trans-scalar}) that the quantities (\ref{subeq:gi-quant}) are
invariant under the general infinitesimal transformation (\ref{func}),
i.e., they are \emph{gauge-invariant}, hence the superscript `gi'.  The
definitions (\ref{subeq:gi-quant}) imply that the gauge-invariant
counterpart $\theta^{\text{gi}}_\een$ of the gauge-dependent variable
$\theta_\een\neq0$ vanishes automatically.  The physical
interpretation of (\ref{theta-gi}) is that, in first-order, the
\emph{global} expansion $\theta_\nul=3H$ is not affected by a
\emph{local} density perturbation.

The quantities (\ref{e-n-gi}) have two essential properties, which
gauge-invariant quantities used in former perturbation theories
\cite{Christopherson:2011ra,Malik:2008im,Knobel:2012wa,Peter:2013woa,c13,kodama1984,
  Ellis1,Ellis2,ellis-1998,mfb1992,Mukhanov-2005} do not have.
Firstly, due to the quotients
$\dot{\varepsilon}_\nul/\dot{\theta}_\nul$ and
$\dot{n}_\nul/\dot{\theta}_\nul$ of time derivatives, the quantities
$\varepsilon_\een^{\text{gi}}$ and $n_\een^{\text{gi}}$ are
\emph{independent} of the definition of time.  As a consequence, the
evolution of $\varepsilon_\een^{\text{gi}}$ and $n_\een^{\text{gi}}$
is only determined by their propagation equations. Secondly, the
quantities (\ref{e-n-gi}) do not contain \emph{spatial} derivatives,
so that unnecessary gradient terms~\cite{2012arXiv1206.1478M} do not
occur in the final equations (\ref{subeq:final}).

The quantities (\ref{e-n-gi}) are completely determined by the
background equations (\ref{subeq:einstein-flrw}) and their perturbed
counterparts (\ref{subeq:pertub-flrw}). In principle, these two sets
can be used to study the evolution of density perturbations in
\textsc{flrw} universes. However, the set (\ref{subeq:pertub-flrw}) is
still too complicated, since it also admits the non-physical
solutions~(\ref{subeq:gauge-dep}). The aim will be a system of
evolution equations for $\varepsilon^{\text{gi}}_\een$ and
$n^{\text{gi}}_\een$ that do not have the gauge modes
(\ref{subeq:gauge-dep}) as solution. In other words, a perturbation
theory will be constructed for which not only the differential
equations are invariant under general infinitesimal coordinate
transformations (\ref{func}), but also their solutions.  Such a theory
will be referred to as a \emph{manifestly covariant and
  gauge-invariant} perturbation theory.  The derivation of this theory
will be the subject of the next subsection.

\subsection{\label{sec:flat-pert}Manifestly Covariant and Gauge-invariant Perturbation Theory}

In this section the derivation of the new perturbation theory is
given.  Firstly, it is observed that the gauge-dependent variable
$\theta_\een$ is not needed in the calculations, since its
gauge-invariant counterpart $\theta^{\text{gi}}_\een$,
(\ref{theta-gi}), vanishes identically.  Eliminating $\theta_\een$
from the differential equations (\ref{FRW6})--(\ref{FRW4a}) with the
help of the (algebraic) constraint equation (\ref{con-sp-1}) yields
the set of four first-order ordinary differential equations
\begin{subequations}
\label{subeq:pertub-gi}
\begin{align}
 &  \dot{R}_\een +
     2HR_\een-
         2\kappa\varepsilon_\nul(1 + w)\vartheta_\een+
     \frac{R_\nul}{3H} \left(\kappa\varepsilon_\een +
 \tfrac{1}{2}R_\een \right)=0,  \label{FRW6gi} \\
 & \dot{\varepsilon}_\een + 3H(\varepsilon_\een + p_\een)+
     \varepsilon_\nul(1 + w)\Bigl[\vartheta_\een +\frac{1}{2H}\left(
 \kappa\varepsilon_\een+\tfrac{1}{2}R_\een\right)\Bigr]=0,
         \label{FRW4gi} \\
 & \dot{\vartheta}_\een+H(2-3\beta^2)\vartheta_\een +
        \frac{1}{\varepsilon_\nul(1+w)}\dfrac{\tilde{\nabla}^2p_\een}{a^2}=0,
\label{FRW5gi}\\
&\dot{n}_\een + 3H n_\een+
       n_\nul \Bigl[\vartheta_\een +\frac{1}{2H}\left(\kappa\varepsilon_\een +
      \tfrac{1}{2}R_\een\right)\Bigr]=0, 
\label{FRW4agi}
\end{align}
\end{subequations}
for the four quantities $\varepsilon_\een$, $n_\een$, $\vartheta_\een$
and~$R_\een$.

Using the background equations (\ref{subeq:einstein-flrw}) to
eliminate all time-derivatives and the first-order constraint equation
(\ref{con-sp-1}) to eliminate $\theta_\een$, the gauge-invariant
quantities (\ref{e-n-gi}) become
\begin{subequations}
\label{subeq:pertub-gi-e-n}
\begin{align}
   & \varepsilon_\een^{\text{gi}}  =
     \dfrac{ \varepsilon_\een R_\nul -
   3\varepsilon_\nul(1 + w) (2H\vartheta_\een +
  \tfrac{1}{2}R_\een) }
  { R_\nul+3\kappa\varepsilon_\nul(1 + w)},
        \label{Egi} \\
   & n_\een^{\text{gi}}  = n_\een-\dfrac{3n_\nul(\kappa\varepsilon_\een+2H\vartheta_\een+
            \tfrac{1}{2}R_\een)}
         {R_\nul+3\kappa\varepsilon_\nul(1+w)}.  \label{nu2}
\end{align}
\end{subequations}
These quantities are completely determined by the background equations
(\ref{subeq:einstein-flrw}) and the first-order
equations~(\ref{subeq:pertub-gi}).  In the study of the evolution of
density perturbations, it is convenient not to use
$\varepsilon^{\text{gi}}_\een$ and $n^{\text{gi}}_\een$ directly, but
instead their corresponding contrast functions $\delta_\varepsilon$
and $\delta_n$
\begin{equation}\label{eq:contrast}
  \delta_\varepsilon(t,\bm{x}) \equiv
      \dfrac{\varepsilon^{\text{gi}}_\een(t,\bm{x})}{\varepsilon_\nul(t)}, \qquad
  \delta_n(t,\bm{x}) \equiv
      \dfrac{n^{\text{gi}}_\een(t,\bm{x})}{n_\nul(t)}.
\end{equation}
The system of equations (\ref{subeq:pertub-gi}) for the four
independent quantities $\varepsilon_\een$, $n_\een$, $\vartheta_\een$
and~$R_\een$ will now be rewritten, using the procedure given in the
Appendix, into a new system of equations for the two independent
quantities $\delta_\varepsilon$ and $\delta_n$.  In this procedure it
is explicitly assumed that $p\not\equiv0$, i.e., the pressure does not
vanish identically.  The case $p\rightarrow0$ will be considered in
Section~\ref{sec:newt-limit} on the non-relativistic limit.  The final
result is the \emph{generalised cosmological perturbation theory} for
the open, flat or closed \textsc{flrw} universe
\begin{subequations}
\label{subeq:final}
\begin{align}
   \ddot{\delta}_\varepsilon + b_1 \dot{\delta}_\varepsilon +
      b_2 \delta_\varepsilon &=
      b_3 \left[\delta_n - \frac{\delta_\varepsilon}{1+w}\right],
              \label{sec-ord}  \\
   \frac{1}{c}\frac{{\text{d}}}{{\text{d}} t}
      \left[\delta_n - \frac{\delta_\varepsilon}{1 + w}\right] & =
     \frac{3Hn_\nul p_n}{\varepsilon_\nul(1 + w)}
     \left[\delta_n - \frac{\delta_\varepsilon}{1 + w}\right].
                 \label{fir-ord}
\end{align}
\end{subequations}
These are two linear differential equations for the two
\emph{independent} and gauge-invariant quantities $\delta_\varepsilon$
and $\delta_n$.  It follows from
$\beta^2\equiv\dot{p}_\nul/\dot{\varepsilon}_\nul$ and equation
(\ref{FRW2}) that the time-derivative of $w\equiv
p_\nul/\varepsilon_\nul$is\footnote{Expression
  (\ref{eq:time-w}) is \emph{independent} of the equation of state,
  since it is derived from the definitions of $\beta$ and $w$, using
  only the energy conservation law (\ref{FRW2}).}
\begin{equation}
  \label{eq:time-w}
  \dot{w}=3H(1+w)(w-\beta^2).
\end{equation}
Defining $p_{nn}\equiv\partial^2p/\partial n^2$ and $p_{\varepsilon
  n}\equiv\partial^2p/\partial\varepsilon\,\partial n$ and using
(\ref{eq:time-w}) the coefficients $b_1$, $b_2$ and~$b_3$ of equation
(\ref{sec-ord}) read
\begin{subequations}
\label{subeq:coeff-contrast}
 \begin{align}
  b_1  = &\, \dfrac{\kappa\varepsilon_\nul(1+w)}{H}
  -2\dfrac{\dot{\beta}}{\beta}-H(2+6w+3\beta^2)
   + R_\nul\left(\dfrac{1}{3H}+
  \dfrac{2H(1+3\beta^2)}
  {R_\nul+3\kappa\varepsilon_\nul(1+w)}\right), \\
  b_2 = & -\tfrac{1}{2}\kappa\varepsilon_\nul(1+w)(1+3w)+
  H^2\left(1-3w+6\beta^2(2+3w)\right)
   +6H\dfrac{\dot{\beta}}{\beta}\left(w+\dfrac{\kappa\varepsilon_\nul(1+w)}
   {R_\nul+3\kappa\varepsilon_\nul(1+w)}\right)  \nonumber \\
   & - R_\nul\left(\tfrac{1}{2}w+
\dfrac{H^2(1+6w)(1+3\beta^2)}{R_\nul+3\kappa\varepsilon_\nul(1+w)}
\right)
-\beta^2\left(\frac{\tilde{\nabla}^2}{a^2}-\tfrac{1}{2}R_\nul
\right), \\
  b_3  =&\,
\Biggl\{\dfrac{-18H^2}{R_\nul+3\kappa\varepsilon_\nul(1+w)}
  \Biggl[\varepsilon_\nul p_{\varepsilon n}(1+w)
   +\dfrac{2p_n}{3H}\dfrac{\dot{\beta}}{\beta}
   +p_n(p_\varepsilon-\beta^2)+n_\nul p_{nn}\Biggr]+
   p_n\Biggr\}\dfrac{n_\nul}{\varepsilon_\nul}
\left(\frac{\tilde{\nabla}^2}{a^2}-\tfrac{1}{2}R_\nul\right).
\label{eq:b3}
\end{align}
\end{subequations}
The equations (\ref{subeq:final}) have been checked (see the attached
\textsc{Maxima} file) using a computer algebra system~\cite{maxima},
as follows.  Substituting the contrast functions (\ref{eq:contrast})
in equations (\ref{subeq:final}), where $\varepsilon_\een^{\text{gi}}$
and $n_\een^{\text{gi}}$ are given by (\ref{subeq:pertub-gi-e-n}), and
subsequently eliminating the time-derivatives of $\varepsilon_\nul$,
$n_\nul$, $H$, $R_\nul$ and $\varepsilon_\een$, $n_\een$,
$\vartheta_\een$, $R_\een$ with the help of equations
(\ref{subeq:einstein-flrw}) and (\ref{subeq:pertub-gi}), respectively,
yields two identities for each of the two equations
(\ref{subeq:final}).

It follows from equation (\ref{fir-ord}) that perturbations in the
particle number density are \emph{gravitationally} coupled to
perturbations in the total energy density if $p_n\equiv(\partial
p/\partial n)_\varepsilon\le0$, or, equivalently,
$p_\varepsilon\equiv(\partial p/\partial\varepsilon)_n\ge\beta^2$, see
(\ref{eq:beta-matter}).  This is the case in a \textsc{flrw} universe
in the radiation-dominated era and in the era after decoupling of
matter and radiation.  This coupling is \emph{independent} of the
nature of the particles, i.e., it holds true for ordinary matter as
well as \textsc{cdm}.  As a consequence, \textsc{cdm} perturbations do
not contract faster than perturbations in ordinary matter do.

The system of equations (\ref{subeq:final}) is equivalent to a system
of \emph{three} first-order differential equations, whereas the
original set (\ref{subeq:pertub-gi}) is a \emph{fourth}-order
system. This difference is due to the fact that the gauge modes
(\ref{subeq:gauge-dep}), which are solutions of the set
(\ref{subeq:pertub-gi}), are completely removed from the solution set
of~(\ref{subeq:final}): one degree of freedom, namely the gauge
function $\xi^0(t,\bm{x})$ in (\ref{eq:trans-scalar}), has disappeared
altogether.

The background equations (\ref{subeq:einstein-flrw}) and the new
perturbation equations (\ref{subeq:final}) constitute a set of
equations which enables one to study the evolution of small
fluctuations in the energy density $\delta_\varepsilon$ and the
particle number density $\delta_n$ in an open, flat or closed
\textsc{flrw} universe with $\Lambda\neq0$ and filled with a perfect
fluid with a non-barotropic equation of state for the pressure
$p=p(n,\varepsilon)$.

\subsection{\label{sec:gi-PT}Gauge-invariant Pressure and Temperature Perturbations}

The gauge-invariant pressure and temperature perturbations, which are
needed in the forthcoming sections, will now be derived.

From the equation of state (\ref{eq:equat-of-state-pressure}) for the
pressure $p=p(n,\varepsilon)$ it follows that
\begin{equation}
\label{eq:p0}
\dot{p}_\nul=p_n\dot{n}_\nul+p_\varepsilon\dot{\varepsilon}_\nul,  \qquad
  p_n\equiv\left(\dfrac{\partial p}{\partial n}\right)_\varepsilon, \quad
p_\varepsilon\equiv\left(\dfrac{\partial p}{\partial \varepsilon}\right)_n.
\end{equation}
Multiplying both sides of this expression by
$\theta_\een/\dot{\theta}_\nul$ and subtracting the result from
$p_\een$ given by (\ref{eq:p1}), one gets, using also
(\ref{e-n-gi}),
\begin{equation}
  \label{eq:def-pgi}
  p_\een-\dfrac{\dot{p}_\nul}{\dot{\theta}_\nul}\theta_\een=
      p_nn^{\text{gi}}_\een+p_\varepsilon\varepsilon^{\text{gi}}_\een.
\end{equation}
Hence, the quantity defined by
\begin{equation}
  \label{eq:gi-p}
  p^{\text{gi}}_\een\equiv p_\een-\dfrac{\dot{p}_\nul}{\dot{\theta}_\nul}\theta_\een,
\end{equation}
is the gauge-invariant pressure perturbation.  Combining
(\ref{eq:def-pgi}) and (\ref{eq:gi-p}) and eliminating $p_\varepsilon$
with the help of (\ref{eq:beta-matter}), one arrives at
\begin{equation}
  \label{eq:p-dia-adia}
  p_\een^{\text{gi}}=\beta^2\varepsilon_\nul\delta_\varepsilon+n_\nul
  p_n\left[\delta_n-\dfrac{\delta_\varepsilon}{1+w}  \right],
\end{equation}
where also (\ref{eq:contrast}) has been used.  The first term in this
expression is the \emph{adiabatic} part of the pressure perturbation
and the second term is the \emph{diabatic} part.

From the equation of state (\ref{eq:es-p-T}) for the energy density
$\varepsilon=\varepsilon(n,T)$ it follows that
\begin{equation}
  \label{eq:e1-dot-e0}
  \dot{\varepsilon}_\nul=\left(\dfrac{\partial\varepsilon}{\partial n}
  \right)_T \dot{n}_\nul+
   \left(\dfrac{\partial\varepsilon}{\partial T} \right)_n\dot{T}_\nul,   \qquad
  \varepsilon_\een=\left(\dfrac{\partial\varepsilon}{\partial n}
  \right)_T  n_\een+
   \left(\dfrac{\partial\varepsilon}{\partial T} \right)_n T_\een.
\end{equation}
Multiplying $\dot{\varepsilon}_\nul$ by
$\theta_\een/\dot{\theta}_\nul$ and subtracting the result from
$\varepsilon_\een$, one finds, using (\ref{e-n-gi}),
\begin{equation}
  \label{eq:time-e}
  \varepsilon^{\text{gi}}_\een=
     \left(\dfrac{\partial\varepsilon}{\partial n}\right)_T n^{\text{gi}}_\een+
\left(\dfrac{\partial\varepsilon}{\partial T}\right)_n
\left[T_\een-\dfrac{\dot{T}_\nul}{\dot{\theta}_\nul}\theta_\een \right],
\end{equation}
implying that the quantity defined by
\begin{equation}
  \label{eq:gi-T}
  T^{\text{gi}}_\een\equiv T_\een-\dfrac{\dot{T}_\nul}{\dot{\theta}_\nul}\theta_\een,
\end{equation}
is the gauge-invariant temperature perturbation.  The expressions
(\ref{eq:gi-p}) and (\ref{eq:gi-T}) are both of the form
(\ref{subeq:gi-quant}).

\subsection{\label{sec:diabatic}Diabatic Density Perturbations}

In this section equations (\ref{subeq:final}) will be linked to
thermodynamics and it will be shown that, in general, density
perturbations evolve diabatically, i.e., they exchange heat with their
environment during their evolution.

The combined First and Second Law of Thermodynamics is given by
\begin{equation}
  \label{eq:combined-fs-thermo}
  {\text{d}}E=T{\text{d}}S-p{\text{d}}V+\mu{\text{d}}N,
\end{equation}
where $E$, $S$ and $N$ are the energy, the entropy and the number of
particles of a system with volume $V$ and pressure $p$, and where
$\mu$, the thermal ---or chemical--- potential, is the energy needed
to add one particle to the system.  In terms of the particle number
density $n=N/V$, the energy per particle $E/N=\varepsilon/n$ and the
entropy per particle $s=S/N$ the law (\ref{eq:combined-fs-thermo}) can
be rewritten
\begin{equation}
  \label{eq:law-rewritten}
  {\text{d}}\left(\dfrac{\varepsilon}{n}N\right)=T{\text{d}}(sN)-
    p{\text{d}}\left(\dfrac{N}{n}\right)+\mu{\text{d}}N,
\end{equation}
where $\varepsilon$ is the energy density.  The system is
\emph{extensive}, i.e., $S(\alpha E, \alpha V, \alpha N)=\alpha
S(E,V,N)$, implying that the entropy of the gas is $S=(E+pV-\mu N)/T$.
Dividing this relation by $N$ one gets the so-called Euler relation
\begin{equation}
  \label{eq:chemical-pot}
  \mu=\dfrac{\varepsilon+p}{n}-Ts.
\end{equation}
Eliminating $\mu$ in (\ref{eq:law-rewritten}) with the help of
(\ref{eq:chemical-pot}), one finds that the combined First and Second
Law of Thermodynamics (\ref{eq:combined-fs-thermo}) can be cast in a
form without $\mu$ and $N$, i.e.,
\begin{equation}
  \label{eq:thermo}
  T{\text{d}} s= {\text{d}}\Bigl(\dfrac{\varepsilon}{n}\Bigr)+
      p{\text{d}}\Bigl(\dfrac{1}{n}\Bigr).
\end{equation}
From the background equations (\ref{subeq:einstein-flrw}) and the
thermodynamic law (\ref{eq:thermo}) it follows that $\dot{s}_\nul=0$,
implying with (\ref{eq:trans-scalar}) that $s_\een=s^{\text{gi}}_\een$
is automatically gauge-invariant.

The thermodynamic relation (\ref{eq:thermo}) can, using
(\ref{eq:contrast}), be rewritten in the form
\begin{equation}
  \label{eq:thermo-een}
  T_\nul s^{\text{gi}}_\een=-\dfrac{\varepsilon_\nul(1+w)}{n_\nul}
   \left[\delta_n-\dfrac{\delta_\varepsilon}{1+w}\right].
\end{equation}
Thus, the right-hand side of (\ref{sec-ord}) is related to local
perturbations in the entropy, and (\ref{fir-ord}) can be considered as
an evolution equation for entropy perturbations.

Adiabatic perturbations do not exchange heat with their
environment, so that $T_\nul s^{\text{gi}}_\een=0$.  This implies with
(\ref{eq:thermo-een}) that $(1+w)\delta_n-\delta_\varepsilon=0$.
Multiplying this expression by $3H\varepsilon_\nul n_\nul$ and
substituting (\ref{eq:contrast}) one finds from the background
conservation laws (\ref{FRW2}) and (\ref{FRW2a}) that the adiabatic
condition $s^{\text{gi}}_\een=0$ reads
$\dot{n}_\nul\varepsilon^{\text{gi}}_\een-\dot{\varepsilon}_\nul
n^{\text{gi}}_\een=0$.  Using that $\varepsilon=\varepsilon(n,T)$ the latter
expression becomes
\begin{equation}
  \label{eq:adia-cond-n-T}
  \left(\dfrac{\partial\varepsilon}{\partial T} \right)_{n}
  \left[\dot{n}_\nul T^{\text{gi}}_\een-n^{\text{gi}}_\een \dot{T}_\nul  \right]=0.
\end{equation}
Since $n$ and $T$ are independent quantities and since in a non-static
universe one has $\dot{n}_\nul\neq0$ and $\dot{T}_\nul\neq0$, the
adiabatic condition (\ref{eq:adia-cond-n-T}) is satisfied if, and only
if,
\begin{equation}
  \label{eq:eps-ind-T}
  \left(\dfrac{\partial\varepsilon}{\partial T}\right)_{n}=0,
\end{equation}
implying that $\varepsilon=\varepsilon(n)$.  In particular, in the
non-relativistic limit, where $\varepsilon=nmc^2$ and $p=0$, density
perturbations are adiabatic.  This is in accordance with the fact that
in the non-relativistic limit, which will be elaborated in the next
subsection, density perturbations do not evolve.  In all other cases
where $p=p(n,\varepsilon)$ local density perturbations evolve
\emph{diabatically}.

Finally, in the limiting case that the particle number density does
not contribute to the pressure, i.e., $p_n\approx0$, it follows from
(\ref{eq:p-dia-adia}) that the pressure perturbation is
\emph{adiabatic}.  In this case the equation of state is barotropic,
i.e., $p\approx p(\varepsilon)$, implying that the coefficient $b_3$,
(\ref{eq:b3}), vanishes so that equation (\ref{sec-ord}) is
homogeneous and equations (\ref{sec-ord}) and (\ref{fir-ord}) are
decoupled.  Consequently, for barotropic equations of state density
perturbations evolve adiabatically.

\subsection{\label{sec:newt-limit}Non-relativistic Limit}

In Section~\ref{sec:unique} it has been shown that the two
gauge-invariant quantities $\varepsilon^{\text{gi}}_\een$ and
$n^{\text{gi}}_\een$ are unique.  It will be demonstrated that in the
non-relativistic limit equations (\ref{subeq:pertub-flrw}) combined
with (\ref{subeq:gi-quant}) reduce to the results (\ref{eq:poisson})
and (\ref{eq:newt-ngi}) of the Newtonian Theory of Gravity and that
the quantities $\varepsilon^{\text{gi}}_\een$ and $n^{\text{gi}}_\een$
become equal to their Newtonian counterparts.

The non-relativistic limit is defined by three requirements,
Carroll~\cite{carroll-2003}:
\begin{itemize}
\item The gravitational field should be weak, i.e., can be considered
  as a perturbation of flat space.
\item The particles are moving slowly with respect to the speed of
  light.
\item The gravitational field of a density perturbation should be
  static, i.e., it does not change with time.
\end{itemize}
This definition of the non-relativistic limit, which is essential to
put an accurate physical interpretation on the quantities
$\varepsilon^{\text{gi}}_\een$ and $n^{\text{gi}}_\een$, has not been
used in former perturbation
theories~\cite{Christopherson:2011ra,Malik:2008im,Knobel:2012wa,Peter:2013woa,c13,
  kodama1984,Ellis1,Ellis2,ellis-1998,mfb1992,Mukhanov-2005} to
explain the meaning of the gauge-invariant quantities.

In a first-order cosmological perturbation theory the gravitational
field is already weak.  In order to meet the first requirement, a flat
($R_\nul=0$) \textsc{flrw} universe is considered.  Using
(\ref{decomp-hij-par}), the local perturbation to the spatial
curvature (\ref{driekrom}) reduces for a flat \textsc{flrw} universe
to
\begin{equation}\label{RnabEE-0}
   R_\een =\dfrac{4}{c^2}\phi^{|k}{}_{|k}=
       -\dfrac{4}{c^2}\dfrac{\nabla^2\phi}{a^2},
\end{equation}
where $\nabla^2$ is the usual Laplace operator.  Substituting this
expression into the perturbation equations (\ref{subeq:pertub-flrw}),
one gets
\begin{subequations}
\label{subeq:pertub-gi-flat}
\begin{align}
    & H(\theta_\een-\vartheta_\een)+
       \dfrac{1}{c^2}\dfrac{\nabla^2\phi}{a^2} = \dfrac{4\pi G_{\text{N}}}{c^4}
        \left[\varepsilon^{\text{gi}}_\een+
         \dfrac{\dot{\varepsilon}_\nul}{\dot{\theta}_\nul}\theta_\een\right],
\label{con-sp-1-flat} \\
    & \dfrac{\nabla^2\dot{\phi}}{a^2}+\dfrac{4\pi G_{\text{N}}}{c^2}
\varepsilon_\nul(1 + w)\vartheta_\een=0, \label{FRW6gi-flat} \\
    &\dot{\varepsilon}_\een +
3H(\varepsilon_\een + p_\een)+
         \varepsilon_\nul(1 + w)\theta_\een=0,  \label{FRW4gi-flat} \\
 & \dot{\vartheta}_\een+H(2-3\beta^2)\vartheta_\een+
   \frac{1}{\varepsilon_\nul(1+w)}\dfrac{\nabla^2p_\een}{a^2}=0, 
  \label{FRW5gi-flat} \\
 & \dot{n}_\een + 3H n_\een +
         n_\nul\theta_\een=0, \label{FRW4agi-flat}
\end{align}
\end{subequations}
where (\ref{e-n-gi}) has been used to eliminate $\varepsilon_\een$
from the constraint equation (\ref{con-sp-1}).

Next, the second requirement will be implemented.  Since the spatial
part $u^i_\een$ of the fluid four-velocity is gauge-dependent with a
physical component and a non-physical gauge part, the second
requirement must be defined by\footnote{Recall that $\bm{u}_\een$ is the
\emph{irrotational} part of the three-space fluid velocity. The
rotational part of $\bm{u}_\een$ is not coupled to density
perturbations and need, therefore, not be considered. See
Section~\ref{sec:decomp-h-u}.}
\begin{equation}
  \label{eq:non-rel-lim}
  u^i_{\een\,\text{physical}} \equiv
    c^{-1}U^i_{\een\,\text{physical}} \rightarrow 0,
\end{equation}
i.e., the \emph{physical} part of the spatial part of the fluid
four-velocity is negligible with respect to the speed of light.  In
this limit, the mean kinetic energy per particle
$\tfrac{1}{2}m\langle{v^2}\rangle=\tfrac{3}{2}k_{\text{B}}T\rightarrow0$
is very small compared to the rest energy $mc^2$ per particle.  This
implies that the pressure $p=nk_{\text{B}}T\rightarrow0$ ($n\neq0$) is
vanishingly small with respect to the rest energy density $nmc^2$.
Therefore, one must take the limits $p_\nul\rightarrow0$ in the
background and $p_\een^{\text{gi}}\rightarrow0$ in the perturbed
universe to arrive at the non-relativistic limit.  With
(\ref{eq:gi-p}) it follows that also $p_\een\rightarrow0$.
Substituting $p_\nul=0$ and $p_\een=0$ into the momentum conservation
laws (\ref{basis-5}) yields, using also the background equation
(\ref{FRW2}) with $w\equiv p_\nul/\varepsilon_\nul\rightarrow0$,
\begin{equation}\label{eq:ui-par-p0}
    \dot{u}^i_\een=-2H u^i_\een.
\end{equation}
Since the physical part of $u_\een^i$ vanishes in the non-relativistic
limit, the general solution of equations (\ref{eq:ui-par-p0}) is
exactly equal to the gauge mode (\ref{drie-ijk})
\begin{equation}\label{eq:ui-gauge-mode}
    \hat{u}^{i}_\een(t,\bm{x})=-\dfrac{1}{a^2(t)}
\tilde{g}^{ik}(\bm{x})\partial_k\psi(\bm{x}),
\end{equation}
where it is used that $H\equiv\dot{a}/a$.  Thus, in the limit
(\ref{eq:non-rel-lim}) one is left with the gauge mode
(\ref{eq:ui-gauge-mode}) only.  Consequently, one may, without losing
any physical information, put the gauge mode $\hat{u}^i_\een$ equal to
zero, implying that $\partial_k\psi=0$, so that $\psi=C$ is an
arbitrary constant in the non-relativistic limit.  Substituting
$\psi=C$ into (\ref{eq:synchronous}) one finds that the
relativistic transformation (\ref{func}) between synchronous
coordinates reduces in the limit (\ref{eq:non-rel-lim}) to the
(infinitesimal) transformation
\begin{equation}
\label{eq:gauge-trans-newt}
    x^0 \rightarrow x^0 - C, \qquad x^i \rightarrow
    x^i-\chi^i(\bm{x}),
\end{equation}
where $C$ is an arbitrary constant and $\chi^i(\bm{x})$ are three
arbitrary functions of the spatial coordinates.  In the
non-relativistic limit time and space transformations are decoupled:
time coordinates may be shifted and spatial coordinates may be chosen
arbitrarily.  The residual gauge freedom $C$ and $\chi^i(\bm{x})$
in the non-relativistic limit may not come as a surprise, since the
Newtonian Theory of Gravity is invariant under the gauge
transformation (\ref{eq:gauge-trans-newt}).

Substituting $\vartheta_\een=0$ and $p=0$ (i.e., $p_\nul=0$ and
$p_\een=0$) into the system (\ref{subeq:pertub-gi-flat}), one gets
\begin{subequations}
\label{subeq:pertub-gi-flat-newt}
\begin{align}
  & \nabla^2\phi = \dfrac{4\pi G_{\text{N}}}{c^2}a^2\varepsilon^{\text{gi}}_\een,
         \label{con-sp-1-flat-newt} \\
  & \nabla^2\dot{\phi}=0, \label{FRW6gi-flat-newt} \\
  & \dot{\varepsilon}_\een + 3H\varepsilon_\een+
         \varepsilon_\nul\theta_\een=0,  \label{FRW4gi-flat-newt} \\
  & \dot{n}_\een + 3H n_\een + n_\nul\theta_\een=0. \label{FRW4agi-flat-newt}
\end{align}
\end{subequations}
The constraint equation (\ref{con-sp-1-flat-newt}) can be found by
subtracting $\tfrac{1}{6}\theta_\een/\dot{H}$ times the
time-derivative of the background constraint equation (\ref{FRW3})
with $R_\nul=0$ from the constraint equation (\ref{con-sp-1-flat}) and
using that $\theta_\nul=3H$.  Note that the cosmological constant
$\Lambda$ need not be zero.

Since $\vartheta_\een=0$ there is no fluid flow so that density
perturbations do not evolve. This implies the basic fact of the
Newtonian Theory of Gravity, namely that the gravitational field is
\emph{static} (\ref{FRW6gi-flat-newt}).  Consequently,
$a^2(t)\varepsilon^{\text{gi}}_\een(t,\bm{x})$ in
(\ref{con-sp-1-flat-newt}) should be replaced by
$a^2(t_0)\varepsilon^{\text{gi}}_\een(t_0,\bm{x})$.  Defining the
potential $\varphi(\bm{x})\equiv\phi(\bm{x})/a^2(t_0)$, equations
(\ref{con-sp-1-flat-newt}) and (\ref{FRW6gi-flat-newt}) imply
\begin{equation}
  \label{eq:poisson}
  \nabla^2\varphi(\bm{x})=4\pi G_{\text{N}} \rho_\een(\bm{x}), \qquad
    \rho_\een(\bm{x})\equiv\dfrac{\varepsilon^{\text{gi}}_\een(t_0,\bm{x})}{c^2},
\end{equation}
which is the Poisson equation of the Newtonian Theory of Gravity.
With (\ref{eq:poisson}) the third requirement for the non-relativistic
limit, i.e., a static gravitational field, has been satisfied.

The expression (\ref{Egi}) reduces in the non-relativistic limit to
$\varepsilon^{\text{gi}}_\een=-R_\een/(2\kappa)$, which is, with
(\ref{RnabEE-0}) and (\ref{FRW6gi-flat-newt}), equivalent to the
Poisson equation (\ref{eq:poisson}).  Using that
$\varepsilon_\nul=n_\nul mc^2$ and $\varepsilon_\een=n_\een mc^2$,
expression (\ref{nu2}) reduces in the non-relativistic limit to the
familiar result
\begin{equation}
  \label{eq:newt-ngi}
    n^{\text{gi}}_\een(\bm{x})=\dfrac{\varepsilon^{\text{gi}}_\een(\bm{x})}{mc^2},
\end{equation}
where it has been used that in the non-relativistic limit
$R_\een=-2\kappa\varepsilon^{\text{gi}}_\een$. 

The universe is in the non-relativistic limit not static, since
$H\neq0$ and $\dot{H}\neq0$, as follows from the background equations
(\ref{subeq:einstein-flrw}) with $w=0$ and $R_\nul=0$.  In the
non-relativistic limit a local density perturbation does not follow
the global expansion of the universe and the system of reference has
become co-moving.  Since density perturbations do not evolve in the
non-relativistic limit, they are essentially \emph{adiabatic}, in
accordance with the conclusion at the end of
Section~\ref{sec:diabatic}.

The gauge modes $\hat{\varepsilon}_\een$, $\hat{n}_\een$ and
$\hat{\theta}_\een$ (\ref{tr-theta}) do \emph{not} vanish, since
$\psi=C$ is an arbitrary constant which cannot be fixed. As a
consequence, the gauge-dependent quantities $\varepsilon_\een$,
$n_\een$ and $\theta_\een$ do \emph{not} become gauge-invariant in the
non-relativistic limit.  In fact, the gauge modes
$\hat{\varepsilon}_\een$, $\hat{n}_\een$ and $\hat{\theta}_\een$ are
solutions of (\ref{FRW4gi-flat-newt}) and (\ref{FRW4agi-flat-newt}).
Since these equations are decoupled from the physical equations
(\ref{con-sp-1-flat-newt}) and (\ref{FRW6gi-flat-newt}) they are not
part of the Newtonian Theory of Gravity and need not be considered.

Finally, the potential $\zeta$ which occurs by (\ref{decomp-hij-par})
in $R_\een$, (\ref{driekrom}), and $\theta_\een$, (\ref{fes5}), in the
general relativistic case, drops from the perturbation theory in the
non-relativistic limit.  Consequently, one is left with one potential
$\varphi(\bm{x})$ only.

It has been shown that equations (\ref{subeq:pertub-flrw}) combined
with the \emph{unique} gauge-invariant quantities
(\ref{subeq:gi-quant}) reduce in the non-relativistic limit to the
Newtonian results (\ref{eq:poisson}) and (\ref{eq:newt-ngi}).
Consequently, $\varepsilon^{\text{gi}}_\een$ and $n^{\text{gi}}_\een$
are the real, physical perturbations to the energy density and
particle number density, respectively.

\section{\label{sec:voorbeeld}Example: the flat \textsc{flrw} Universe}

In this section analytic solutions of equations~(\ref{subeq:final})
are derived for a flat ($R_\nul=0$) \textsc{flrw}
universe with a vanishing cosmological constant ($\Lambda=0$) in its
radiation-dominated phase and in the era after decoupling of matter
and radiation.

\subsection{\label{sec:rad-dom-era}Radiation-dominated Era}

In the radiation-dominated epoch one has
$\varepsilon=a_{\text{B}}T_\gamma^4$, where $a_{\text{B}}$ is the
black body constant and $T_\gamma$ the radiation temperature. The
pressure is given by the ultra-relativistic equation of state
$p=\tfrac{1}{3}\varepsilon$, so that $p_n=0$,
$p_\varepsilon=\tfrac{1}{3}$, implying, with (\ref{eq:beta-matter}),
that $\beta^2=\tfrac{1}{3}$. The perturbation equations
(\ref{subeq:final}) reduce to
\begin{subequations}
\label{subeq:final-rad}
\begin{align}
 &  \ddot{\delta}_\varepsilon-H\dot{\delta}_\varepsilon-
  \left[\frac{1}{3}\frac{\nabla^2}{a^2}-
   \tfrac{2}{3}\kappa\varepsilon_\nul\right]
   \delta_\varepsilon  = 0,   \label{eq:delta-rad} \\
 &   \frac{1}{c}\frac{{\text{d}}}{{\text{d}} t}
        \left(\delta_n-\tfrac{3}{4}\delta_\varepsilon\right) = 0.
            \label{eq:entropy-rad}
\end{align}
\end{subequations}
Since $p_n=0$ density perturbations evolve adiabatically, see
Section~\ref{sec:diabatic}\@.  Equation (\ref{eq:entropy-rad})
expresses the fact that perturbations in the particle number density
are \emph{gravitationally} coupled to perturbations in the energy
density.

Equation (\ref{eq:delta-rad}) will now be rewritten in dimensionless
quantities.  The solutions of the background equations
(\ref{subeq:einstein-flrw}) are given by
\begin{equation}
  \label{eq:exact-sol-rad}
   H\propto t^{-1}, \qquad  \varepsilon_\nul\propto t^{-2}, \qquad
   n_\nul\propto t^{-3/2}, \qquad a\propto t^{1/2},
\end{equation}
implying that $T_{\nul\gamma}\propto a^{-1}$.  The dimensionless time
$\tau$ is defined by $\tau\equiv t/t_0$. Since $H\equiv\dot{a}/a$, one
finds that
\begin{equation}
   \frac{{\text{d}}^k}{c^k{\text{d}}
t^k}=\left[\frac{1}{ct_0}\right]^k\frac{{\text{d}}^k}{{\text{d}}\tau^k}=
   \left[2H(t_0)\right]^k
   \frac{{\text{d}}^k}{{\text{d}}\tau^k}, \qquad k=1,2.  \label{dtau-n}
\end{equation}
Substituting $\delta_\varepsilon(t,\bm{x})=
\delta_\varepsilon(t,\bm{q})\exp(\text{i}\bm{q}\cdot\bm{x})$ into
equation (\ref{eq:delta-rad}) and using (\ref{dtau-n}) yields
\begin{equation}
  \label{eq:new-rad}
  \delta_\varepsilon^{\prime\prime}-\dfrac{1}{2\tau}\delta_\varepsilon^\prime+
  \left[\dfrac{\mu_{\text{r}}^2}{4\tau}+\dfrac{1}{2\tau^2}\right]\delta_\varepsilon=0,
\qquad \tau\ge 1,
\end{equation}
where a prime denotes differentiation with respect to $\tau$. The
parameter $\mu_{\text{r}}$ is given by
\begin{equation}
     \mu_\text{r} \equiv
\frac{2\pi}{\lambda_0}\frac{1}{H(t_0)}\frac{1}{\sqrt{3}}, \qquad
    \lambda_0\equiv\lambda a(t_0),  
\label{xi}
\end{equation} 
with $\lambda a(t_0)$ the physical scale of a perturbation at time
$t_0$, and $|\bm{q}|=2\pi/\lambda$.  To solve equation
(\ref{eq:new-rad}), replace $\tau$ by
$x\equiv\mu_{\text{r}}\sqrt{\tau}$. After transforming back to $\tau$,
one finds
\begin{equation}
 \delta_\varepsilon(\tau,\bm{q}) =
      \Bigl[A_1(\bm{q})\sin\left(\mu_{\text{r}}\sqrt{\tau}\right) +
A_2(\bm{q})\cos\left(\mu_{\text{r}}\sqrt{\tau}\right)\Bigr]\sqrt{\tau},
\label{nu13}
\end{equation}
where the `constants' of integration
$A_1(\bm{q})$ and $A_2(\bm{q})$ are given by
\begin{equation}
\label{subeq:C1-C2}
   A_{1\atop2}(\bm{q}) =
   \delta_\varepsilon(t_0,\bm{q}) {\sin\mu_{\text{r}}\atop\cos\mu_{\text{r}}}\mp
  \frac{1}{\mu_{\text{r}}}{\cos\mu_{\text{r}}\atop\sin\mu_{\text{r}}}
\left[\delta_\varepsilon(t_0,\bm{q})-\frac{\dot{\delta}
_\varepsilon(t_0,\bm{q})}{H(t_0)}\right].
\end{equation}
For large-scale perturbations ($\lambda\rightarrow\infty$), it follows
from (\ref{nu13}) and (\ref{subeq:C1-C2}) that
\begin{equation}
    \delta_\varepsilon(t) = -\left[\delta_\varepsilon(t_0)-
       \frac{\dot{\delta}_\varepsilon(t_0)}{H(t_0)}\right]
       \frac{t}{t_0}
    +\left[2\delta_\varepsilon(t_0)
    - \frac{\dot{\delta}_\varepsilon(t_0)}{H(t_0)}\right]
     \left(\frac{t}{t_0}\right)^{\tfrac{1}{2}}. \label{delta-H-rad}
\end{equation}
The energy density contrast has two contributions to the growth rate,
one proportional to $t$ and one proportional to $t^{1/2}$.  These two
solutions have been found, with the exception of the precise factors
of proportionality, by a large number of authors, see Lifshitz and
Khalatnikov \cite{c15}, (8.11), Adams and Canuto
\cite{adams-canuto1975}, (4.5b), Olson~\cite{olson1976}, page 329,
Peebles~\cite{c11}, (86.20), Kolb and Turner~\cite{kolb}, (9.121) and
Press and Vishniac~\cite{C12}, (33).  Consequently, the generalised
cosmological perturbation theory corroborates for large-scale
perturbations the results of the literature.

A new result is that small-scale perturbations ($\lambda\rightarrow0$)
oscillate with an \emph{increasing} amplitude according to
\begin{equation}\label{dc-small}
   \delta_\varepsilon(t,\bm{q}) \approx
\delta_\varepsilon(t_0,\bm{q})
    \left(\frac{t}{t_0}\right)^{\tfrac{1}{2}}
   \cos\left[\mu_{\text{r}}-
\mu_{\text{r}}\left(\frac{t}{t_0}\right)^{\tfrac{1}{2}}\right],
\end{equation}
as follows from (\ref{nu13}) and (\ref{subeq:C1-C2}).  In the standard
equation (\ref{eq:delta-standard}) $\vartheta_\een=0$, i.e., no fluid
flow, so that this equation yields oscillating density perturbations
with a \emph{constant} amplitude.  In contrast, the new theory
(\ref{subeq:final-rad}) has $\vartheta_\een\neq0$, implying that
growing perturbations are found.  This is explained in detail in
Section~\ref{sec:stand-theory}\@.

By virtue of equation (\ref{eq:entropy-rad}) particle number density
fluctuations $\delta_n$ are coupled to fluctuations
$\delta_\varepsilon$ in the energy density.  Since equation
(\ref{eq:entropy-rad}) is independent of the nature of the particles
and \textsc{cdm} interacts only via gravity with ordinary matter and
radiation, the fluctuations in \textsc{cdm} are gravitationally
coupled to fluctuations in the energy density.  In other words,
$\delta_n=\tfrac{3}{4}\delta_\varepsilon$ holds true for ordinary
matter as well as \textsc{cdm}. Consequently, in the
radiation-dominated universe \textsc{cdm} does not contract faster
than ordinary matter, so that star formation can commence only after
decoupling.

\subsection{\label{sec:mat-dom-era}Era after Decoupling of Matter and Radiation}

Once protons and electrons combine to yield hydrogen, the radiation
pressure becomes negligible, and the equations of state
(\ref{eq:es-p-T}) become those of a non-relativistic monatomic perfect
gas
\begin{equation}
  \varepsilon(n,T) = nmc^2+\tfrac{3}{2}nk_{\text{B}}T, \qquad
  p(n,T) = nk_{\text{B}}T,    \label{state-mat}
\end{equation}
where $k_{\text{B}}$ is Boltzmann's constant, $m$ the mean particle
mass, and $T$ the temperature of the matter.  It is assumed that the
\textsc{cdm} particle mass is larger than or equal to the proton mass,
${m_{\text{CDM}}\ge m_{\text{H}}}$, implying that for the mean
particle mass $m$ one has $mc^2\gg k_{\text{B}}T$, so that $w\equiv
p_\nul/\varepsilon_\nul\ll1$.  Therefore, as follows from the
background equations (\ref{FRW3}) and (\ref{FRW2}), one may neglect
the pressure $nk_{\text{B}}T$ and the kinetic energy density
$\tfrac{3}{2}nk_{\text{B}}T$ with respect to the rest-mass energy
density $nmc^2$ in the \emph{un}perturbed universe.  However,
neglecting the pressure in the perturbed universe yields non-evolving
density perturbations with a \emph{static} gravitational field, as has
been demonstrated in Section~\ref{sec:newt-limit}\@.  Consequently, it
is important to take the pressure into account in the perturbed
universe.

Eliminating $T$ from (\ref{state-mat}) yields
$p(n,\varepsilon)=\tfrac{2}{3}(\varepsilon-nmc^2)$, so that
$p_\varepsilon\equiv(\partial p/\partial\varepsilon)_n=\tfrac{2}{3}$
and $p_n\equiv(\partial p/\partial
n)_\varepsilon=-\tfrac{2}{3}mc^2$. Substituting $p_n$, $p_\varepsilon$
and (\ref{state-mat}) into (\ref{eq:beta-matter}) on finds, using
$mc^2\gg k_{\text{B}}T$,
\begin{equation}
     \beta(t)\approx \frac{v_{\text{s}}(t)}{c}=\sqrt{\frac{5}{3}
        \frac{k_{\text{B}}T_\nul(t)}{mc^2}}, \qquad 
     w(t)\approx\frac{k_{\text{B}}T_\nul(t)}{mc^2}\approx\tfrac{3}{5}\beta^2(t),
      \qquad T_\nul\propto a^{-2},
\label{coef-nu1}
\end{equation}
with $v_{\text{s}}$ the adiabatic speed of sound and $T_\nul$ the
matter temperature.  The fact that $T_\nul\propto a^{-2}$ follows from
the equations of state (\ref{state-mat}) and the conservation laws
(\ref{FRW2}) and (\ref{FRW2a}). This, in turn, implies with
(\ref{coef-nu1}) that $\dot{\beta}/\beta=-H$.  The system
(\ref{subeq:final}) can now be rewritten as
\begin{subequations}
  \label{final-dust}
  \begin{align}
   & \ddot{\delta}_\varepsilon + 3H\dot{\delta}_\varepsilon-
  \left[\beta^2\frac{\nabla^2}{a^2}+
   \tfrac{5}{6}\kappa\varepsilon_\nul\right]
   \delta_\varepsilon=-\frac{2}{3}\frac{\nabla^2}{a^2}(\delta_n-\delta_\varepsilon),
   \label{dde-dn-de}\\
   & \frac{1}{c}\frac{{\text{d}}}{{\text{d}} t}
   \left(\delta_n-\delta_\varepsilon\right)=
   -2H\left(\delta_n-\delta_\varepsilon\right), \label{eq:dn-de}
  \end{align}
\end{subequations}
where $w\ll1$ and $\beta^2\ll1$ have been neglected with respect to constants
of order unity.

From equation (\ref{eq:dn-de}) it follows that
\begin{equation}
  \label{eq:dn-dn-a-2}
  \delta_n-\delta_\varepsilon \propto a^{-2},
\end{equation}
where it is used that $H\equiv\dot{a}/a$.

Using that $k_{\text{B}}T_\nul\ll mc^2$, one finds for the perturbed
counterparts of (\ref{state-mat})
\begin{equation}
  \label{eq:dn-de-dT}
  \delta_n-\delta_\varepsilon\approx
        -\dfrac{3}{2}\dfrac{k_{\text{B}}T_\nul}
        {mc^2}\delta_T, \qquad
   \delta_p = \delta_n+\delta_T,
\end{equation}
where $\delta_p$ is the relative pressure perturbation defined by
$\delta_p\equiv p^{\text{gi}}_\een/p_\nul$ and $\delta_T$ is the relative
matter temperature perturbation defined by $\delta_T\equiv
T^{\text{gi}}_\een/T_\nul$, see (\ref{eq:gi-p}) and (\ref{eq:gi-T}).
Combining (\ref{coef-nu1}) and (\ref{eq:dn-dn-a-2}) one finds from
(\ref{eq:dn-de-dT}) that $\delta_T$ is nearly constant, i.e.,
\begin{equation}
  \label{eq:dT-constant}
  \delta_T(t,\bm{x})\approx\delta_T(t_0,\bm{x}),
\end{equation}
to a very good approximation.

From (\ref{coef-nu1}) and (\ref{eq:dn-de-dT}) one may infer that the
source term of equation (\ref{dde-dn-de}) is of the same order of
magnitude as the term with coefficient $\beta^2$.  It follows from
(\ref{eq:b3}) that the source term vanishes for barotropic equations
of state $p=p(\varepsilon)$.  As will become clear in
Section~\ref{sec:pop-iii-stars}, the source term of equation
(\ref{dde-dn-de}) is crucial for the understanding of star formation
in the early universe after decoupling.  Therefore, the realistic
equation of state $p=p(n,\varepsilon)$ has been incorporated from the
outset in the perturbation theory (\ref{subeq:final}), so that
pressure perturbations in the perturbed universe can be taken into
account.

Equation (\ref{dde-dn-de}) will now be rewritten in dimensionless
quantities.  The solutions of the background equations
(\ref{subeq:einstein-flrw}) are given by
\begin{equation}
  \label{eq:exact-sol-mat}
   H\propto t^{-1}, \qquad  \varepsilon_\nul\propto t^{-2}, \qquad
   n_\nul\propto t^{-2}, \qquad a\propto t^{2/3},
\end{equation}
where the kinetic energy density and pressure have been neglected with
respect to the rest-mass energy density.  The dimensionless time
$\tau$ is defined by $\tau\equiv t/t_0$.  Using that
$H\equiv\dot{a}/a$, one gets
\begin{equation}
   \frac{{\text{d}}^k}{c^k{\text{d}}
t^k}=\left[\frac{1}{ct_0}\right]^k\frac{{\text{d}}^k}{{\text{d}}\tau^k}=
   \left[\tfrac{3}{2}H(t_0)\right]^k
   \frac{{\text{d}}^k}{{\text{d}}\tau^k}, \qquad k=1,2.    \label{dtau-n-dust}
\end{equation}
Substituting
$\delta_\varepsilon(t,\bm{x})=\delta_\varepsilon(t,\bm{q})\exp(\text{i}\bm{q}\cdot\bm{x})$
and
$\delta_n(t,\bm{x})=\delta_n(t,\bm{q})\exp(\text{i}\bm{q}\cdot\bm{x})$
into equations (\ref{final-dust}) and using (\ref{eq:dn-de-dT}),
(\ref{eq:dT-constant}) and (\ref{dtau-n-dust}) one finds that
equations (\ref{final-dust}) can be combined into one equation
\begin{equation}\label{eq:dust-dimless}
    \delta_\varepsilon^{\prime\prime}+\frac{2}{\tau}\delta_\varepsilon^\prime+
\left[\dfrac{4}{9}\dfrac{\mu_{\text{m}}^2}{\tau^{8/3}}-\frac{10}{9\tau^2}
\right]\delta_\varepsilon=
-\dfrac{4}{15}\dfrac{\mu^2_{\text{m}}}{\tau^{8/3}}
\delta_T(t_0,\bm{q}), \qquad \tau\ge 1,
\end{equation}
where a prime denotes differentiation with respect to~$\tau$. The
parameter $\mu_{\text{m}}$ is given~by
\begin{equation}\label{eq:const-mu}
\mu_{\text{m}}\equiv\frac{2\pi}{\lambda_0}\frac{1}{H(t_0)}\frac{
v_{\text{s}}(t_0)}{c},  \qquad \lambda_0\equiv\lambda a(t_0),
\end{equation}
with $\lambda a(t_0)$ the physical scale of a perturbation at time
$t_0$, and $|\bm{q}|=2\pi/\lambda$.  To solve equation
(\ref{eq:dust-dimless}) replace $\tau$ by
$x\equiv2\mu_{\text{m}}\tau^{-1/3}$. After transforming back to
$\tau$, one finds for the general solution of (\ref{eq:dust-dimless})
\begin{equation}\label{eq:matter-physical}
 \delta_\varepsilon(\tau,\bm{q}) =
    \Bigl[B_1(\bm{q})
      J_{+\frac{7}{2}}\bigl(2\mu_{\text{m}}\tau^{-1/3}\bigr)
          + B_2(\bm{q})J_{-\frac{7}{2}}
       \bigl(2\mu_{\text{m}}\tau^{-1/3}\bigr)\Bigr]\tau^{-1/2}
      -\dfrac{3}{5}\left[1+
   \dfrac{5\tau^{2/3}}{2\mu_{\text{m}}^2}\right]\delta_T(t_0,\bm{q}),   
\end{equation}
where $J_{\pm7/2}(x)$ are Bessel functions of the first kind and $B_1(\bm{q})$
and $B_2(\bm{q})$ are the `constants' of integration, calculated with
the help of \textsc{Maxima}~\cite{maxima}:
\begin{align}
\label{subeq:B1-B2}
  B_{1\atop2}(\bm{q}) =&\; \dfrac{3\sqrt{\pi}}{20\mu_{\text{m}}^{3/2}}
   \left[\bigl(4\mu_{\text{m}}^2-5\bigr){\cos2\mu_{\text{m}}\atop\sin2\mu_{\text{m}}}
    \mp10\mu_{\text{m}}{\sin2\mu_{\text{m}}\atop\cos2\mu_{\text{m}}}\right]\delta_T(t_0,\bm{q})\;+
  \nonumber \\
  & \; \dfrac{\sqrt{\pi}}{8\mu_{\text{m}}^{7/2}}\left[\bigl(8\mu_{\text{m}}^4-
     30\mu_{\text{m}}^2+15\bigr){\cos2\mu_{\text{m}}\atop\sin2\mu_{\text{m}}}
       \mp\bigl(20\mu_{\text{m}}^3-30\mu_{\text{m}}\bigr){\sin2\mu_{\text{m}}\atop\cos2\mu_{\text{m}}}\right]
  \delta_\varepsilon(t_0,\bm{q})\;+ \nonumber \\
   & \; \dfrac{\sqrt{\pi}}{8\mu_{\text{m}}^{7/2}}\left[\bigl(24\mu_{\text{m}}^2-15\bigr)
       {\cos2\mu_{\text{m}}\atop\sin2\mu_{\text{m}}}\pm
      \bigl(8\mu_{\text{m}}^3-30\mu_{\text{m}}\bigr){\sin2\mu_{\text{m}}\atop\cos2\mu_{\text{m}}}
     \right]\dfrac{\dot{\delta}_\varepsilon(t_0,\bm{q})}{H(t_0)}.
\end{align}
In the large-scale limit $\lambda\rightarrow\infty$ terms with
$\nabla^2$ vanish, so that the general solution of equation
(\ref{eq:dust-dimless}) is
\begin{equation}
    \delta_\varepsilon(t) =
    \dfrac{1}{7}\left[5\delta_\varepsilon(t_0)+\dfrac{2\dot{\delta}_\varepsilon(t_0)}{H(t_0)}\right]
 \left(\frac{t}{t_0}\right)^{\tfrac{2}{3}}
     + \dfrac{2}{7}\left[\delta_\varepsilon(t_0)-\dfrac{\dot{\delta}_\varepsilon(t_0)}{H(t_0)}\right]
      \left(\frac{t}{t_0}\right)^{-\tfrac{5}{3}}.
  \label{eq:new-dust-53-adiabatic}
\end{equation}
Thus, for large-scale perturbations the initial value
$\delta_T(t_0,\bm{q})$ does not play a role during the evolution:
large-scale perturbations evolve only under the influence of gravity.
These perturbations are so large that heat exchange does not play a
role during their evolution in the linear phase.  For perturbations
much larger than the Jeans scale (i.e., the peak value in
Figure~\ref{fig:collapse}), gravity alone is insufficient to explain
star formation within $13.82\,\text{Gyr}$, since they grow as
$\delta_\varepsilon\propto t^{2/3}$.  The solution proportional to
$t^{2/3}$ is a standard result.  Since $\delta_\varepsilon$ is
gauge-invariant, the standard non-physical gauge mode proportional to
$t^{-1}$ is absent from the new theory. Instead, a physical mode
proportional to $t^{-5/3}$ is found.  This mode has also been found by
Bardeen~\cite{c13}, Table~I, and by Mukhanov
\emph{et~al.}~\cite{mfb1992}, expression (5.33).  In order to arrive
at the $t^{-5/3}$ mode, Bardeen has to use the `uniform expansion
gauge.' In the generalised cosmological perturbation theory the Hubble
function is \emph{automatically} uniform, (\ref{theta-gi}), without
any additional gauge condition.  Consequently,
(\ref{eq:new-dust-53-adiabatic}) is in agreement with results given in
the literature.

In the small-scale limit $\lambda\rightarrow0$, one finds
\begin{equation}
  \label{eq:small-scale-dust}
  \delta_\varepsilon(t,\bm{q})\approx -\tfrac{3}{5}\delta_T(t_0,\bm{q})+
     \left(\dfrac{t}{t_0}\right)^{-\tfrac{1}{3}}
   \Bigl[\tfrac{3}{5}\delta_T(t_0,\bm{q})+\delta_\varepsilon(t_0,\bm{q})\Bigr]
      \cos\left[2\mu_{\text{m}}-2\mu_{\text{m}}\left(\dfrac{t}{t_0} \right)^{-\tfrac{1}{3}}  \right].
\end{equation}
Thus, density perturbations with scales much smaller than the Jeans
scale oscillate with a decaying amplitude which is smaller than unity:
these perturbations are so small that gravity is insufficient to let
perturbations grow. Heat loss alone is not enough for the growth of
density perturbations.  Consequently, perturbations with scales much
smaller than the Jeans scale will never reach the non-linear regime.

In the next section it is shown that for density perturbations with
scales of the order of the Jeans scale, the action of both gravity and
heat loss together may result in massive stars several hundred million
years after decoupling of matter and radiation.

\section{\label{sec:pop-iii-stars}Evolution of Small-Scale
  Inhomogeneities after Decoupling}

In this section it is demonstrated that the generalised cosmological
perturbation theory based on the General Theory of Relativity combined
with thermodynamics and a realistic equation of state for the pressure
$p=p(n,\varepsilon)$, yields that in the era after decoupling of
matter and radiation small-scale inhomogeneities may grow very fast.
As in Section~\ref{sec:voorbeeld}, a flat ($R_\nul=0$) \textsc{flrw}
universe with vanishing cosmological constant ($\Lambda=0$) is
considered.

\subsection{\label{sec:p-obs-q}Introducing Observable Quantities}

The parameter $\mu_{\text{m}}$ (\ref{eq:const-mu}) will be expressed
in observable quantities, namely the present values of the background
radiation temperature, $T_{\nul\gamma}(t_{\text{p}})$, the Hubble
parameter, $\mathcal{H}(t_{\text{p}})=cH(t_{\text{p}})$, and the
redshift at decoupling, $z(t_{\text{dec}})$.

The redshift $z(t)$ as a function of the scale factor $a(t)$ is given
by
\begin{equation}
  \label{eq:redshift}
  z(t)=\dfrac{a(t_{\text{p}})}{a(t)}-1,
\end{equation}
where $a(t_{\text{p}})$ is the present value of the scale factor. For
a flat \textsc{flrw} universe one may take $a(t_{\text{p}})=1$.

Substituting (\ref{coef-nu1}) into (\ref{eq:const-mu}), one gets
\begin{equation}
  \label{eq:H-dec-wmap-T-dec}
  \mu_{\text{m}}=\dfrac{2\pi}{\lambda_{\text{dec}}}\dfrac{1}{H(t_{\text{dec}})}
     \sqrt{\dfrac{5}{3}
    \dfrac{k_{\text{B}}T_{\nul}(t_{\text{dec}})}{mc^2}},
        \qquad \lambda_{\text{dec}}\equiv\lambda a(t_{\text{dec}}),
\end{equation}
where $t_{\text{dec}}$ is the time when a perturbation starts to
contract and $\lambda_{\text{dec}}$ the physical scale of a
perturbation at time $t_{\text{dec}}$.  Using (\ref{eq:exact-sol-mat})
and (\ref{eq:redshift}), one finds
\begin{equation}
  \label{eq:H-dec-wmap}
  \mu_{\text{m}}=\dfrac{2\pi}{\lambda_{\text{dec}}}
     \dfrac{1}{\mathcal{H}(t_{\text{p}})\bigl[z(t_{\text{dec}})+1\bigr]}
\sqrt{\dfrac{5}{3}\dfrac{k_{\text{B}}T_{\nul\gamma}(t_{\text{p}})}{m}},
        \qquad \lambda_{\text{dec}}\equiv\lambda a(t_{\text{dec}}),
\end{equation}
where it is used that
$T_\nul(t_{\text{dec}})=T_{\nul\gamma}(t_{\text{dec}})$, and that
$T_{\nul\gamma}\propto a^{-1}$ after decoupling, as follows from
(\ref{eq:exact-sol-rad}).  With (\ref{eq:H-dec-wmap}) the parameter
$\mu_{\text{m}}$ is expressed in observable quantities.

\subsection{\label{sec:wmap}Initial Values from Planck Satellite}

The physical quantities measured by Planck \cite{2013arXiv1303.5076P}
and needed in the parameter $\mu_{\text{m}}$ (\ref{eq:H-dec-wmap}) of
the generalised cosmological perturbation theory are the redshift at
decoupling, the present values of the Hubble function and the
background radiation temperature, the age of the universe and the
fluctuations in the background radiation temperature.  The numerical
values of these quantities are
\begin{subequations}
\label{subeq:wmap}
\begin{align}
   z(t_{\text{dec}})& = 1090.43,  \label{eq:init-wmap-z} \\
   cH(t_{\text{p}})=\mathcal{H}(t_{\text{p}})& =
   67.3\,\text{km/sec/Mpc}=2.18\times10^{-18}\,\text{sec}^{-1},\label{eq:init-wmap-T}\\
  T_{\nul\gamma}(t_\text{p})& = 2.725\,\text{K}, \label{eq:back-T-tp}\\
  t_{\text{p}} & =13.82\,\text{Gyr}, \\
  \delta_{T_\gamma}(t_{\text{dec}}) & \lesssim 10^{-5}. \label{eq:T-gamma}
\end{align}
\end{subequations}
Substituting the observed values
(\ref{eq:init-wmap-z})--(\ref{eq:back-T-tp}) into
(\ref{eq:H-dec-wmap}), one finds
\begin{equation}\label{eq:nu-m-lambda}
    \mu_{\text{m}}=\dfrac{16.56}{\lambda_{\text{dec}}}, \qquad
\lambda_{\text{dec}} \text{ in pc},
\end{equation}
where it is used that the proton mass is
$m=m_{\text{H}}=1.6726\times10^{-27}\,\text{kg}$, 
$1\,\text{pc}=3.0857\times10^{16}\,\text{m}=3.2616\;\text{ly}$, the
speed of light $c=2.9979\times10^8\,\text{m/s}$ and
Boltzmann's constant $k_{\text{B}}=1.3806\times10^{-23}\,\text{J}\,\text{K}^{-1}$.

The Planck observations of the fluctuations
$\delta_{T_\gamma}(t_{\text{dec}})$, (\ref{eq:T-gamma}), in the
background radiation temperature yield for the fluctuations in the
energy density
\begin{equation}
  \label{eq:fluct-en}
|\delta_\varepsilon(t_{\text{dec}},\bm{q})|\lesssim10^{-5}.
\end{equation}
In addition, it is assumed that
\begin{equation}
  \label{eq:d-fluct-en-dt}
  \dot{\delta}_\varepsilon(t_{\text{dec}},\bm{q}) \approx 0,
\end{equation}
i.e., during the transition from the radiation-dominated era to the
era after decoupling, perturbations in the energy density are
approximately constant with respect to time.  It follows from
(\ref{eq:dn-de-dT}) that during the linear phase of the evolution one
has $\delta_n(t,\bm{q})\approx\delta_\varepsilon(t,\bm{q})$, so that
the initial values $\delta_n(t_{\text{dec}},\bm{q})$ and
$\dot{\delta}_n(t_{\text{dec}},\bm{q})$ need not be specified.

\subsection{\label{sec:hier-struc}Star Formation in the Early Universe}

At the moment of decoupling of matter and radiation, photons could not
ionise matter any more and the two constituents fell out of thermal
equilibrium. As a consequence, the pressure drops from a very high
radiation pressure $p=\tfrac{1}{3}a_{\text{B}}T^4_\gamma$ just before
decoupling to a very low gas pressure $p=nk_{\text{B}}T$ after
decoupling. This fast and chaotic transition from a high pressure
epoch to a very low pressure era may result in large relative pressure
perturbations $\delta_p\equiv p^{\text{gi}}_\een/p_\nul$.  With
(\ref{eq:dn-de-dT}) and (\ref{eq:fluct-en}) it follows that
$\delta_T\equiv T^{\text{gi}}_\een/T_\nul$ could be large.  As will be shown,
relative initial pressure perturbations
\begin{equation}
  \label{eq:large-pp}
   \delta_p(t_{\text{dec}},\bm{q}) \approx
      \delta_T(t_{\text{dec}},\bm{q}) \lesssim -0.005,
\end{equation}
may result in primordial stars, the so-called (hypothetical)
Population~\textsc{iii} stars, several hundred million years after the
Big Bang.
\begin{figure}
\begin{center}
\includegraphics[scale=0.5]{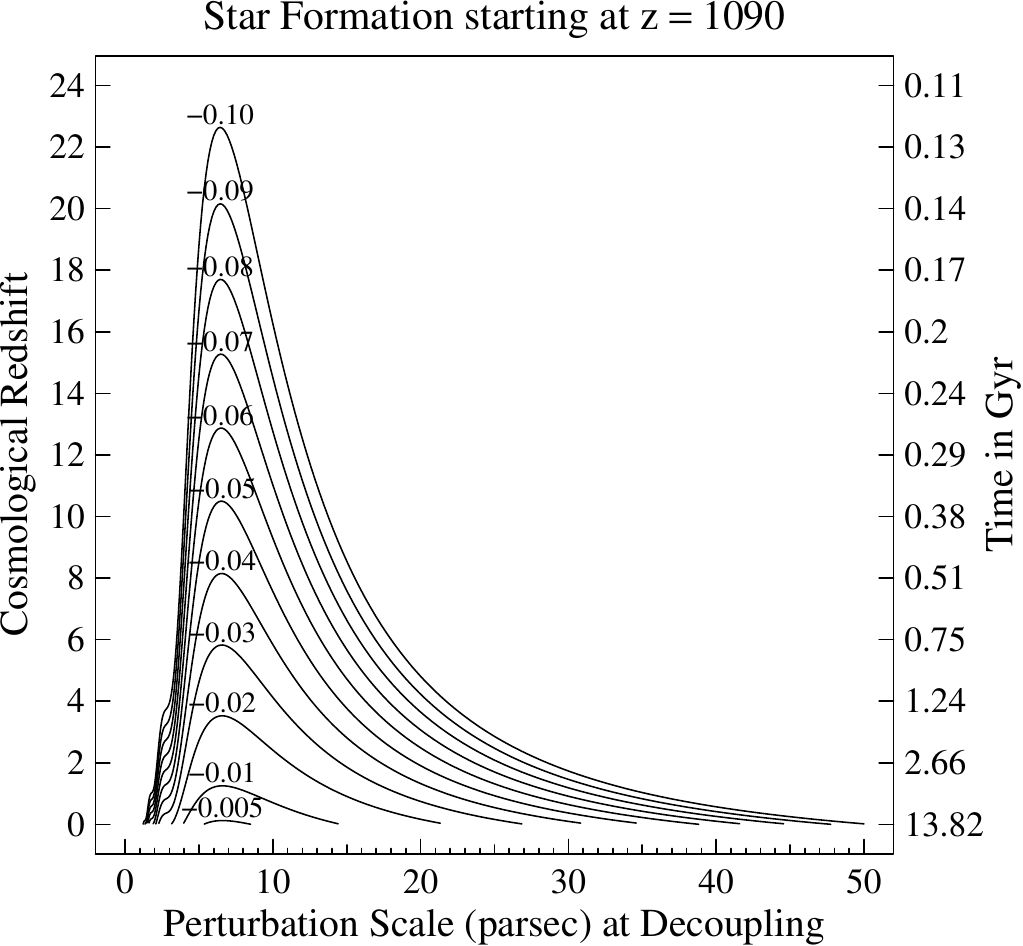}
\caption{The curves give the redshift and time, as a function of
  $\lambda_{\text{dec}}$, when a linear perturbation in the energy
  density with initial values
  $\delta_\varepsilon(t_{\text{dec}},\bm{q})\lesssim10^{-5}$ and
  $\dot{\delta}_\varepsilon(t_{\text{dec}},\bm{q})\approx0$ starting
  to grow at an initial redshift of $z(t_{\text{dec}})=1090$ becomes
  non-linear, i.e., $\delta_\varepsilon(t,\bm{q})=1$.  The numbers at
  each of the curves are the initial relative perturbations
  $\delta_T(t_{\text{dec}},\bm{q})$ in the matter temperature. For
  each curve, the Jeans scale (i.e., the peak value) is at
  $6.5\,\text{pc}$.}
\label{fig:collapse}
\end{center}
\end{figure}
The evolution equation (\ref{eq:dust-dimless}) is solved
numerically~\cite{soetaert-2010} (see attached file) and the results
are summarised in Figure~\ref{fig:collapse}, which is constructed as
follows.  For each choice of $\delta_T(t_{\text{dec}},\bm{q})$
equation (\ref{eq:dust-dimless}) is integrated for a large number of
values for the initial perturbation scale $\lambda_{\text{dec}}$ using
the initial values (\ref{eq:fluct-en}) and
(\ref{eq:d-fluct-en-dt}). The integration starts at $\tau\equiv
t/t_{\text{dec}}=1$, i.e., at $z(t_{\text{dec}})=1090$ and will be
halted if either $z=0$ (i.e., $\tau=[z(t_{\text{dec}})+1]^{3/2}$), or
$\delta_\varepsilon(t,\bm{q})=1$ for $z>0$ has been reached.  One
integration run yields one point on the curve for a particular choice
of the scale $\lambda_{\text{dec}}$ if
$\delta_\varepsilon(t,\bm{q})=1$ has been reached for $z>0$.  If the
integration halts at $z(t_{\text{p}})=0$ and still
$\delta_\varepsilon(t_{\text{p}},\bm{q})<1$, then the perturbation
belonging to that particular scale $\lambda_{\text{dec}}$ has not yet
reached its non-linear phase today, i.e., at
$t_{\text{p}}=13.82\,\text{Gyr}$. On the other hand, if the
integration is stopped at $\delta_\varepsilon(t,\bm{q})=1$ and $z>0$,
then the perturbation has become non-linear within
$13.82\,\text{Gyr}$.  This procedure has been performed for
$\delta_T(t_{\text{dec}},\bm{q})$ in the range $-0.005$, $-0.01$,
$-0.02$, \ldots, $-0.1$.  Each curve denotes the time and scale for
which $\delta_\varepsilon(t,\bm{q})=1$ for a particular value of
$\delta_T(t_{\text{dec}},\bm{q})$.

The growth of a perturbation is governed by both gravity as well as
heat loss.  From Figure~\ref{fig:collapse} one may infer that the
optimal scale for growth is around
$6.5\,\text{pc}\approx21\;\text{ly}$.  At this scale, which is
independent of the initial value of the matter temperature
perturbation $\delta_T(t_{\text{dec}},\bm{q})$, heat loss and gravity
work together perfectly, resulting in a fast growth.  Perturbations
with scales smaller than $6.5\,\text{pc}$ reach their non-linear phase
at a later time, because their internal gravity is weaker than for
large-scale perturbations. On the other hand, perturbations with
scales larger than $6.5\,\text{pc}$ cool down slower because of their
large scales, resulting also in a smaller growth rate.  Since the
growth rate decreases rapidly for perturbations with scales below
$6.5\,\text{pc}$, this scale will be considered as the
\emph{relativistic} counterpart of the classical \emph{Jeans scale}.
The relativistic Jeans scale $6.5\,\text{pc}$ is much smaller than the
horizon size at decoupling, given by
$d_{\text{H}}(t_{\text{dec}})=3ct_{\text{dec}}\approx3.5\times10^5\,\text{pc}
\approx1.1\times10^6\,\text{ly}$.

\subsection{\label{sec:heat-loss}Heat Loss of a Density Perturbation
  during its Contraction}

The heat loss of a density perturbation during its contraction after
decoupling can be calculated from the combined First and Second Law of
Thermodynamics (\ref{eq:thermo-een}) rewritten in the form
\begin{equation}
  \label{eq:fs-therm-law}
  T_\nul
  s^{\text{gi}}_\een=-\dfrac{\varepsilon_\nul}{n_\nul}(\delta_n-\delta_\varepsilon)-
  \dfrac{p_\nul}{n_\nul}\delta_n,
\end{equation}
where it is used that $w\equiv p_\nul/\varepsilon_\nul$.  Substituting
expressions (\ref{state-mat}) and (\ref{eq:dn-de-dT}) into
(\ref{eq:fs-therm-law}) and using also (\ref{eq:dT-constant}), one
finds the entropy per particle of a density perturbation:
\begin{equation}
  \label{eq:heat-exchange}
  s^{\text{gi}}_\een(t,\bm{x}) \approx \tfrac{1}{2}k_{\text{B}}
  \bigl[3\delta_T(t_0,\bm{x})-2\delta_n(t,\bm{x})\bigr],
\end{equation}
where it is used that $mc^2\gg k_{\text{B}}T_\nul$.  From
(\ref{eq:heat-exchange}) it follows that for $\delta_T\le0$ and
$\delta_n>0$ the entropy perturbation is negative,
$s^{\text{gi}}_\een<0$.  Since for growing perturbations one has
$\dot{\delta}_n>0$ the entropy perturbation decreases, i.e.,
$\dot{s}^{\text{gi}}_\een\approx-k_{\text{B}}\dot{\delta}_n<0$, during
contraction.  This is to be expected, since a local density
perturbation is not isolated from its environment.  Only for an
isolated system the entropy never decreases.

\subsection{\label{sec:jeans-mass}Relativistic Jeans Mass}

The Jeans mass at decoupling, $M_{\text{J}}(t_{\text{dec}})$, can be
estimated by assuming that a density perturbation has a spherical
symmetry with diameter the relativistic Jeans scale
$\lambda_{\text{J,dec}}\equiv\lambda_{\text{J}}
a(t_{\text{dec}})$. The relativistic Jeans mass at decoupling is then
given by
\begin{equation}
  \label{eq:M-dec}
  M_{\text{J}}(t_{\text{dec}})=
     \dfrac{4\pi}{3}\left[\tfrac{1}{2}\lambda_{\text{J,dec}}\right]^3
     n_\nul(t_{\text{dec}})m.
\end{equation}
The particle number density $n_\nul(t_{\text{dec}})$ can be calculated from its
value $n_\nul(t_{\text{eq}})$ at the end of the radiation-dominated
era.  By definition, at the end of the radiation-domination era the
matter energy density $n_\nul mc^2$ equals the energy density of the
radiation:
\begin{equation}\label{eq:mat-en-eq}
n_\nul(t_{\text{eq}})mc^2=a_{\text{B}}T_{\nul\gamma}^4(t_{\text{eq}}).
\end{equation}
Since $n_\nul\propto a^{-3}$ and $T_{\nul\gamma}\propto a^{-1}$, one
finds, using (\ref{eq:redshift}) and (\ref{eq:mat-en-eq}), for the
Jeans mass (\ref{eq:M-dec}) at time~$t_{\text{dec}}$
\begin{equation}
  \label{eq:M-dec-n-dec}
  M_{\text{J}}(t_{\text{dec}})=
  \tfrac{1}{6}\pi\lambda^3_{\text{J,dec}}
 \dfrac{a_{\text{B}}T_{\nul\gamma}^4(t_{\text{p}})}{c^2}
   \bigl[z(t_{\text{eq}})+1\bigr]\bigl[z(t_{\text{dec}})+1\bigr]^3.
\end{equation}
Using (\ref{eq:init-wmap-z}), the black body constant
$a_{\text{B}}=7.5657\times10^{-16}\,\text{J}/\text{m}^3/\text{K}^4$,
the red-shift at matter-radiation equality, $z(t_{\text{eq}})=3391$,
Planck~\cite{2013arXiv1303.5076P}, and the speed of light
$c=2.9979\times10^8\,\text{m/s}$, one finds for the Jeans mass at
decoupling
\begin{equation}
  \label{eq:jeans-mass-decoup}
  M_{\text{J}}(t_{\text{dec}})\approx4.4\times10^3\,\text{M}_\odot,
\end{equation}
where it is used that one solar mass
$1\,\text{M}_\odot=1.9889\times10^{30}\,\text{kg}$ and the
relativistic Jeans scale
$\lambda_{\text{J,dec}}=6.5\,\text{pc}$, the peak value in
Figure~\ref{fig:collapse}.

\section{\label{sec:stand-theory}Why the Newtonian Perturbation Theory
  is inadequate}

The standard evolution equation for relative density perturbations
$\delta(t,\bm{x})$ in a flat, $R_\nul=0$, \textsc{flrw} universe with
vanishing cosmological constant, $\Lambda=0$, reads
\begin{equation}\label{eq:delta-standard}
  \ddot{\delta} + 2H\dot{\delta}-
  \left[\beta^2\frac{\nabla^2}{a^2}+
   \tfrac{1}{2}\kappa\varepsilon_\nul(1+w)(1+3w)\right]
   \delta =0.
\end{equation}
This equation is derived from the Newtonian Theory of Gravity using
that $\beta^2=w=\tfrac{1}{3}$ in the radiation-dominated era and in
the epoch after decoupling $w\ll1$ and with $\beta^2$ given by
(\ref{coef-nu1}).

In order to investigate the validity of the standard equation, the
\emph{relativistic} analogue of (\ref{eq:delta-standard}) pertaining
to a barotropic equation of state $p=p(\varepsilon)$ will now be
derived from the background equations (\ref{subeq:einstein-flrw}) and
perturbation equations (\ref{subeq:pertub-gi}).  Since $p_n=0$,
equations (\ref{FRW2a}) and (\ref{FRW4agi}) need not be considered.
Moreover, $p_\varepsilon=\beta^2$, (\ref{eq:beta-matter}), implying
that the perturbed pressure is given by
$p_\een=\beta^2\varepsilon_\een$. With $R_\nul=0$, $\Lambda=0$ and
$\delta\equiv\varepsilon_\een/\varepsilon_\nul$, equation
(\ref{FRW4gi}) becomes
\begin{equation}
  \label{eq:dot-delta}
  \dot{\delta} + 3H\delta\Bigl[\beta^2+\tfrac{1}{2}(1-w)\Bigr]+
     (1 + w)\left[\vartheta_\een + \dfrac{R_\een}{4H}\right]=0,
   \qquad w\equiv\dfrac{p}{\varepsilon}, \quad
       \beta^2\equiv\dfrac{\dot{p}}{\dot{\varepsilon}},
\end{equation}
where $\kappa\varepsilon_\nul=3H^2$, (\ref{FRW3}), has been used.
Differentiating (\ref{eq:dot-delta}) with respect to time and
eliminating the time-derivatives of $H$, $\vartheta_\een$ and $R_\een$
with the help of the background equations (\ref{subeq:einstein-flrw})
and perturbation equations (\ref{subeq:pertub-gi}), respectively, and,
subsequently, eliminating $R_\een$ with the help of
(\ref{eq:dot-delta}), one finds, using \textsc{Maxima}~\cite{maxima},
that the set of equations (\ref{subeq:pertub-gi}) reduces to the
system
\begin{subequations}
\label{subeq:algemene-vergelijkingen}
\begin{align}
 & \ddot{\delta}+2H\dot{\delta}\Bigl[1+3\beta^2-3w\Bigr]-\biggl[\beta^2\dfrac{\nabla^2}{a^2}+
   \tfrac{1}{2}\kappa\varepsilon_\nul\Bigl((1+w)(1+3w) \Bigr. \Bigr. \nonumber \\
 &    \Bigl. \biggl. \phantom{\ddot{\delta}+2H\dot{\delta}\Bigl[1}
        +4w-6w^2+12\beta^2w-4\beta^2-6\beta^4 \Bigr)-6\beta\dot{\beta}H\biggr]\delta  
    =-3H\beta^2(1+w)\vartheta_\een, \\
 &  \dot{\vartheta}_\een+H(2-3\beta^2)\vartheta_\een+
           \dfrac{\beta^2}{1+w}\dfrac{\nabla^2\delta}{a^2}=0,
\end{align}
\end{subequations}
where it is used that the time-derivative of $w$ is given by
(\ref{eq:time-w}). 

The system (\ref{subeq:algemene-vergelijkingen}) consists of two
\emph{relativistic} equations for two unknown quantities, namely the
density fluctuation~$\delta$ and the divergence $\vartheta_\een$ of
the spatial part of the fluid four-velocity.  Thus, the relativistic
perturbation equations (\ref{subeq:pertub-gi}) which are derived for a
general equation of state for the pressure $p=p(n,\varepsilon)$ reduce
to the relativistic system (\ref{subeq:algemene-vergelijkingen}) for a
barotropic equation of state $p=p(\varepsilon)$.  Since the gauge
modes (\ref{subeq:gauge-dep}) are solutions of the set
(\ref{subeq:pertub-gi}), the gauge modes
\begin{equation}
  \label{eq:gauge-modes-standard}
  \hat{\delta}(t,\boldsymbol{x})=
     \dfrac{\psi(\boldsymbol{x})\dot{\varepsilon}_\nul(t)}{\varepsilon_\nul(t)}=
     -3H(t)\psi(\boldsymbol{x})\bigl[1+w(t)\bigr],
  \qquad \hat{\vartheta}_\een(t,\boldsymbol{x})=
        -\dfrac{\nabla^2\psi(\boldsymbol{x})}{a^2(t)},
\end{equation}
are solutions of equations (\ref{subeq:algemene-vergelijkingen}) with
$\dot{w}=3H(1+w)(w-\beta^2)$, (\ref{eq:time-w}).

The relativistic equations (\ref{subeq:algemene-vergelijkingen}) are
exact for first-order perturbations in a fluid described by a
barotropic equation of state $p=p(\varepsilon)$. This fact has the
following consequences for the Newtonian equation
(\ref{eq:delta-standard}):

\paragraph{Radiation-dominated Era.}  In this era,
the pressure is given by a linear barotropic equation of state
$p=w\varepsilon$, so that $p_n=0$ and $p_\varepsilon=w$.  Since
$p_\varepsilon=\beta^2$, (\ref{eq:beta-matter}), it follows from
(\ref{eq:time-w}) that $\beta^2=w$ is constant. In the case of a
radiation-dominated universe this constant is
$w=\beta^2=\tfrac{1}{3}$.  Consequently, equations
(\ref{subeq:algemene-vergelijkingen}) reduce to
\begin{subequations}
\label{subeq:standard}
\begin{align}
 & \ddot{\delta} + 2H\dot{\delta}-
  \left[w\frac{\nabla^2}{a^2}+
   \tfrac{1}{2}\kappa\varepsilon_\nul(1+w)(1+3w)\right]
   \delta =-3Hw(1+w)\vartheta_\een, 
        \label{eq:delta-standard-genrel}  \\
 &  \dot{\vartheta}_\een+H(2-3w)\vartheta_\een+
           \dfrac{w}{1+w}\dfrac{\nabla^2\delta}{a^2}=0.
  \label{eq:continuity}
\end{align}
\end{subequations}
The gauge modes (\ref{eq:gauge-modes-standard}) are solutions of the
system (\ref{subeq:standard}) for $\dot{w}=0$.

Comparing the standard equation (\ref{eq:delta-standard}) with the
relativistic equations (\ref{subeq:standard}), it follows that the
standard equation implies that $\vartheta_\een=0$ and
$\nabla^2\delta=0$.  Since $\nabla^2\delta$ can be large for
small-scale perturbations, the standard equation
(\ref{eq:delta-standard}) is inadequate to study density perturbations
in the radiation-dominated era.

The generalised cosmological perturbation theory (\ref{subeq:final})
takes $\vartheta_\een$ into account, so that (\ref{subeq:final-rad})
yields oscillating density perturbations with an \emph{increasing}
amplitude.  In contrast, the standard equation
(\ref{eq:delta-standard}) for which $\vartheta_\een=0$ yields
oscillating perturbations with a \emph{constant} amplitude.

\paragraph{Era after Decoupling of Matter and Radiation.} In this era
one has $w\ll1$, and $\beta^2\ll1$.  Since $\beta^2$ is given by
(\ref{coef-nu1}) it follows that $\dot{\beta}/\beta=-H$.  Using that
$3H^2=\kappa\varepsilon_\nul$, (\ref{FRW3}), one gets
$6\beta\dot{\beta}H=-2\kappa\varepsilon_\nul\beta^2$.  Neglecting $w$
and $\beta^2$ with respect to constants of order unity, the system
(\ref{subeq:algemene-vergelijkingen}) reduces to
\begin{subequations}
\label{subeq:algemene-vergelijkingen-mat-dom}
\begin{align}
 & \ddot{\delta}+2H\dot{\delta}-\left[\beta^2\dfrac{\nabla^2}{a^2}+
   \tfrac{1}{2}\kappa\varepsilon_\nul\right]\delta=-3H\beta^2\vartheta_\een,
      \label{eq:alg-mat-dom-delta} \\
 &  \dot{\vartheta}_\een+2H\vartheta_\een+
           \beta^2\dfrac{\nabla^2\delta}{a^2}=0.
\end{align}
\end{subequations}
The gauge modes (\ref{eq:gauge-modes-standard}) are solutions of the
system (\ref{subeq:algemene-vergelijkingen-mat-dom}) for $w\ll1$ and
$\nabla^2\psi=0$.  Consequently, for the system
(\ref{subeq:algemene-vergelijkingen-mat-dom}) $\psi$ is an arbitrary
infinitesimal constant so that $\vartheta_\een$ is a purely physical
quantity. However, $\delta$ is still gauge-dependent, implying that
one cannot impose \emph{physical} initial conditions
$\delta(t_0,\boldsymbol{x})$ and $\dot{\delta}(t_0,\boldsymbol{x})$.
These facts are in line with (\ref{eq:gauge-trans-newt}) in the
non-relativistic limit, since a cosmological fluid for which $w\ll1$
and $\beta^2\ll1$ can be described by a non-relativistic equation of
state.

Just as in the radiation-dominated era, the standard equation
(\ref{eq:delta-standard}) implies that $\vartheta_\een=0$ and
$\nabla^2\delta=0$.  However, since $\beta^2\ll1$, the source term of
(\ref{eq:alg-mat-dom-delta}) is very small, so that the influence of
$\vartheta_\een$ on the evolution of a density perturbation is,
although non-zero, rather small.  This explains the fact that both the
standard equation (\ref{eq:delta-standard}) as well as the homogeneous
part of equation (\ref{dde-dn-de}) yield oscillating solutions with a
\emph{decreasing} amplitude, as can be inferred from
(\ref{eq:matter-physical}) with $\delta_T=0$.  The main disadvantage
of the standard equation (\ref{eq:delta-standard}) is that it is only
adapted to a barotropic equation of state $p=p(\varepsilon)$.
Therefore, the important phenomenon of heat exchange of a density
perturbation with its environment is not taken into account by
equation (\ref{eq:delta-standard}). The \emph{generalised cosmological
  perturbation theory} (\ref{subeq:final}) is adapted to the more
realistic equation of state $p=p(n,\varepsilon)$, so that heat
exchange is taken into account.  As a consequence, the new
perturbation theory (\ref{final-dust}) may explain the existence of
the so-called (hypothetical) Population~\textsc{iii} stars, as has
been demonstrated in Section~\ref{sec:pop-iii-stars}\@.

Is has to be concluded that the standard equation
(\ref{eq:delta-standard}) is inadequate to study the evolution of
density perturbations in the universe in the era after decoupling of
matter and radiation.

\appendix*

\section{Derivation of the Generalised Cosmological Perturbation Theory
  using Computer Algebra}

In this Appendix the perturbation equations (\ref{subeq:final}) of the
main text will be derived from the basic perturbation equations
(\ref{subeq:pertub-gi}) and the definitions (\ref{eq:contrast}).  This
will be done by first deriving the evolution equations for the
gauge-invariant quantities $\varepsilon^{\text{gi}}_\een$ and
$n^{\text{gi}}_\een$ (\ref{e-n-gi}), or, equivalently, (\ref{subeq:pertub-gi-e-n}):
\begin{subequations}
\label{subeq:eerste}
\begin{align}
  \ddot{\varepsilon}^{\text{gi}}_\een+a_1\dot{\varepsilon}^{\text{gi}}_\een+
  a_2\varepsilon^{\text{gi}}_\een & = a_3 \left(n_\een^{\text{gi}} -
  \frac{n_\nul}{\varepsilon_\nul(1+w)}\varepsilon_\een^{\text{gi}}\right),
      \label{eq:vondst2}  \\
  \dfrac{1}{c}\dfrac{\text{d}}{\text{d}t}\left(n_\een^{\text{gi}} -
  \frac{n_\nul}{\varepsilon_\nul(1+w)}\varepsilon_\een^{\text{gi}}\right)&=
    -3H\left(1-\frac{n_\nul
  p_n}{\varepsilon_\nul(1+w)}\right) \left(n_\een^{\text{gi}} -
  \frac{n_\nul}{\varepsilon_\nul(1+w)}\varepsilon_\een^{\text{gi}}\right).   \label{eq:vondst1}
\end{align}
\end{subequations}
The coefficients $a_1$, $a_2$ and $a_3$ occurring in equation
(\ref{eq:vondst2}) are given~by
\begin{subequations}
\label{subeq:coeff}
\begin{align}
  a_1 & = \dfrac{\kappa\varepsilon_\nul(1+w)}{H}
  -2\dfrac{\dot{\beta}}{\beta}+H(4-3\beta^2)
    +R_\nul\left(\dfrac{1}{3H} + \dfrac{2H(1+3\beta^2)}
  {R_\nul+3\kappa\varepsilon_\nul(1+w)}\right), \label{eq:alpha-1} \\
  a_2 & = \kappa\varepsilon_\nul(1+w)-
  4H\dfrac{\dot{\beta}}{\beta}+2H^2(2-3\beta^2)
  +R_\nul\left(\dfrac{1}{2}+
\dfrac{5H^2(1+3\beta^2)-2H\dfrac{\dot{\beta}}{\beta}}
  {R_\nul+3\kappa\varepsilon_\nul(1+w)}\right)
-\beta^2\left(\frac{\tilde{\nabla}^2}{a^2}-\tfrac{1}{2}R_\nul \right), \label{eq:alpha-2} \\
  a_3 & = \Biggl\{\dfrac{-18H^2}{R_\nul+3\kappa\varepsilon_\nul(1+w)}
  \Biggl[\varepsilon_\nul p_{\varepsilon n}(1+w)+
   \dfrac{2p_n}{3H}\dfrac{\dot{\beta}}{\beta} + p_n(p_\varepsilon-\beta^2)+n_\nul
   p_{nn}\Biggr]+p_n\Biggr\}
\left(\frac{\tilde{\nabla}^2}{a^2}-\tfrac{1}{2}R_\nul\right).
       \label{eq:alpha-3}
\end{align}
\end{subequations}
In calculating the coefficients $a_1$, $a_2$ and~$a_3$,
(\ref{subeq:coeff}), it is used that the time derivative of the
quotient $w\equiv p_\nul/\varepsilon_\nul$ is given by
(\ref{eq:time-w}).  Moreover, it is convenient \emph{not} to expand
the function $\beta^2\equiv\dot{p}_\nul/\dot{\varepsilon}_\nul$ since
this will considerably complicate the expressions for the
coefficients~$a_1$, $a_2$ and~$a_3$.

\subsection{Derivation of the Evolution Equation for the Energy
  Density Perturbation}

\begin{table}
\caption{\label{eq:aij}The coefficients $\alpha_{ij}$ figuring in
the equations~(\ref{subeq:nieuw}).}  \normalsize   
\[  \begin{array}{c|cccc} 
   & \varepsilon_\een & n_\een & \vartheta_\een & R_\een  \\ 
 \hline 
& & & & \\
 \dot{\varepsilon}_\een & 3H(1+p_\varepsilon)+\dfrac{\kappa\varepsilon_\nul(1+w)}{2H} &
      3Hp_n & \varepsilon_\nul(1+w) &
      \dfrac{\varepsilon_\nul(1+w)}{4H} \\  
& & & & \\
 \dot{n}_\een & \dfrac{\kappa n_\nul}{2H} & 3H & n_\nul &
  \dfrac{n_\nul}{4H} \\  
& & & & \\
 \dot{\vartheta}_\een & \dfrac{p_\varepsilon}{\varepsilon_\nul(1+w)}\dfrac{\tilde{\nabla}^2}{a^2} &
       \dfrac{p_n}{\varepsilon_\nul(1+w)}\dfrac{\tilde{\nabla}^2}{a^2} &
       H(2-3\beta^2) & 0 \\  
& & & & \\
 \dot{R}_\een & \dfrac{\kappa R_\nul}{3H} & 0 &
  -2\kappa\varepsilon_\nul(1+w) &
        2H+\dfrac{R_\nul}{6H} \\  
& & & & \\
 \varepsilon^{\text{gi}}_\een  & \dfrac{-R_\nul}{R_\nul+3\kappa\varepsilon_\nul(1+w)} & 0 &
     \dfrac{6\varepsilon_\nul
H(1+w)}{R_\nul+3\kappa\varepsilon_\nul(1+w)} &
     \dfrac{\tfrac{3}{2}\varepsilon_\nul(1+w)}
     {R_\nul+3\kappa\varepsilon_\nul(1+w)}  \\  
\end{array}  \]
\end{table}

In order to derive equation (\ref{eq:vondst2}), the system
(\ref{subeq:pertub-gi}) and expression (\ref{Egi}) will be rewritten
in the form, using (\ref{eq:p1}),
\begin{subequations}
\label{subeq:nieuw}
\begin{align}
  \dot{\varepsilon}_\een+\alpha_{11}\varepsilon_\een+
    \alpha_{12}n_\een+
   \alpha_{13}\vartheta_\een+\alpha_{14}R_\een & = 0,
\label{nieuw1} \\
   \dot{n}_\een+\alpha_{21}\varepsilon_\een +
    \alpha_{22}n_\een+
   \alpha_{23}\vartheta_\een+\alpha_{24}R_\een & = 0,
\label{nieuw2} \\
  \dot{\vartheta}_\een +\alpha_{31}\varepsilon_\een +
     \alpha_{32}n_\een+
   \alpha_{33}\vartheta_\een+\alpha_{34}R_\een & = 0,
\label{nieuw3} \\
    \dot{R}_\een+\alpha_{41}\varepsilon_\een+
     \alpha_{42}n_\een+
   \alpha_{43}\vartheta_\een+\alpha_{44}R_\een & = 0,
\label{nieuw4} \\
  \varepsilon^{\text{gi}}_\een+\alpha_{51}\varepsilon_\een+\alpha_{52}n_\een+
    \alpha_{53}\vartheta_\een+\alpha_{54}R_\een & = 0,
\label{nieuw5}
\end{align}
\end{subequations}
where the coefficients $\alpha_{ij}$ are given in Table~\ref{eq:aij}.

\paragraph{Step 1.} First the quantity $R_\een$ will be eliminated
from equations (\ref{subeq:nieuw}). Differentiating equation
(\ref{nieuw5}) with respect to time and eliminating the time
derivatives $\dot{\varepsilon}_\een$, $\dot{n}_\een$,
$\dot{\vartheta}_\een$ and~$\dot{R}_\een$ with the help of equations
(\ref{nieuw1})--(\ref{nieuw4}), one arrives at the equation
\begin{equation}\label{eq:equiv}
   \dot{\varepsilon}^{\text{gi}}_\een + p_1\varepsilon_\een+p_2 n_\een+
   p_3\vartheta_\een+p_4 R_\een=0,
\end{equation}
where the coefficients $p_1,\ldots,p_4$ are given by
\begin{equation}\label{eq:coef-pi}
  p_i=\dot{\alpha}_{5i}-\alpha_{51}\alpha_{1i}-
         \alpha_{52}\alpha_{2i}-\alpha_{53}\alpha_{3i}-\alpha_{54}\alpha_{4i}.
\end{equation}
From equation (\ref{eq:equiv}) it follows that
\begin{equation}\label{eq:sol-3R1}
   R_\een=-\dfrac{1}{p_4}\dot{\varepsilon}^{\text{gi}}_\een-
     \dfrac{p_1}{p_4}\varepsilon_\een-\dfrac{p_2}{p_4}n_\een-
     \dfrac{p_3}{p_4}\vartheta_\een.
\end{equation}
In this way the quantity $R_\een$ has been expressed as a linear
combination of the quantities $\dot{\varepsilon}^{\text{gi}}_\een$,
$\varepsilon_\een$, $n_\een$ and $\vartheta_\een$. Upon replacing
$R_\een$ in equations (\ref{subeq:nieuw}) by the right-hand side of
(\ref{eq:sol-3R1}), one arrives at the system of equations
\begin{subequations}
\label{subeq:tweede}
\begin{align}
 \dot{\varepsilon}_\een+q_1\dot{\varepsilon}_\een^{\text{gi}}+
   \gamma_{11}\varepsilon_\een+\gamma_{12}n_\een+
   \gamma_{13}\vartheta_\een & = 0, \label{tweede1} \\
 \dot{n}_\een+q_2\dot{\varepsilon}_\een^{\text{gi}}+
    \gamma_{21}\varepsilon_\een+\gamma_{22}n_\een+
    \gamma_{23}\vartheta_\een & = 0, \label{tweede2} \\
 \dot{\vartheta}_\een+q_3\dot{\varepsilon}_\een^{\text{gi}}+
   \gamma_{31}\varepsilon_\een+\gamma_{32}n_\een+
   \gamma_{33}\vartheta_\een & = 0, \label{tweede3} \\
     \dot{R}_\een+
   q_4\dot{\varepsilon}^{\text{gi}}_\een+\gamma_{41}\varepsilon_\een+\gamma_{42}n_\een+
   \gamma_{43}\vartheta_\een & = 0, \label{tweede4} \\
 \varepsilon^{\text{gi}}_\een+
   q_5\dot{\varepsilon}^{\text{gi}}_\een+\gamma_{51}\varepsilon_\een+\gamma_{52}n_\een+
   \gamma_{53}\vartheta_\een & = 0, \label{tweede5}
\end{align}
\end{subequations}
where the coefficients $q_i$ and $\gamma_{ij}$ are given by
\begin{equation}\label{eq:betaij}
  q_i=-\dfrac{\alpha_{i4}}{p_4}, \qquad
   \gamma_{ij}=\alpha_{ij}+q_i p_j.
\end{equation}
It has now been achieved that the quantity $R_\een$ occurs
\emph{explicitly} only in equation (\ref{tweede4}), whereas $R_\een$
occurs \emph{implicitly} in the remaining equations. Therefore,
equation (\ref{tweede4}) is not needed anymore.  Equations
(\ref{tweede1})--(\ref{tweede3}) and (\ref{tweede5}) are four ordinary
differential equations for the four unknown quantities
$\varepsilon_\een$, $n_\een$, $\vartheta_\een$ and
$\varepsilon^{\text{gi}}_\een$.

\paragraph{Step 2.} In the same way as in Step~1, the explicit
occurrence of the quantity $\vartheta_\een$ will be eliminated from the
system of equations~(\ref{subeq:tweede}).  Differentiating equation
(\ref{tweede5}) with respect to time and eliminating the time
derivatives $\dot{\varepsilon}_\een$, $\dot{n}_\een$
and $\dot{\vartheta}_\een$ with the help of equations
(\ref{tweede1})--(\ref{tweede3}), one arrives at
\begin{equation}
\label{eq:ddot-egi}
  q_5\ddot{\varepsilon}^{\text{gi}}_\een+r\dot{\varepsilon}^{\text{gi}}_\een+
     s_1\varepsilon_\een+s_2n_\een+s_3\vartheta_\een=0,
\end{equation}
where the coefficients $s_i$ and $r$ are given by
\begin{subequations}
\label{eq:coef-qi}
\begin{align}
  s_i & = \dot{\gamma}_{5i}-\gamma_{51}\gamma_{1i}-\gamma_{52}\gamma_{2i}-
       \gamma_{53}\gamma_{3i}, \\
  r & = 1+\dot{q}_5-\gamma_{51}q_1-\gamma_{52}q_2-\gamma_{53}q_3.
\end{align}
\end{subequations}
From equation (\ref{eq:ddot-egi}) it follows that
\begin{equation}\label{eq:sol-theta1}
  \vartheta\een=-\dfrac{q_5}{s_3}\ddot{\varepsilon}^{\text{gi}}_\een-
     \dfrac{r}{s_3}\dot{\varepsilon}^{\text{gi}}_\een-
     \dfrac{s_1}{s_3}\varepsilon_\een-\dfrac{s_2}{s_3}n_\een.
\end{equation}
In this way the quantity $\vartheta_\een$ is expressed as a linear
combination of the quantities $\ddot{\varepsilon}^{\text{gi}}_\een$,
$\dot{\varepsilon}^{\text{gi}}_\een$, $\varepsilon_\een$ and
$n_\een$. Upon replacing $\vartheta_\een$ in equations
(\ref{subeq:tweede}) by the right-hand side of (\ref{eq:sol-theta1}),
one arrives at the system of equations
\begin{subequations}
\label{subeq:derde}
\begin{align}
\dot{\varepsilon}_\een-\gamma_{13}\dfrac{q_5}{s_3}\ddot{\varepsilon}^{\text{gi}}_\een+
   \left(q_1-\gamma_{13}\dfrac{r}{s_3}\right)\dot{\varepsilon}^{\text{gi}}_\een
  +\left(\gamma_{11}-\gamma_{13}\dfrac{s_1}{s_3}\right)\varepsilon_\een
  +\left(\gamma_{12}-\gamma_{13}\dfrac{s_2}{s_3}\right)n_\een & =0, \label{derde1}
\\
\dot{n}_\een-\gamma_{23}\dfrac{q_5}{s_3}\ddot{\varepsilon}^{\text{gi}}_\een+
   \left(q_2-\gamma_{23}\dfrac{r}{s_3}\right)\dot{\varepsilon}^{\text{gi}}_\een
    +\left(\gamma_{21}-\gamma_{23}\dfrac{s_1}{s_3}\right)\varepsilon_\een
  +\left(\gamma_{22}-\gamma_{23}\dfrac{s_2}{s_3}\right)n_\een & =0, \label{derde2}
\\
\dot{\vartheta}_\een-\gamma_{33}\dfrac{q_5}{s_3}\ddot{\varepsilon}^{\text{gi}}_\een+
   \left(q_3-\gamma_{33}\dfrac{r}{s_3}\right)\dot{\varepsilon}^{\text{gi}}_\een
   +\left(\gamma_{31}-\gamma_{33}\dfrac{s_1}{s_3}\right)\varepsilon_\een
  +\left(\gamma_{32}-\gamma_{33}\dfrac{s_2}{s_3}\right)n_\een & =0, \label{derde3}
\\
  \dot{R}_\een-\gamma_{43}\dfrac{q_5}{s_3}\ddot{\varepsilon}^{\text{gi}}_\een+
   \left(q_4-\gamma_{43}\dfrac{r}{s_3}\right)\dot{\varepsilon}^{\text{gi}}_\een
  + \left(\gamma_{41}-\gamma_{43}\dfrac{s_1}{s_3}\right)\varepsilon_\een
  +\left(\gamma_{42}-\gamma_{43}\dfrac{s_2}{s_3}\right)n_\een & =0, \label{derde4}
\\
\varepsilon^{\text{gi}}_\een-\gamma_{53}\dfrac{q_5}{s_3}\ddot{\varepsilon}^{\text{gi}}_\een+
   \left(q_5-\gamma_{53}\dfrac{r}{s_3}\right)\dot{\varepsilon}^{\text{gi}}_\een
    +\left(\gamma_{51}-\gamma_{53}\dfrac{s_1}{s_3}\right)\varepsilon_\een
  +\left(\gamma_{52}-\gamma_{53}\dfrac{s_2}{s_3}\right)n_\een & =0. \label{derde5}
\end{align}
\end{subequations}
It has now been achieved that the quantities $\vartheta_\een$ and
$R_\een$ occur \emph{explicitly} only in equations (\ref{derde3}) and
(\ref{derde4}), whereas they occur \emph{implicitly} in the remaining
equations.  Therefore, equations (\ref{derde3}) and (\ref{derde4}) are
not needed anymore.  Equations (\ref{derde1}), (\ref{derde2}) and
(\ref{derde5}) are three ordinary differential equations for the three
unknown quantities $\varepsilon_\een$, $n_\een$ and
$\varepsilon^{\text{gi}}_\een$.

\paragraph{Step 3.} At first sight, the next steps would be to
eliminate, successively, the quantities $\varepsilon_\een$ and
$n_\een$ from equation (\ref{derde5}) with the help of equations
(\ref{derde1}) and~(\ref{derde2}).   One would then end up with a
fourth-order differential equation for the unknown quantity
$\varepsilon^{\text{gi}}_\een$.  This, however, is impossible, since
the gauge-dependent quantities $\varepsilon_\een$ and
$n_\een$ do \emph{not} occur explicitly in equation
(\ref{derde5}), as will now be shown.  Firstly, it is observed that
equation (\ref{derde5}) can be rewritten in the form
\begin{equation}\label{eq:eindelijk}
  \ddot{\varepsilon}^{\text{gi}}_\een+a_1\dot{\varepsilon}^{\text{gi}}_\een+
    a_2\varepsilon^{\text{gi}}_\een=
    a_3\left(n_\een+\dfrac{\gamma_{51}s_3-\gamma_{53}s_1}
    {\gamma_{52}s_3-\gamma_{53}s_2} \varepsilon_\een\right),
\end{equation}
where the coefficients $a_1$, $a_2$ and $a_3$ are given by
\begin{equation}
\label{subeq:vierde}
  a_1  = -\dfrac{s_3}{\gamma_{53}}+\dfrac{r}{q_5}, \qquad
  a_2  = -\dfrac{s_3}{\gamma_{53}q_5}, \qquad
  a_3  = \dfrac{\gamma_{52}s_3}{\gamma_{53}q_5}-\dfrac{s_2}{q_5}.
\end{equation}
These are precisely the coefficients (\ref{subeq:coeff}).  Secondly,
one finds
\begin{equation}\label{eq:check}
  \dfrac{\gamma_{51}s_3-\gamma_{53}s_1}{\gamma_{52}s_3-\gamma_{53}s_2}=
      -\dfrac{n_\nul}{\varepsilon_\nul(1+w)}.
\end{equation}
Finally, using the definitions (\ref{e-n-gi}) and the conservation
laws (\ref{FRW2}) and (\ref{FRW2a}), it is found that
\begin{equation}
  \label{eq:equal-gi-non-gi}
  n_\een-\dfrac{n_\nul}{\varepsilon_\nul(1+w)}\varepsilon_\een=
  n^{\text{gi}}_\een-\dfrac{n_\nul}{\varepsilon_\nul(1+w)}\varepsilon^{\text{gi}}_\een.
\end{equation}
Thus, the right-hand side of (\ref{eq:eindelijk}) does not explicitly
contain the gauge-dependent quantities $\varepsilon_\een$ and
$n_\een$.  With the help of expression (\ref{eq:equal-gi-non-gi}) one
can rewrite equation (\ref{eq:eindelijk}) in the
form~(\ref{eq:vondst2}).

The derivation of the coefficients (\ref{subeq:coeff})
from~(\ref{subeq:vierde}) and the proof of the
equality~(\ref{eq:check}) is straightforward, but extremely
complicated.  The computer algebra system
\textsc{Maxima}~\cite{maxima} has been used to perform this algebraic task.

\subsection{Derivation of the Evolution Equation for the Entropy
  Perturbation}

The basic set of equations (\ref{subeq:pertub-gi}) from which the
generalised cosmological perturbation theory is derived is of
fourth-order.  From this system a second-order equation
(\ref{eq:vondst2}) for $\varepsilon^{\text{gi}}_\een$ has been
extracted.  Therefore, the remaining system from which an evolution
equation for $n^{\text{gi}}_\een$ can be derived is at most of second
order.  Since gauge-invariant quantities
$\varepsilon^{\text{gi}}_\een$ and $n^{\text{gi}}_\een$ have been
used, one degree of freedom, namely the gauge function
$\xi^0(t,\bm{x})$ in (\ref{eq:trans-scalar}) has disappeared.  As a
consequence, only a first-order evolution equation for
$n^{\text{gi}}_\een$ can be derived.  Instead of deriving an equation
for $n^{\text{gi}}_\een$, an evolution equation (\ref{eq:vondst1}) for
the entropy perturbation, which contains $n^{\text{gi}}_\een$, will be
derived.

From the combined First and Second Law of Thermodynamics
(\ref{eq:thermo}) it follows that
\begin{equation}
  \label{eq:thermo-n0-e0}
  T_\nul s_\een=-\dfrac{\varepsilon_\nul(1+w)}{n^2_\nul}\left[n_\een-
       \dfrac{n_\nul}{\varepsilon_\nul(1+w)}\varepsilon_\een\right],
\end{equation}
where the right-hand side is gauge-invariant by virtue of
(\ref{eq:equal-gi-non-gi}), so that $s_\een=s^{\text{gi}}_\een$ is
gauge-invariant, in accordance with the remark below
(\ref{eq:thermo}).  Differentiating the term between square brackets in
(\ref{eq:thermo-n0-e0}) with respect to time and using the
background equations (\ref{FRW2}) and (\ref{FRW2a}), the first-order
equations (\ref{FRW4gi}) and (\ref{FRW4agi}) and the definitions
$w\equiv p_\nul/\varepsilon_\nul$ and
$\beta^2\equiv\dot{p}_\nul/\dot{\varepsilon}_\nul$, one finds
\begin{equation}
  \label{eq:vondst-gauge}
    \dfrac{1}{c}\dfrac{\text{d}}{\text{d}t}\left(n_\een -
  \frac{n_\nul}{\varepsilon_\nul(1+w)}\varepsilon_\een\right)=
    -3H\left(1-\frac{n_\nul
  p_n}{\varepsilon_\nul(1+w)}\right) \left(n_\een -
  \frac{n_\nul}{\varepsilon_\nul(1+w)}\varepsilon_\een\right),
\end{equation}
where \textsc{Maxima}~\cite{maxima} has been used to perform the
algebraic task.  By virtue of (\ref{eq:equal-gi-non-gi}), one may in
this equation replace $n_\een$ and $\varepsilon_\een$ by
$n^{\text{gi}}_\een$ and $\varepsilon^{\text{gi}}_\een$, respectively,
thus obtaining equation (\ref{eq:vondst1}).

\subsection{Evolution Equations for the Contrast Functions}
\label{app:contrast}

First the entropy equation (\ref{fir-ord}) will be derived. From the
definitions~(\ref{eq:contrast}) it follows that
\begin{equation}\label{eq:sgi-contrast}
  n_\een^{\text{gi}}-\frac{n_\nul}{\varepsilon_\nul(1+w)}\varepsilon_\een^{\text{gi}}=
       n_\nul\left(\delta_n-\dfrac{\delta_\varepsilon}{1+w}\right).
\end{equation}
Differentiating this expression with respect to $ct$ yields
\begin{equation}\label{eq:diff-sgi}
  \dfrac{1}{c}\dfrac{\text{d}}{\text{d}t}\left(n_\een^{\text{gi}} -
  \frac{n_\nul}{\varepsilon_\nul(1+w)}\varepsilon_\een^{\text{gi}}\right)=
   \dot{n}_\nul\left(\delta_n-\dfrac{\delta_\varepsilon}{1+w}
\right)+
  n_\nul\dfrac{1}{c}\dfrac{\text{d}}{\text{d} t}
  \left(\delta_n-\dfrac{\delta_\varepsilon}{1+w}\right).
\end{equation}
Using equations (\ref{FRW2a}) and (\ref{eq:vondst-gauge}), one arrives
at equation (\ref{fir-ord}) of the main text.

Finally, equation (\ref{sec-ord}) is derived. Upon substituting the
expressions
\begin{equation}
\label{subeq:afgeleiden}
\varepsilon^{\text{gi}}_\een  = \varepsilon_\nul\delta_\varepsilon, \qquad
\dot{\varepsilon}^{\text{gi}}_\een =\dot{\varepsilon}_\nul\delta_\varepsilon+
      \varepsilon_\nul\dot{\delta}_\varepsilon, \qquad
\ddot{\varepsilon}^{\text{gi}}_\een=\ddot{\varepsilon}_\nul\delta_\varepsilon+
      2\dot{\varepsilon}_\nul\dot{\delta}_\varepsilon+
      \varepsilon_\nul\ddot{\delta}_\varepsilon,
\end{equation}
into equation (\ref{eq:vondst2}), and dividing by $\varepsilon_\nul$,
one finds
\begin{equation}
\label{subeq:tilde-alpha}
 b_1=2\dfrac{\dot{\varepsilon}_\nul}{\varepsilon_\nul}+a_1, \qquad
 b_2=\dfrac{\ddot{\varepsilon}_\nul}{\varepsilon_\nul}+
      a_1\dfrac{\dot{\varepsilon}_\nul}{\varepsilon_\nul}+a_2, \qquad
 b_3=a_3\dfrac{n_\nul}{\varepsilon_\nul},
\end{equation}
where also (\ref{eq:sgi-contrast}) has been used. Using
\textsc{Maxima}, one arrives at the
coefficients~(\ref{subeq:coeff-contrast}) of the main text.

\newpage







\section*{Supplementary material}

# ☐ Relativistic Cosmological Perturbation Theory and the Evolution of Small-Scale Inhomogeneities

P. G. Miedema

Calculation of the coefficients (45) of the perturbation equations (43)
for delta_e and delta_n using the procedure described in the Appendix

Maxima 5.29.1    (http://maxima.sourceforge.net)

The Maxima file will be sent to the reader upon request:
pieter.miedema@gmail.com

++++++++++++++++++++++++++++++++++++++++++++++++++++++

Clean start of Maxima:

(%i94) reset()\$ kill(all)\$ ratfac:true\$

Inform Maxima that beta [in equation (28d)] depends on time, see
the remark below equations (A.2):

(%i2) depends(%beta,t);

(%o2) [β(t)]

(%i12) depends(%beta,t);

(%o10) $\dfrac{n0}{4\,H}$

(%i10) %alpha[2,4]:n0/(4*H);

(%i9) %alpha[2,3]:n0;

(%o9) n0

(%i3) 3 (pe+1) H + $\dfrac{\kappa\,e0\,(w+1)}{2\,H}$

(%i8) %alpha[2,2]:3*H;

(%o8) 3 H

(%i7) %alpha[2,1]:%kappa*n0/(2*H);

(%o7) $\dfrac{\kappa\,n0}{2\,H}$

---

# ☐ 1 Coefficients alpha[i,j] in Table I

Delta is the generalised Laplace operator and kappa=8*pi*G/c^4

(%i13) %alpha[1,1]:3*H+(1+pe)*%kappa*e0*(1+w)/(2*H);

(%o11) $\dfrac{\Delta\,pe}{a^2\,e0\,(w+1)}$

(%i11) %alpha[3,1]:pe*%Delta/(e0*(1+w)*a^2);

(%o12) $\dfrac{\Delta\,pn}{a^2\,e0\,(w+1)}$

(%i12) %alpha[3,2]:pn*%Delta/(e0*(1+w)*a^2);

(%o13) $(2-3\,\beta^2)\,H$

(%i13) %alpha[3,3]:H*(2-3*%beta^2);

(%o14) 0

(%i14) %alpha[3,4]:0;

(%o15) $\dfrac{\kappa\,R0}{3\,H}$

(%i15) %alpha[4,1]:%kappa*R0/(3*H);

(%o16) 0

(%i16) %alpha[4,2]:0;

(%o17) $-2\,\kappa\,e0\,(w+1)$

(%i17) %alpha[4,3]:-2*%kappa*e0*(1+w);

(%o18) $\dfrac{R0}{6\,H}+2\,H$

(%i18) %alpha[4,4]:2*H+R0/(6*H);

(%i4) %alpha[1,2]:3*H*pn;

(%i5) %alpha[1,3]:e0*(w+1);

(%o5) e0 (w+1)

(%i6) %alpha[1,4]:e0*(1+w)/(4*H);

(%o6) $\dfrac{e0\,(w+1)}{4\,H}$

(%i19) %alpha[5,1]:-R0/(R0+3*%kappa*e0*(1+w));

(%o19)
$$-\frac{R0}{R0+3\kappa e0(w+1)}$$

(%i20) %alpha[5,2]:0;

(%o20) 0

(%i21) %alpha[5,3]:6*e0*H*(1+w)/(R0+3*%kappa*e0*(1+w));

(%o21)
$$\frac{6\, e0(w+1)H}{R0+3\kappa e0(w+1)}$$

(%i22) %alpha[5,4]:3/2*e0*(1+w)/(R0+3*%kappa*e0*(1+w));

(%o22)
$$\frac{3\, e0(w+1)}{2(R0+3\kappa e0(w+1))}$$

## 2 Background equations (7)

(%i23) gradef(R0, t, -2*H*R0)$ diff(R0, t);

(%o24) $-2\,H\,R0$

(%i25) gradef(e0, t, -3*H*e0*(1+w))$ diff(e0, t);

(%o26) $-3\, e0(w+1)H$

(%i27) gradef(n0, t, -3*H*n0)$ diff(n0, t);

(%o28) $-3\, n0\, H$

Definition of the scale factor of the universe:

(%i29) gradef(a, t, a*H)$ diff(a, t);

(%o30) $a\, H$

Definition of the Hubble function H(t) via the time-derivative of (7a). Eliminating the time-derivatives of R0 and e0 with (7b) and (7c), respectively, one gets:

(%i31) gradef(H, t, -(1/6)*R0-(1/2)*%kappa*e0*(1+w))$ diff(H, t);

(%o32)
$$-\frac{R0}{6}-\frac{\kappa e0(w+1)}{2}$$

Time-derivative of w given by (44), see remark below equations (A.2):

(%i33) gradef(w, t, 3*H*(1+w)*(w-%beta^2))$ diff(w, t);

(%o34) $3(w+1)(w-\beta^2)H$

## 3 Definition of the background and perturbed pressure p(n,e)

Time-derivative of the background pressure (46):

(%i35) gradef(p0, t, pe*diff(e0,t)+pn*diff(n0,t))$ diff(p0, t);

(%o36) $-3\, e0\, pe(w+1)H-3\, n0\, pn\, H$

Time-derivatives of the partial derivatives of the pressure:

(%i37) depends([pee,pen,pnn], t); pne:pen$

(%o37) $[pee(t), pen(t), pnn(t)]$

(%i39) gradef(pe, t, pee*diff(e0,t)+pen*diff(n0,t))$ diff(pe, t);

(%o40) $-3\, e0\, pee(w+1)H-3\, n0\, pen\, H$

(%i41) gradef(pn, t, pen*diff(e0,t)+pnn*diff(n0,t))$ diff(pn, t);

(%o42) $-3\, e0\, pen(w+1)H-3\, n0\, pnn\, H$

## 4 Calculation of the coefficients (45) of equation (43a)

Expressions (A.5):

(%i43) for i:1 step 1 thru 4 do p[i]:diff(%alpha[5,i], t)-%alpha[5,1]*%alpha[1,i]-%alpha[5,2]*%alpha[2,i]-%alpha[5,3]*%alpha[3,i]-%alpha[5,4]*%alpha[4,i]$

Expressions (A.8):

(%i44) for i:1 thru 5 do q[i]:-%alpha[i,4]/p[4]$

(%i45) for i:1 step 1 thru 5 do for j:1 step 1 thru 4 do %gamma[i,j]:%alpha[i,j]+q[i]*p[j]$

Expressions (A.10):

(%i46) for i:1 step 1 thru 3 do s[i]:diff(%gamma[5,i], t)-
%gamma[5,1]*%gamma[1,i]-%gamma[5,2]*%gamma[2,i]-
%gamma[5,3]*%gamma[3,i]$

(%i47) r:1+diff(q[5], t)-%gamma[5,1]*q[1]-
%gamma[5,2]*q[2]-%gamma[5,3]*q[3]$

Coefficients a1, a2 and a3 given by (A.14):

(%i48) a1:ratsimp(-s[3]/%gamma[5,3]+r/q[5])$

(%i49) a2:ratsimp(-s[3]/(%gamma[5,3]*q[5]))$

(%i50) a3:ratsimp(%gamma[5,2]*s[3]/(%gamma[5,3]*q[5])-s[2]/q[5])$

Check of the identity (A.15):

(%i51) A15:ratsimp((%gamma[5,1]*s[3]-%gamma[5,3]*s[1])/
(%gamma[5,2]*s[3]-%gamma[5,3]*s[2]))$

Substitute the definition of beta:

(%i52) A15_1:ratsimp(subst(sqrt(diff(p0,t)/diff(e0,t)), %beta, A15))$

Perform differentiation:

(%i53) A15_2:ratsimp(ev(A15_1, diff))$

Substitute the definition of w:

(%i54) A15_3:ratsimp(subst(p0/e0, w, A15_2))$

Substitute the definition of beta:

(%i55) A15_4:ratsimp(subst(sqrt(diff(p0,t)/diff(e0,t)), %beta, A15_3))$

Finally, substitute the definition of w:

(%i56) A15_5:ratsimp(subst(p0/e0, w, A15_4));

$$(\%o56) \quad -\frac{n0}{p0+e0}$$

Coefficients b1, b2 and b3 given by (A.22).

After calculating the coefficients, the Maxima expressions are made manageable by replacing the term R0+3*kappa*e0*(1+w) by N:

(%i57) b1:2*diff(e0, t)/e0+a1$
ratexpand(ratsubst(N, R0+3*%kappa*e0*(1+w), b1));

$$(\%o58) \quad \frac{3\,H}{N} - \frac{18\,\beta^2\,\kappa\,e0\,w\,H}{N} - \frac{6\,\kappa\,e0\,H}{N} - 6\,w\,H + 3\,\beta^2$$

$$H - \frac{2\left(\frac{d}{d\,t}\beta\right)}{\beta}$$

(%i59) b2:ratsimp(diff(e0, t, 2)/e0+a1*diff(e0, t)/e0+a2$
ratexpand(ratsubst(N, R0+3*%kappa*e0*(1+w), b2));

$$(\%o60) \quad \frac{6\left(\frac{d}{d\,t}\beta\right)\kappa\,e0\,w\,H^2}{w\,N+N} - \frac{24\,\kappa\,e0\,w\,H^2}{w\,N+N} + \frac{12\left(\frac{d}{d\,t}\beta\right)\kappa\,e0\,w\,H}{\beta\,w\,N+\beta\,N}$$

$$\frac{72\,\beta^2\,\kappa\,e0\,w\,H^2}{w\,N+N} - \frac{18\,\kappa\,e0\,w^2\,H^2}{w\,N+N} - \frac{117\,\beta^2\,\kappa\,e0\,w\,H^2}{\beta\,w\,N+\beta\,N} + \frac{39\,\kappa\,e0\,w\,H^2}{\beta\,w\,N+\beta\,N} +$$

$$\frac{54\,\beta^2\,\kappa\,e0\,w\,H^2}{\beta\,w\,N+\beta\,N} + \frac{9\,\beta^2\,H^2}{w+1} + \frac{9\,\beta^2\,H^2}{w+1} + \frac{2\,e0\,w^2\,e0}{n0\,pn\,N} + \frac{6\left(\frac{d}{d\,t}\beta\right)}{\beta\,w^2\,H}$$

$$\frac{pe\,w\,N}{2\,w+2} - \frac{w\,N}{2\,w+2} + \frac{pe\,N}{w+1} - \frac{9\,\beta^2\,H^2}{w+1} + \frac{9\,\beta^2\,H^2}{w+1} + \frac{9\,\kappa\,e0\,H^2}{2\,w+2} - \frac{w^2}{\kappa\,e0\,w^2} +$$

$$\frac{6\left(\frac{d}{d\,t}\beta\right)w\,H}{2\,w+2} + \frac{a^2\,e0\,w+a^2\,e0}{\Delta\,n0\,pn} + \frac{\Delta\,pe\,w}{2\,w+2} + \frac{3\,\kappa\,e0\,pe\,w^2}{2\,w+2} - \frac{w^2}{2\,w+2} -$$

$$\frac{3\,\kappa\,n0\,pn\,w}{2\,w+2} + \frac{a^2\,e0}{2\,w+2} - \frac{3\,\kappa\,n0\,pe}{2\,w+2} + \frac{\Delta\,pe}{w+1} + \frac{a^2\,w+a^2}{w+1} - \frac{3\,\kappa\,e0\,pe\,w}{w-1} +$$

$$\frac{3\,\kappa\,n0\,pn\,w}{2\,w+2} + \frac{a^2\,e0}{2\,w+2} + \frac{3\,\kappa\,pe\,w}{2\,w+2} - \frac{\kappa\,e0\,w}{w-1}$$

(%i61) b3:a3*n0/e0$
       ratexpand(ratsubst(N, R0+3*%kappa*e0*(1+w), b3));

(%o62)

$$\frac{n0\,pn\,N}{2\,e0} - \frac{27\,\kappa\,e0\,n0\,pen\,w\,H^2}{N} - \frac{27\,\kappa\,n0^2\,pmn\,w\,H^2}{N} - \frac{27\,\kappa\,n0\,pe\,pn\,w\,H^2}{N} +$$

$$\frac{18\,\beta^2\,\Delta\,n0\,pn\,w\,H^2}{N} - \frac{27\,\beta^2\,\kappa\,n0\,pen\,w\,H^2}{N} - \frac{54\,\kappa\,n0^2\,pmn\,w\,H^2}{N} - \frac{18\,\Delta\,n0\,pe\,pn\,H^2}{N} -$$

$$\frac{18\,\Delta\,n0\,pen\,H^2}{N} + \frac{18\,\Delta\,n0^2\,pmn\,H^2}{N} - \frac{27\,\kappa\,n0^2\,pmn\,H^2}{N} - \frac{27\,\beta^2\,\kappa\,n0\,pen\,H^2}{N} +$$

$$\frac{18\,\frac{\left(\frac{d}{dt}\beta\right)}{\beta}\kappa\,n0\,pn\,w\,H^2}{N} + \frac{12\,\frac{\left(\frac{d}{dt}\beta\right)}{\beta}\kappa\,\Delta\,n0\,pn\,H^2}{N} + \frac{18\,\frac{\left(\frac{d}{dt}\beta\right)}{\beta}\kappa\,n0\,pen\,H^2}{a^2\,N} +$$

$$\frac{9\,n0^2\,pnn\,w\,H^2}{\beta\,N} - \frac{9\,\beta^2\,e0\,N}{e0} - \frac{9\,n0\,pn\,w\,H^2}{a^2\,N} + \frac{18\,\frac{\left(\frac{d}{dt}\beta\right)}{\beta}\kappa\,n0\,pn\,H}{a^2\,N} +$$

$$\frac{6\,\left(\frac{d}{dt}\beta\right)n0\,pn\,H}{a^2\,e0} + \frac{9\,n0\,pen\,w\,H^2}{e0} + \frac{\Delta\,n0\,pn}{a^2\,e0} -$$

$$\frac{3\,\kappa\,n0\,pn}{2} - \frac{3\,\kappa\,n0\,pn\,w}{2}$$

(%i64) ngi:(n1-3*n0*(%kappa*e1+2*H*%theta+
       1/2*R1))/(R0+3*%kappa*e0*(1+w));

(%o64)

$$n1 - \frac{3\,n0\left(\frac{R1}{2} + 2\,\theta\,H + \kappa\,e1\right)}{R0 + 3\,\kappa\,e0\,(w+1)}$$

---

# ☐ 5 Check of the coefficients b1, b2 and b3

In the present file the above calculated coefficients will now be checked.

The above calculated coefficients should be recast by hand to obtain expressions (45), using N=R0+3*%kappa*e0*(1+w). The thus-obtained coefficients are checked in a separate file.

Expressions (41):

(%i63) egi:(e1*R0-3*e0*(1+w)*(2*H*%theta+
       1/2*R1))/(R0+3*%kappa*e0*(1+w)):

(%o63)

$$\frac{e1\,R0 - 3\,\kappa\,e0\,(w+1)\left(\frac{R1}{2} + 2\,\theta\,H\right)}{R0 + 3\,\kappa\,e0\,(w+1)}$$

Expressions (42):

(%i65) delta_e:egi/e0;

(%o65)

$$\frac{e1\,R0 - 3\,\kappa\,e0\,(w+1)\left(\frac{R1}{2} + 2\,\theta\,H\right)}{e0\left(R0 + 3\,\kappa\,e0\,(w+1)\right)}$$

(%i66) delta_n:ngi/n0;

(%o66)

$$\frac{n1 - \dfrac{3\,n0\left(\frac{R1}{2} + 2\,\theta\,H + \kappa\,e1\right)}{R0 + 3\,\kappa\,e0\,(w+1)}}{n0}$$

Equations (40), rewritten in the form (A.3a)-(A.3d):

(%i67) gradef(e1, t, -%alpha[1,1]*e1-%alpha[1,2]*n1-
       %alpha[1,3]*%theta-%alpha[1,4]*R1)$

(%i68)

$$-\frac{e0\,(w+1)\,R1}{4\,H} + e1\left(-3\,(pe+1)\,H - \frac{\kappa\,e0\,(w+1)}{2\,H}\right) - 3\,n1\,pn\,H - \theta\,e0$$

(%o68)

(%i69) gradef(n1, t, -%alpha[2,1]*e1-%alpha[2,2]*n1-
       %alpha[2,3]*%theta-%alpha[2,4]*R1)$

(%i70)

$$-\frac{n0\,R1}{4\,H} - 3\,n1\,H - \frac{\kappa\,e1\,n0}{2\,H} - \theta\,n0$$

(%i71) gradef(%theta, t, -%alpha[3,1]*e1-%alpha[3,2]*n1-
       %alpha[3,3]*%theta-%alpha[3,4]*R1)$

(%i72)

$$-(2-3\,\beta^2)\,\theta\,H - \frac{\Delta\,n1\,pn}{a^2\,e0} - \frac{\Delta\,e1\,pe}{a^2\,e0\,(w+1)}$$

(%i73) gradef(R1, t, _%alpha[4,1]*e1-%alpha[4,2]*n1-
%alpha[4,3]*%theta-%alpha[4,4]*R1)$
diff(R1,t);

$$(\%o74)\quad \left(-\frac{R0}{6\,H}-2\,H\right)R1-\frac{\kappa\,e1\,R0}{3\,H}+2\,\kappa\,\theta\,e0\,(w+1)$$

Check of equation (43b) [left-hand side minus right-hand side]:

(%i75) eq_43b:ratexpand(diff(delta_n-delta_e/(1+w), t)-
3*H*n0*pn/(e0^(1+w))*(delta_n-delta_e/(1+w)))$

(%o84) 0

Substitute the definition of beta:

(%i76) eq_43b_1:subst(sqrt(diff(p0,t)/diff(e0,t)), %beta, eq_43b)$

Finally, substitute the definition of w:

(%i77) ratsimp(subst(p0/e0, w, eq_43a_1));

(%o77) 0

Check of equation (43a) [left-hand side minus right-hand side]:

(%i78) eq_43a:ratsimp(diff(diff(delta_e,t,2) + b1*diff(delta_e,t) +
b2*delta_e - b3 * (delta_n-delta_e/(1+w)))$

Substitute the definition of beta:

(%i79) eq_43a_1:ratsimp(subst(sqrt(diff(p0,t)/diff(e0,t)), %beta, eq_43a))$

Substitute the definition of w:

(%i80) eq_43a_2:ratsimp(subst(p0/e0, w, eq_43a_1))$

Perform differentiation:

(%i81) eq_43a_3:ratsimp(ev(eq_43a_2, diff))$

Finally, substitute the definition of w:

(%i82) ratsimp(subst(p0/e0, w, eq_43a_3));

(%o82) 0

Check of (A.16):

(%i83) ratsimp((n1-n0/(e0^(1+w))*e1)-(ngi-n0/(e0^(1+w))*egi));

(%o83) 0

Check of equation (A.13) [left-hand side minus right-hand side]:

(%i84) ratsimp(diff(egi,t,2)+a1*diff(egi,t)+a2*egi-
a3*(ni+(%gamma[5,1]*s[3]-%gamma[5,3]*s[1])/(%gamma[5,2]*s[3-
%gamma[5,3]*s[2])*e1));

(%o84) 0

Replacing in (A.13) n1 by ngi and e1 by egi, see (A.16),
yields the same result:

(%i85) A13:ratsimp(diff(egi,t,2)+a1*diff(egi,t)+a2*egi-
a3*(ngi+(%gamma[5,1]*s[3]-%gamma[5,3]*s[1])/(%gamma[5,2]*s[3]-
%gamma[5,3]*s[2])*egi))$

(%i86) A13_1:subst(sqrt(diff(p0,t)/diff(e0,t)), %beta, A13)$

(%i87) A13_2:ratsimp(ev(A13_1, diff))$

(%i88) A13_3:ratsimp(subst(p0/e0,w,A13_2))$

(%i89) A13_4:ratsimp(subst(sqrt(diff(p0,t)/diff(e0,t)), %beta, A13_3))$

(%i90) A13_5:ratsimp(subst(p0/e0,w,A13_4));

(%o90) 0

Proof of equation (A.18):

(%i91) diabatic:n1-n0*e1/(e0^(1+w));

$$(\%o91)\quad n1-\frac{e1\,n0}{e0\,(w+1)}$$

Substitute the definition of beta:

(%i92) factor_A18:ratsimp(diff(diabatic,t)/diabatic);

(%i93) factor_A18_1:factor(subst
(sqrt(diff(p0,t)/diff(e0,t)),%beta,factor_A18));

$$(\%o93)\quad -\frac{3\,(e0\,w-n0\,pn+e0)\,H}{e0\,(w+1)}$$

# Relativistic Cosmological Perturbation Theory and the Evolution of Small-Scale Inhomogeneities

P. G. Miedema

Consistency check of equations (43) with coefficients (45)

Maxima 5.29.1 (http://maxima.sourceforge.net)

The Maxima file will be sent to the reader upon request:
pieter.miedema@gmail.com

++++++++++++++++++++++++++++++++++++++++++++++++++++++++++++++++++++++++++++

Clean start of Maxima:

(%i48) reset()$ kill(all)$ ratfac:true$

Inform Maxima that beta [in equation (28d)] and w [in equation (7c)] are time-dependent quantities:

(%i2) depends([%beta,w],t);

(%o2) [β(t), w(t)]

## 1 Coefficients (45)

Delta is the generalised Laplace operator and kappa=8*pi*G/c^4

(%i3) b1:%kappa*e0*(1+w)/H-2*diff(%beta,t)/%beta-
H*(2+6*w-3*%beta^2)+R0*(1/(3*H)+
2*H*(1+3*%beta^2)/(R0+3*%kappa*e0*(1+w));

(%o3) $R0 \left( \frac{2\left(3\,\beta^2+1\right)H}{R0+3\,\kappa\,e0\,(w+1)} + \frac{1}{3\,H} \right) - \left(6\,w+3\,\beta^2+2\right)H + \frac{\kappa\,e0\,(w+1)}{H} - \frac{2\left(\frac{d}{dt}\,\beta\right)}{\beta}$

(%i4) b2:-1/2*%kappa*e0*(1+w)*(1+3*w)+2*H^2*(1-3*w+
6*%beta^2*w*(2+3*w)+6*H*H*diff(%beta,t)/%beta*(w+
%kappa*e0*(1+w)/(R0+3*%kappa*e0*(1+w))-
R0*(1/2*w+H^2*(1-6*w)*(1+3*%beta^2)/(R0+
3*%kappa*e0*(1+w)))-%beta^2*H^2*(1-1/2*R0));

(%o4) $-R0 \left( \frac{\left(3\,\beta^2+1\right)\left(6\,w+1\right)H^2}{R0+3\,\kappa\,e0\,(w+1)} + \frac{w}{2} \right) + \left(6\,\beta^2\left(3\,w+2\right)w - 3\,w+1\right)H^2 - \frac{\kappa\,e0\,(w+1)\left(3\,w+1\right)}{2}$

(%i5) b3:-18*H^2*%*(e0*pen*(1+w)+
2*pn*diff(%beta,t)/%beta/(3*H)+pn*(pe-%beta^2)+
n0*pnn)/(R0+3*%kappa*e0*(1+w))+
pn)*(%Delta/a^2-1/2*R0)  *(n0/e0);

(%o5) $e0 \left( pn - \frac{18\left(\frac{2\left(\frac{d}{dt}\,\beta\right)pn}{3\,\beta\,H} + e0\,pen\,(w+1) + n0\,pnn + \left(pe - \beta^2\right)pn\right)H^2}{R0+3\,\kappa\,e0\,(w+1)} \right) \frac{\frac{\Delta}{a^2} - \frac{R0}{2}}{e0}$

## 2 Background equations (7)

Definition of the scale factor of the universe:

(%i6) gradef(a, t, H*a)$

(%o7) $a\,H$

Definition of the Hubble function H(t) via the time-derivative of R0 and e0 with (7b) and (7c), respectively, one gets:

Eliminating the time-derivatives of (7a).

(%i8) gradef(H, t, -1/6*R0-1/2*%kappa*e0*(1+w))$
diff(H,t);

(%o9) $-\frac{R0}{6} - \frac{\kappa\,e0\,(w+1)}{2}$

(%i10) gradef(R0, t, -2*H*R0)$
diff(R0,t);

(%o11) $-2\,H\,R0$

## 3 Definition of the background and perturbed pressure p(n,e)

(%i12) grade f(e0, t, -3*H*e0*(1+w))$
diff(e0,t);

(%o13) $-3\,e0\,(w+1)\,H$

(%i14) grade f(n0, t, -3*H*n0)$
diff(n0,t);

(%o15) $-3\,n0\,H$

Time-derivative of the background pressure (46):

(%i16) grade f(p0, t, pn*diff(n0,t)+pe*diff(e0,t))$
diff(p0,t);

(%o17) $-3\,e0\,(w+1)\,H-3\,n0\,pn\,H$

First-order pressure perturbation (11):

(%i18) p1:pe*e1+pn*n1;
(%o18) $n1\,pn+e1\,pe$

Time-derivatives of the partial derivatives of the perturbed pressure:

(%i19) depends([pee,pen,pne,pnn],t); pne:pen$
[ pee(t), pen(t), pne(t), pnn(t)]

(%i21) grade f(pe, t, pen*diff(n0,t)+pee*diff(e0,t))$
diff(pe,t);

(%o22) $-3\,e0\,pee\,(w+1)\,H-3\,n0\,pen\,H$

(%i23) grade f(pn, t, pnn*diff(n0,t)+pen*diff(e0,t))$
diff(pn,t);

(%o24) $-3\,e0\,pen\,(w+1)\,H-3\,n0\,pnn\,H$

## 4 Perturbation equations (40)

(%i25) grade f(e1, t, -3*H*(e1+p1)-e0*(1+w)*(%theta+
(%kappa*e1+1/2*R1)/(2*H)))$
diff(e1,t);

(%o26) $-e0\,(w+1)\left(\dfrac{\frac{R1}{2}+\kappa\,e1}{2\,H}+\theta\right)-3\,(n1\,pn+e1\,pe)\,H$

(%i27) grade f(n1, t, -3*H*n1-n0*(%theta+(%kappa*e1+
1/2*R1)/(2*H)))$
diff(n1,t);

(%o28) $-n0\left(\dfrac{\frac{R1}{2}+\kappa\,e1}{2\,H}+\theta\right)-3\,n1\,H$

(%i29) grade f(%theta, t, -H*(2-3*%beta^2)*%theta-
%Delta/a^2*p1/(%kappa*e0*(1+w)))$
diff(%theta,t);

(%o30) $-(2-3\,\beta^2)\,\theta\,H-\dfrac{\Delta\,(n1\,pn+e1\,pe)}{a^2\,e0\,(w+1)}$

(%i31) grade f(R1, t, -2*H*R1+2*%kappa*e0*(1+w)*%theta-
R0/(3*H)*(%kappa*e1+1/2*R1))$
diff(R1,t);

(%o32) $-2\,H\,R1-\dfrac{R0\left(\frac{R1}{2}+\kappa\,e1\right)}{3\,H}+2\,\kappa\,\theta\,e0\,(w+1)$

## 5 Gauge-invariant quantities (41)

(%i33) egi:(e1*R0-3*e0*(1+w)*(2*H*%theta+
1/2*R1))/(R0+3*%kappa*e0*(1+w));

(%o33) $\dfrac{e1\,R0-3\,e0\,(w+1)\left(\frac{R1}{2}+2\,\theta\,H\right)}{R0+3\,\kappa\,e0\,(w+1)}$

(%i34) ngi:n1-3*n0*(%kappa*e1+2*H*%theta+
1/2*R1)/(R0+3*%kappa*e0*(1+w));

(%o34) $n1-\dfrac{3\,n0\left(\frac{R1}{2}+2\,\theta\,H+\kappa\,e1\right)}{R0+3\,\kappa\,e0\,(w+1)}$

## 6 Contrast functions (42)

(%i35) delta_e:egi/e0;

(%o35) $\dfrac{e1\,R0-3\,e0\,(w+1)\left(\frac{R1}{2}+2\,\theta\,H\right)}{e0\,(R0+3\,\kappa\,e0\,(w+1))}$

(%i36) `delta_n:ngi/n0;`

(%o36)
$$n1 - \frac{3\,n0\left(\frac{R1}{2}+2\,\theta\,H+\kappa\,e1\right)}{R0+3\,\kappa\,e0\,(w+1)}$$
$$n0$$

## 7 Consistency check of equation (43b)

Equation (43b) [left-hand side minus right-hand side]:

(%i37) `eq_43b:ratsimp(diff(delta_n-delta_e/(1+w),t)-` `3*H*n0*pn/(e0*(1+w))*` `(delta_n-delta_e/(1+w)));`

(%o37)
$$\frac{e1\left(\left(3\,e0\,w^2+(3\,e0-3\,e0\,pe)\,w-3\,n0\,pn-3\,e0\,pe\right)H-e0\left(\frac{d}{dt}w\right)\right)}{e0^2\,(w+1)^2}$$

Substitute the definition of w:

(%i38) `eq_43b_1:ratsimp(subst(p0/e0, w, eq_43b));`

(%o38)
$$\frac{e1\left(\left(3\,e0\,n0\,pn+(3\,e0\,p0+3\,e0^2)\,pe-3\,p0^2-3\,e0\,p0\right)H+e0^2\left(\frac{d}{dt}\frac{p0}{e0}\right)\right)}{e0\,(p0+e0)^2}$$

Perform differentiation:

(%i39) `eq_43b_2:ratsimp(ev(eq_43b_1,diff));`

(%o39)
$$\frac{3\,e1\,(e0\,pe-p0)\,H}{e0\,(p0+e0)^2}$$

Finally, substitute the definition of w:

(%i40) `ratsimp(subst(p0/e0, w, eq_43b_2));`

(%o40) 0

## 8 Consistency check of equation (43a)

Equation (43a) [left-hand side minus right-hand side]:

(%i41) `eq_43a:ratsimp(diff(delta_e,t,2)+b1*diff(delta_e,t)+` `b2*delta_e-b3*(delta_n-delta_e/(1+w)))$`

(%i42) `eq_43a_1:ratsimp(subst(sqrt(diff(p0,t)/diff(e0,t)), %beta, eq_43a))$`

Substitute the definition of beta:

(%i43) `eq_43a_2:ratsimp(subst(p0/e0, w, eq_43a_1))$`

Perform differentiation:

(%i44) `eq_43a_3:ratsimp(ev(eq_43a_2,diff))$`

Substitute the definition of w:

(%i45) `eq_43a_4:ratsimp(subst(p0/e0, w, eq_43a_3));`

(%o45)
$$(9\,(e0\,pe-p0))\left(\left(3\,e0\,pn+(3\,e0\,p0+3\,e0^2)\,pe-3\,p0^2-3\,e0\,p0\right)H+e0^2\left(\frac{d}{dt}\frac{p0}{e0}\right)\right)R0\Big/(R1+4\,\theta\,H+2\,\kappa\,e1))/(2\,e0^3\,(R0+3\,\kappa\,p0+3\,\kappa\,e0)^2)$$

Perform differentiation:

(%i46) `eq_43a_5:ratsimp(ev(eq_43a_4,diff));`

(%o46)
$$\frac{27\,(e0\,pe-p0)^2\,(e0\,w-p0)\,H^2\,R0\,(R1+4\,\theta\,H+2\,\kappa\,e1)}{2\,e0^3\,(R0+3\,\kappa\,p0+3\,\kappa\,e0)^2}$$

Finally, substitute the definition of w:

(%i47) `ratsimp(subst(p0/e0, w, eq_43a_5));`

(%o47) 0

```r
###############################################################
#
#   Relativistic Cosmological Perturbation Theory and
#   the Evolution of Small-Scale Inhomogeneities
#
#   P.G. Miedema
#
#   Program to calculate figure 1 in the main text
#   The R file will be send to the reader upon request:
#   pieter_miedema@gmail.com
#
#   The R Project for Statistical Computing: http://www.r-project.org
#
###############################################################

library(deSolve)  # load package "deSolve" to use the solver "lsodar" at line 110

m <- 1.6726e-27  # proton mass in kg
c <- 2.9979e8    # speed of light in m/s
parsec <- 3.0857e16  # parsec (pc) in m
K_B <- 1.3806e-23  # Boltzmann's constant in J/K
T_gamma <- 2.725  # present value of the background radiation temperature in K
H_0 <- ...  # present value of the Hubble parameter in km/s/Mpc
H_0sec <- H_0 * 1000 / (parsec * 1e6)  # present value of the Hubble parameter in 1/s
H_m <- H_0sec / c  # present value of the Hubble parameter in 1/m
H_pc <- H_m * parsec  # present value of the Hubble parameter in 1/pc
z_LP <- ...  # present value of ...
z_dec <- 1090.43  # redshift at decoupling
tau_dec <- ...  # value of dimensionless time tau at decoupling
tau_0 <- ...  # value of dimensionless time tau at 13.82 Gyr, end of integration
t_dec <- ...  # time of decoupling
factor <- 2*pi/(z_dec+1) ...

###############################################################

equation.86 <- function (tau, y, parms)
{
  ydot <- vector (len=2)
  aux <- H_pc^2/tau^(8/3)
  ydot[1] <- y[2]
  ydot[2] <- (2/tau)*y[2] - (4/9) * aux - (10/9)/tau^2) * y[1] - (4/15) * aux * delta_T
  return(list(ydot))
}

stop.conditions <- function (tau, y, parms)
{
  stop <- vector (len=2)
  stop[1] <- y[1] - delta1        # delta1
  stop[2] <- tau - tau_0          # z=0
  return(stop)
}

###############################################################

pdf(file="fig1.pdf", family="Times")  # open a plotfile in pdf-format

par(mar=c(3.2,3.2,4), cex=1.2, cex.axis=1.2, pty="s")
plot.new()
plot.window(xlim=c(0, 50), ylim=c(0, 24))
title(main=expression(paste("Star Formation starting at ", z==1090)),
      cex.main = 1.4, font.main=1, col.main="black", line=1.0)

pc <- seq(0,50,by=10)
axis(1, las=1, at=pc, tick=TRUE, label=pc, tcl=0.4, mgp=c(2, 0.3, 0))
tussen <- seq(5,45,by=10)
axis(1, las=1, at=tussen, tick=TRUE, label=FALSE, tcl=0.25, mgp=c(2, 0.3, 0))
eenheden <- seq(1,50,by=1)
axis(1, las=1, at=eenheden, tick=TRUE, label=FALSE, tcl=0.15, mgp=c(2, 0.3, 0))
mtext("Perturbation Scale (parsec) at Decoupling", cex=1.6, side=1, line=1.5)

zt <- seq(0, 24, by=2)
axis(2, at=zt, labels=TRUE, las=1, tcl=0.4, mgp=c(2, 0.3, 0))
mtext("Cosmological Redshift", cex=1.6, side=2, line=1.7)

axis(4, at=zt, labels=round(t_dec * (z_dec+1)/(zt+1))^(1.5, 2),
     las=2, tcl=0.4, mgp=c(2, 0.3, 0))
mtext("Time in Gyr", cex=1.6, side=4, line=2.5)
box()

# perturbations with scales outside the interval [0.5, 60] parsec do not become
# non-linear within 13.82 Gyr:
# scale_min < 0.5; scale_max <- 60; increment <- 0.01
# initially the increment should be small, since the line is steep:
# range <- length(seq(log10(scale_min), log10(scale_max), increment))
Jeans.scale <- vector()
for (k in 1:11)
{
  z <- vector(); lambda.nonlin <- vector()
  tau.end <- result[2,1]; delta <- result[2,2]
  for (lambda.dec in range.lambda.dec)
  {
    mu_m <- factor/lambda.dec  # see (94)
    y <- c(delta_e, dot.delta_e)  # initial values at tau_dec (start of integration)
    result <- lsodar(y, tau.start:tau.end, fun=equation.86, rootfun=stop.conditions, parms)
    # Only the end values, i.e., result[2,...], are needed:
    tau.end <- result[2,1]; delta <- result[2,2]
    if (round(delta)==1)
    {
      i <- i+1
      lambda.nonlin[i] <- lambda.dec
      z[i] <- (z_dec+1) / tau.end^(2/3)-1.0
    }
  }

  z_max <- max(z)
  lambda.nonlin_max <- lambda.nonlin[z==z_max]; Jeans.scale[k] <- lambda.nonlin_max

  if (k==1)  text(lambda.nonlin_max, z_max, "-0.005", adj=c(0.5,-0.15))
  if (k==2)  text(lambda.nonlin_max, z_max, "-0.01",  adj=c(0.5,-0.15))
  if (k==3)  text(lambda.nonlin_max, z_max, "-0.02",  adj=c(0.5,-0.15))
  if (k==4)  text(lambda.nonlin_max, z_max, "-0.03",  adj=c(0.5,-0.15))
  if (k==5)  text(lambda.nonlin_max, z_max, "-0.04",  adj=c(0.5,-0.15))
  if (k==6)  text(lambda.nonlin_max, z_max, "-0.05",  adj=c(0.5,-0.15))
  if (k==7)  text(lambda.nonlin_max, z_max, "-0.06",  adj=c(0.5,-0.15))
  if (k==8)  text(lambda.nonlin_max, z_max, "-0.07",  adj=c(0.5,-0.15))
  if (k==9)  text(lambda.nonlin_max, z_max, "-0.08",  adj=c(0.5,-0.15))
  if (k==10) text(lambda.nonlin_max, z_max, "-0.09",  adj=c(0.5,-0.15))
  if (k==11) text(lambda.nonlin_max, z_max, "-0.10",  adj=c(0.5,-0.15))

  points(z ~ lambda.nonlin, type="l")
}

dev.off()  # close the plotfile

###############################################################
# Calculation of the Jeans mass expressed in sun's mass:
# a_B <- ...  # redshift at matter-radiation equality
# m_Sun <- 1.9889e30  # sun's mass in kg
m_J <- ...  # Jeans scale in pc
JS <- mean(Jeans.scale)  # mean Jeans scale in pc
M_J <- (1/6)*pi^1.5*parsec^3*a_B*T_gamma^4/(c^2*(z_eq+1)^2*(z_dec+1)^3 / m_sun  # (104)
```



\end{document}